\documentclass[notitlepage,openbib,12pt]{article}
\UseRawInputEncoding
\usepackage{amsfonts}
\usepackage{amsmath}
\usepackage{amssymb}
\usepackage{graphicx}
\usepackage{color}
\usepackage[colorlinks,citecolor=blue]{hyperref}

\usepackage{mathpazo}
\usepackage{amssymb}

\newtheorem{theo}{Theorem}
\newtheorem{prop}{Proposition}

\newtheorem{define}{Definition}

\newtheorem{lemma}{Lemma}

\makeindex
\input epsf

\begin{document}

\title{\textbf{Mechanism Design with Sequential-Move Games: Revelation
Principle}}
\author{Siyang Xiong\thanks{
Department of Economics, University of California, Riverside, United States,
siyang.xiong@ucr.edu}}
\maketitle

\begin{abstract}
Traditionally, mechanism design focuses on simultaneous-move games (e.g., 
\cite{rm1981}). In this paper, we study mechanism design with
sequential-move games, and provide two results on revelation principles for
general solution concepts (e.g., perfect Bayesian equilibrium, obvious
dominance, strong-obvious dominance). First, if a solution concept is
additive, implementation in sequential-move games is equivalent to
implementation in simultaneous-move games. Second, for any solution concept $%
\rho $ and any social choice function $f$, we identify a canonical operator $%
\gamma ^{\left( \rho ,f\right) }$, which is defined on primitives. We prove
that, if $\rho $ is monotonic, $f$ can be implemented by a sequential-move
game if and only if $\gamma ^{\left( \rho ,f\right) }$ is achievable, which
translates a complicated mechanism design problem into checking some
conditions defined on primitives. Most of the existing solution concepts are
either additive or monotonic.
\end{abstract}

\newpage

\baselineskip= 20pt

\section{Introduction}

\label{sec:intro}

Traditional mechanism design theory imposes an implicit assumption: only a
simultaneous-move mechanism can be adopted. For instance, \cite{rm1981} is
the first paper that finds the optimal simultaneous-bid mechanism for a
general auction setup.

However, sequential-move mechanisms are adopted in many of our usual
practices, e.g., English auctions, Dutch auctions, run-off elections, FIFA
world cup bids. This immediately begs the question: does the traditional
mechanism design theory (which focuses on simultaneous-move mechanisms)
suffer loss of generality? Specifically, 
\begin{equation}
\left( 
\begin{array}{c}
\text{does the optimal simultaneous-bid mechanism in \cite{rm1981} remain
the optimal one} \\ 
\text{when we allow for sequential-bid mechanisms?}%
\end{array}%
\right)  \label{aab1}
\end{equation}%
In sequential-bid mechanisms, payoff-relevant information is partially
disclosed at each round of bidding, e.g., we may disclose all of buyers'
bids in previous rounds, or we may disclose some or none of them. Do these
different rules change bidders' strategic behaviors and the final outcome?

Another critical dimension is solution concept. We may adopt different
solution concepts for different problems. For instance, we usually adopt
Bayesian Nash equilibrium in auction setups (e.g., \cite{rm1981}), while we
adopt weak dominance (or equivalently, strategyproofness) in matching
problems (e.g., the deferred acceptance mechanism in \cite{dgls1962}).%
\footnote{%
The cooperative game-theory approach is usually adopted in matching
problems, and non-cooperative solution concepts (e.g., Bayesian Nash
equilibrium) do not apply to such problems.} Recently, \cite{sli2017} and 
\cite{mppt2023} propose two new solution concepts: obvious dominance and
strong-obvious dominance, and they prove that, for these solution concepts,
implementation by sequential-move mechanisms differs substantially from
implementation by simultaneous-move mechanisms. This result implies that (%
\ref{aab1}) is a non-trivial question, and begs an answer for it.

Interestingly, when we compare implementation in weak dominance to
implementation in obvious dominance, some authors (e.g., \cite{iayg2018})
define the previous one on\ simultaneous-move games only, while define the
latter one on sequential-move games. Does this suffer loss of generality? If
no, in what sense?

We divide solution concepts into two groups: Category I and Category II.
Category I includes all of the solution concepts on which implementation by
a sequential-move mechanism is equivalent to implementation by a
simultaneous-move mechanism, and Category II includes all the others. \cite%
{sli2017} and \cite{mppt2023} prove that obvious dominance and
strong-obvious dominance are in Category II, while folk wisdom seems to
suggest that weak dominance and Bayesian Nash equilibrium are in Category I.
This leads to an important question: what property the last two solution
concepts possess (but the first two do not) makes them in Category I? For
instance, consider a new solution concept: max-min equilibrium (see below).%
\footnote{%
An max-min equilibrium describes strategic behaviors of uncertain-averse
players, who assess strategies by the worst possible scenarios.} Is it in
Category I or Category II?

\begin{eqnarray*}
\text{Bayesian Nash Equilibrium}\text{: } &&\int\limits_{\Theta _{-i}}\left(
u_{i}^{\theta _{i}}\left[ f\left( \theta _{i},\theta _{-i}\right) \right]
-u_{i}^{\theta _{i}}\left[ f\left( \theta _{i}^{\prime },\theta _{-i}\right) %
\right] \right) \mu ^{\theta _{i}}\left[ d\theta _{-i}\right] \geq 0\text{, }%
\forall \left( i,\theta \right) \in \mathcal{I}\times \Theta \text{,} \\
\text{obvious dominance}\text{: } &&\min_{\theta _{-i}\in \Theta
_{-i}}u_{i}^{\theta _{i}}\left[ f\left( \theta _{i},\theta _{-i}\right) %
\right] \geq \max_{\theta _{-i}\in \Theta _{-i}}u_{i}^{\theta _{i}}\left[
f\left( \theta _{i}^{\prime },\theta _{-i}\right) \right] \text{, }\forall
\left( i,\theta \right) \in \mathcal{I}\times \Theta \text{,} \\
\text{max-min equilibrium}\text{: } &&\min_{\theta _{-i}\in \Theta
_{-i}}u_{i}^{\theta _{i}}\left[ f\left( \theta _{i},\theta _{-i}\right) %
\right] \geq \min_{\theta _{-i}\in \Theta _{-i}}u_{i}^{\theta _{i}}\left[
f\left( \theta _{i}^{\prime },\theta _{-i}\right) \right] \text{, }\forall
\left( i,\theta \right) \in \mathcal{I}\times \Theta \text{.}
\end{eqnarray*}

A powerful tool in traditional mechanism design is revelation principle: a
social choice function can be implemented by a general simultaneous-move
mechanism if and only if it can be implemented by the induced direct
mechanism. General mechanisms are complicated and they are not primitives,
while direct mechanisms are simple and they are primitives.\footnote{%
A direct mechanisms is equivalent to a social choice function.} \emph{The
revelation principle translates a complicated problem involving
non-primitives to a simple problem defined on primitives.}

In this paper, we study mechanism design with sequential-move mechanisms for
general solution concepts.\footnote{%
In mechanism design, there are two paradigms: static mechanism design (or
equivalently, one-period mechanism design) and dynamic mechanism design (or
equivalently, multi-period mechanism design). This paper belongs to static
mechanism design. That is, we design a mechanism in one single period, but
within this period, our mechanism is a sequential-move game rather than a
simultaneous-move game. Revelation principle for dynamic mechanism design is
studied in \cite{rm1986}, \cite{tsaw2021}.} In this setup, we aim to
establish an analogous revelation principle, which translates a complicated
problem involving mechanisms to a simple problem defined on primitives.

Given simultaneous-move games, even though a social choice function could
possibly be implemented by different equilibria in different mechanisms, all
of them induce the same "outcome": a direct mechanism, and incentive
compatibility on this unique direct mechanism defines the simple problem in
revelation principle. However, given sequential-move games, players disclose
their types sequentially and gradually, and the protocol of disclosing is
not \emph{a priori} fixed. Hence, a social choice function could possibly be
implemented by different equilibria in different mechanisms, which lead to
multiple different "outcomes." We thus face two difficulties: (1) what is
the counterpart of "direct mechanism" for each implemented outcome? and (2)
which "outcome" should we focus on, when we define the simple problem in
revelation principle?

Due to this problem (i.e., indeterminacy of implemented outcomes), current
revelation principles for obvious dominance and strong-obvious dominance
have the following form: a social choice function $f$ can be implemented by
a general sequential-move mechanism if and only if $f$ can be implemented by
some mechanism in \emph{a particular set of potential sequential-move
mechanisms} (e.g., \cite{sbyg2017}, \cite{iayg2018}, \cite{am2020}, \cite%
{mppt2023}). This translates a complicated mechanism design problem into a
simpler mechanism design problem, which is not yet defined on primitives.

In order to achieve our goal, we propose a device called \emph{operator},
which is defined on primitives. If a social choice function is implemented
by a general sequential-move mechanism, this mechanism induces a particular
operator, which describes the dynamic process of players disclosing their
types sequentially. We show that a social choice function can be implemented
by a general sequential-move mechanism if and only if the corresponding
operator satisfies some properties. In particular, the traditional
revelation principle for simultaneous-move mechanisms is a degenerate case
of our revelation principle. A conceptual description of our revelation
principle is provided in Section \ref{sec:reveal}.

Based on this, we provide two results for general solution concepts. First,
we identify a property of solution concepts: additivity (Definition \ref%
{def:solution:dissect}). We prove that additive solution concepts (e.g.,
weak dominance, perfect Bayesian equilibrium, max-min equilibrium) are in
Category I, and as a result, the traditional revelation principle remains
valid when we allow for sequential-move mechanisms.

Second, we consider non-additive solution concepts, e.g., obvious dominance,
strong-obvious dominance. Nevertheless, these solution concepts share
another property: monotonicity (Definition \ref{def:solution:monotonic}).

For any social choice function $f:\Theta \longrightarrow \mathcal{X}$, the
value of $f\left( \theta \right) $ hinges critically on the value of $\theta 
$, which is privately observed by the players. Thus, if a mechanism
implements $f$, all of the players must reveal their types fully in the
end---this is formalized as a condition defined on primitives: the operator
induced by the mechanism (which implements $f$) is \emph{achievable}
(Definition \ref{def:solution:achieve}).

Our second result is that we identify a canonical operator $\gamma ^{\left(
\rho ,f\right) }$ for any solution concept $\rho $ and any social choice
function $f$. If $\rho $ is monotonic, we prove that $\gamma ^{\left( \rho
,f\right) }$ is a lower bound for any operator induced by a mechanism which
implements $f$ under $\rho $. This leads to a revelation principle for
monotonic solution concepts: $f$ can be implemented by a general
sequential-move mechanism if and only \ if $\gamma ^{\left( \rho ,f\right) }$
is achievable, which is a condition defined on primitives.

The remainder of the paper proceeds as follows. We describe our revelation
principle in Section \ref{sec:reveal}. We define the model in Section \ref%
{sec:model}. In Section \ref{sec:perfect-recall}, we identify a set of games
and strategy profiles which substantially simplify our analysis. We propose
new tools in Section \ref{sec:tool}. We study additive and monotonic
solution concepts in Sections \ref{sec:revelation-principle} and \ref%
{sec:monotonic}, respectively. We conclude in Section \ref{sec:conclude}.

\section{Revelation principle: a conceptual summary}

\label{sec:reveal}

In this section, we describe both the traditional revelation principle and
our revelation principle for sequential-move mechanisms. The former is a
degenerate case of the latter.

Given simultaneous-move games and Bayesian Nash equilibrium, the traditional
revelation principle says that a social choice function $f:\Theta
\longrightarrow \mathcal{X}$ is implemented by a mechanism $G$ and an
equilibrium $S$ if and only if (i)\ $f=G\circ S$ is the induced direct
mechanism, and (ii) (Bayesian) incentive compatibility holds, i.e.,

\begin{equation}
\int\limits_{\Theta _{-i}}\left( u_{i}^{\theta _{i}^{\ast }}\left[ f\left(
\theta _{i}^{\ast },\theta _{-i}\right) \right] -u_{i}^{\theta _{i}^{\ast }}%
\left[ f\left( \theta _{i}^{\prime },\theta _{-i}\right) \right] \right) \mu
^{\theta _{i}^{\ast }}\left[ d\theta _{-i}\right] \geq 0\text{, }\forall
\left( i,\theta ^{\ast }\right) \in \mathcal{I}\times \Theta \text{, }%
\forall \theta ^{\prime }\in \Theta \diagdown \left\{ \theta ^{\ast
}\right\} \text{.}  \label{khh3}
\end{equation}

We now consider sequential-move games, and for simplicity, we focus on
perfect-information games only in this section. Suppose that $f:\Theta
\longrightarrow \mathcal{X}$ is implemented by a sequential-move mechanism $%
G $ and a strategy profile $S$. In order to establish a revelation
principle, we need to generalize two ideas: (i)\ what object (defined on
primitives)\ does $G\circ S$ induce? (ii) how should incentive compatibility
be defined in such a setup?

Regarding (i), we propose to use an "operator" (Definition \ref%
{def:operator:semi}) to represent $G\circ S$. An operator $\gamma $ is a
function which maps any set $\widehat{\Theta }$ in $\left( \times _{i\in 
\mathcal{I}}\left[ 2^{\Theta _{i}}\diagdown \left\{ \varnothing \right\} %
\right] \right) $ to a partition of $\widehat{\Theta }$, or precisely, 
\begin{equation*}
\gamma :\left( \times _{i\in \mathcal{I}}\left[ 2^{\Theta _{i}}\diagdown
\left\{ \varnothing \right\} \right] \right) \times \Theta \longrightarrow
\times _{i\in \mathcal{I}}\left[ 2^{\Theta _{i}}\diagdown \left\{
\varnothing \right\} \right] \text{ and }\left( 
\begin{array}{c}
\text{given }\theta ,\theta ^{\prime }\in \widehat{\Theta }\text{, } \\ 
\text{we have }\theta \in \gamma \left[ \widehat{\Theta },\theta \right] 
\text{ and} \\ 
\theta ^{\prime }\in \gamma \left[ \widehat{\Theta },\theta \right]
\Longrightarrow \gamma \left[ \widehat{\Theta },\theta ^{\prime }\right]
=\gamma \left[ \widehat{\Theta },\theta \right]%
\end{array}%
\right) \text{.}
\end{equation*}%
We use $\gamma ^{G\circ S}$ to represent the operator induced by $G\circ S$.
Suppose the true state is $\theta ^{\ast }$. At the beginning of the game,
players follow $S\left( \theta ^{\ast }\right) $ to proceed to the next
information set, and $\gamma ^{G\circ S}\left[ \Theta ,\theta ^{\ast }\right]
$ denotes the set of states that take the same strategy (i.e., $S\left(
\theta ^{\ast }\right) $); at the second information set, players follow $%
S\left( \theta ^{\ast }\right) $ to proceed to the next information set, and 
$\gamma ^{G\circ S}\left[ \gamma ^{G\circ S}\left[ \Theta ,\theta ^{\ast }%
\right] ,\theta ^{\ast }\right] $ denotes the set of states that take the
same strategy (i.e., $S\left( \theta ^{\ast }\right) $);.... i.e., $G\circ S$
describes how players sequentially disclose their types, which is recorded
as:%
\begin{equation}
\Theta \longrightarrow \gamma ^{G\circ S}\left[ \Theta ,\theta ^{\ast }%
\right] \longrightarrow \gamma ^{G\circ S}\left[ \gamma ^{G\circ S}\left[
\Theta ,\theta ^{\ast }\right] ,\theta ^{\ast }\right] \longrightarrow
\gamma ^{G\circ S}\left[ \gamma ^{G\circ S}\left[ \gamma ^{G\circ S}\left[
\Theta ,\theta ^{\ast }\right] ,\theta ^{\ast }\right] ,\theta ^{\ast }%
\right] ...  \label{khh1}
\end{equation}%
Since $f:\Theta \longrightarrow \mathcal{X}$ is implemented by $G\circ S$,
players must disclose the true state $\theta ^{\ast }$ in the end, i.e., the
process described in (\ref{khh1}) must lead to $\left\{ \theta ^{\ast
}\right\} $. --- This condition is called \emph{achievability} (Definition %
\ref{def:solution:achieve}).

Regarding (ii), we break the incentive compatibility condition in a
sequential-move game into two parts. First, at any particular information
set, the adopted solution concept dictates whether it is incentive
compatible to falsely report $\theta ^{\prime }$ at the true state $\theta
^{\ast }$. We use an abstract function (called solution notion, see
Definition \ref{def:solution}) to describe it:%
\begin{equation*}
\rho :\left( \times _{i\in \mathcal{I}}\left[ 2^{\Theta _{i}}\diagdown
\left\{ \varnothing \right\} \right] \right) \times \Theta \times \Theta
\times \mathcal{I}\longrightarrow \left\{ 0,1\right\} \text{.}
\end{equation*}%
The interpretation is:%
\begin{equation*}
\rho \left[ \widehat{\Theta },\theta ^{\ast },\theta ^{\prime },i\right]
=1\Longleftrightarrow \left[ 
\begin{array}{c}
\text{at a history }h\text{ in a game }G\text{,} \\ 
\text{with }\widehat{\Theta }\text{ being the set of states in which }S\text{
leads to }h\text{, } \\ 
\text{agent }i\text{ prefers truthfully revealing }\theta ^{\ast }\text{ }
\\ 
\text{to falsely reporting }\theta ^{\prime }%
\end{array}%
\right] \text{.}
\end{equation*}%
For instance, Perfect Bayesian equilibrium could be represented by $\rho
^{PBE}$ as follows.%
\begin{equation*}
\rho ^{PBE}\left[ \widehat{\Theta },\theta ^{\ast },\theta ^{\prime },i%
\right] \equiv \left\{ 
\begin{tabular}{ll}
$1\text{,}$ & if $\int\limits_{\widehat{\Theta }_{-i}}\left( u_{i}^{\theta
_{i}^{\ast }}\left[ f\left( \theta _{i}^{\ast },\widehat{\theta }%
_{-i}\right) \right] -u_{i}^{\theta _{i}^{\ast }}\left[ f\left( \theta
_{i}^{\prime },\widehat{\theta }_{-i}\right) \right] \right) \mu ^{\theta
_{i}^{\ast }}\left[ d\widehat{\theta }_{-i}\right] \geq 0$, \\ 
&  \\ 
$0\text{,}$ & otherwise.%
\end{tabular}%
\right. \text{.}
\end{equation*}

Second, since players disclose their types gradually, we require that,
whenever a player discloses her type partially at an information set, she
must be incentive compatible to do so.---The contrapositive statement is
called $\left( \rho ,f\right) $-consistency (Definition \ref%
{def:solution:consistent}), which is formalized as follows. 
\begin{equation}
\rho \left[ \widehat{\Theta },\theta ^{\ast },\theta ^{\prime },i\right]
=0\Longrightarrow \theta _{i}^{\prime }\in \gamma _{i}^{G\circ S}\left[ 
\widehat{\Theta },\theta ^{\ast }\right] \text{, }\forall i\in \mathcal{I}%
\text{,}  \label{khh2}
\end{equation}%
That is, if it is not incentive compatible to falsely report $\theta
^{\prime }$ (i.e., $\rho \left[ \widehat{\Theta },\theta ^{\ast },\theta
^{\prime },i\right] =0$), then player $i$ must bundle $\theta _{i}^{\prime }$
with $\theta _{i}^{\ast }$ (i.e., $\theta _{i}^{\prime }\in \gamma
_{i}^{G\circ S}\left[ \widehat{\Theta },\theta ^{\ast }\right] $).

Given these new notions, it is easy to show a conceptual revelation
principle:%
\begin{equation*}
\left( 
\begin{array}{c}
f\text{ is implemented by} \\ 
\text{a sequential-move mechanism }G \\ 
\text{and a strategy profile }S \\ 
\text{under the solution concept }\rho \text{ }%
\end{array}%
\right) \Longleftrightarrow \left( 
\begin{array}{c}
\text{the induced operator }\gamma ^{G\circ S}\text{ } \\ 
\text{is both achievable and }\left( \rho ,f\right) \text{-consistent}%
\end{array}%
\right) \text{.}
\end{equation*}%
This revelation principle generalizes the traditional revelation principle
for simultaneous-move games. To see this, suppose that $f$ is implemented by
a simultaneous-move mechanism $G$ and a strategy profile $S$. Thus, the
induced operator $\gamma ^{G\circ S}$ satisfies%
\begin{equation*}
\gamma ^{G\circ S}\left[ \Theta ,\theta ^{\ast }\right] =\left\{ \theta
^{\ast }\right\} \text{, }\forall \theta ^{\ast }\in \Theta \text{,}
\end{equation*}%
i.e., players fully disclosed $\theta ^{\ast }$ in $G$ at each true state $%
\theta ^{\ast }\in \Theta $, and achievability of $\gamma ^{G\circ S}$
holds. Furthermore, this implies

\begin{equation*}
\theta _{i}^{\prime }\notin \gamma _{i}^{G\circ S}\left[ \Theta ,\theta
^{\ast }\right] \text{, }\forall \left( i,\theta ^{\ast }\right) \in 
\mathcal{I}\times \Theta \text{, }\forall \theta ^{\prime }\in \Theta
\diagdown \left\{ \theta ^{\ast }\right\} \text{, }
\end{equation*}%
and hence, $\left( \rho ,f\right) $-consistency (i.e., (\ref{khh2}))
immediately implies%
\begin{equation*}
\rho \left[ \Theta ,\theta ^{\ast },\theta ^{\prime },i\right] =1\text{, }%
\forall \left( i,\theta ^{\ast }\right) \in \mathcal{I}\times \Theta \text{, 
}\forall \theta ^{\prime }\in \Theta \diagdown \left\{ \theta ^{\ast
}\right\} \text{,}
\end{equation*}%
which becomes the usual Baysian incentive compatibility condition (i.e., (%
\ref{khh3})), if $\rho $ is Bayesian Nash equilibrium.

\section{Model}

\label{sec:model}

There is a finite set of agents, denoted by $\mathcal{I}$. Let $\Theta
\equiv \times _{i\in \mathcal{I}}\Theta _{i}$ denote a set of states, which
could be either finite or infinite. Let $\mathcal{X}$ denote a set of social
outcomes. At each state $\left( \theta _{i}\right) _{i\in \mathcal{I}}\in
\Theta $, agent $i$ observes $\theta _{i}$ privately, and agent $i$'s
utility depends only on $\theta _{i}$, i.e., her utility function is
described $u_{i}^{\theta _{i}}:\mathcal{X}\longrightarrow 
\mathbb{R}
$. Our goal is to implement a social choice function (hereafter, SCF) $%
f:\Theta \longrightarrow \mathcal{X}$.

Let $\mathbb{N}$ denote the set of positive integers. We use $E\subset
E^{\prime }$ to denote that $E$ is a weak subset of $E^{\prime }$, and use $%
E\subsetneqq E^{\prime }$ to denote that $E$ is a strict subset of $%
E^{\prime }$. Throughout the paper, we use $-i$ to denote $\mathcal{I}%
\diagdown \left\{ i\right\} $. For any $E\subset \times _{j\in \mathcal{I}%
}X_{j}$ and any $i\in \mathcal{I}$, define%
\begin{equation*}
E_{i}\equiv \left\{ e_{i}\in X_{i}:\exists e_{-i}\in X_{-i}\text{, }\left(
e_{i},e_{-i}\right) \in E\right\} \text{.}
\end{equation*}

\subsection{Mechanisms}

\label{sec:model:mechanism}

A mechanism (or equivalently, a sequential-move game), denoted by $G$, is a
tuple:%
\begin{equation*}
G\equiv \left[ 
\begin{array}{c}
\mathcal{I}\text{, }\mathcal{A}\text{, }\mathcal{H}\text{, }\mathcal{T}\text{%
, }K\in \mathbb{N}\text{, }\phi :\mathcal{H}\diagdown \mathcal{T}%
\longrightarrow \mathcal{I}\text{, } \\ 
A:\mathcal{H}\diagdown \mathcal{T}\longrightarrow 2^{\mathcal{A}}\diagdown
\left\{ \varnothing \right\} \text{, }\left( \zeta _{i}:\mathcal{H}%
\longrightarrow \mathcal{H}\right) _{i\in \mathcal{I}}\text{, \ }g:\mathcal{T%
}\longrightarrow \mathcal{X}%
\end{array}%
\right] \text{,}
\end{equation*}%
where $\mathcal{A}$ is a set of actions, and $\mathcal{H}$ is a set of
histories such that%
\begin{equation*}
\sigma \in \mathcal{H}\text{ and }\mathcal{H\subset }\left\{ \sigma \right\}
\cup \left[ \cup_{k=1}^{K}\mathcal{A}^{k}\right] \text{,}
\end{equation*}%
where $\sigma $ denotes the initial history, at which no action has been
taken by any player yet. For each $h\in \mathcal{H}$, let $\left\vert
h\right\vert $ denote the length of $h$, i.e., $\left\vert \sigma
\right\vert =0$ and $\left\vert \left( a^{1},...,a^{k}\right) \right\vert =k$%
. We require $K\geq \left\vert h\right\vert $ for any $h\in \mathcal{H}$,
i.e., $K$ is an upper bound for lengths of histories.\footnote{%
One technical difficulty induced by unbounded history lengths is that a
strategy profile may result in a terminal history with infinitely-many
moves. For simplicity, we impose the upper bound to avoid this technical
difficulty.} However, the upper bound (i.e., $K$) can be arbitrarily large
for all of the games we consider.

For any history $h\in \mathcal{H}\diagdown \left\{ \sigma \right\} $, let
Path$\left[ h\right] $ denote the set of sub-histories of $h$, and
rigorously, 
\begin{equation*}
\text{Path}\left[ h\right] \equiv \left\{ \sigma \right\} \cup \left\{
\left( a^{1},...,a^{L^{\prime }}\right) :L^{\prime }\leq L\right\} \text{, }%
\forall h=\left( a^{1},...,a^{L}\right) \in \mathcal{H}\diagdown \left\{
\sigma \right\} \text{.}
\end{equation*}%
We require Path$\left[ h\right] \subset \mathcal{H}$ for any $h\in \mathcal{H%
}$. A history is terminal if and only if it is not a sub-history of a
different history. We use $\mathcal{T}$ to denote the set of terminal
histories.

At each non-terminal history $h\in \mathcal{H}\diagdown \mathcal{T}$, $\phi
\left( h\right) \in \mathcal{I}$ denotes the agent who will choose the next
action, and the set of actions available for $\phi \left( h\right) $ at $h$
is $A\left( h\right) $. For any $a\in A\left( h\right) $, let $\left[ h,a%
\right] $ denote the history of "$a$ following $h$," i.e., $\left[ \sigma ,a%
\right] =\left( a\right) \in \mathcal{H}$ and%
\begin{equation*}
\left[ h=\left( a^{1},...,a^{k}\right) ,\text{ }a\right] =\left(
a^{1},...,a^{k},a\right) \in \mathcal{H}\text{,}
\end{equation*}
\begin{equation*}
\text{and }A\left[ h\right] \equiv \left\{ a\in \mathcal{A}:\left[ h,\text{ }%
a\right] \in \mathcal{H}\right\} \text{, }\forall h\in \mathcal{H}\diagdown 
\mathcal{T}\text{.}
\end{equation*}%
For each agent $i\in \mathcal{I}$, the function $\zeta _{i}:\mathcal{H}%
\longrightarrow \mathcal{H}$ describes $i$'s information sets. For any two
of $i$'s histories, $h,h^{\prime }\in \phi ^{-1}\left( i\right) $, agent $i$
cannot distinguish them if and only if $\zeta _{i}\left( h\right) =\zeta
_{i}\left( h^{\prime }\right) $, which defines an equivalence relation on $%
\mathcal{H}\diagdown \mathcal{T}$:%
\begin{equation*}
h\overset{G}{\sim }h^{\prime }\Longleftrightarrow \left( 
\begin{array}{c}
\exists i\in \mathcal{I}\text{, }\phi \left( h\right) =\phi \left( h^{\prime
}\right) =i\text{,} \\ 
\zeta _{i}\left( h\right) =\zeta _{i}\left( h^{\prime }\right)%
\end{array}%
\right) \text{, }\forall \left( h,h^{\prime }\right) \in \left[ \mathcal{H}%
\diagdown \mathcal{T}\right] \times \left[ \mathcal{H}\diagdown \mathcal{T}%
\right] \text{.}
\end{equation*}%
At a history $h\in \phi ^{-1}\left( i\right) $, agent $i$'s information set
is $\left\{ h^{\prime }\in \mathcal{H}\diagdown \mathcal{T}:h\overset{G}{%
\sim }h^{\prime }\right\} $. We require%
\begin{equation*}
h\overset{G}{\sim }h^{\prime }\Longrightarrow A\left( h\right) =A\left(
h^{\prime }\right) \text{, }\forall \left( h,h^{\prime }\right) \in \left[ 
\mathcal{H}\diagdown \mathcal{T}\right] \times \left[ \mathcal{H}\diagdown 
\mathcal{T}\right] \text{.}
\end{equation*}%
Finally, $g:\mathcal{T}\longrightarrow \mathcal{X}$ maps each terminal
history $h\in \mathcal{T}$ to a social outcome $g\left( h\right) \in 
\mathcal{X}$.

Let $\mathcal{G}$ denote the set of all such mechanisms. In particular, a
mechanism $G$ is a perfect-information game if and only if 
\begin{equation*}
h\overset{G}{\sim }h^{\prime }\Longrightarrow h=h^{\prime }\text{, }\forall
\left( h,h^{\prime }\right) \in \left[ \mathcal{H}\diagdown \mathcal{T}%
\right] \times \left[ \mathcal{H}\diagdown \mathcal{T}\right] \text{,}
\end{equation*}%
i.e., every player's information set contains one single history. Let $%
\mathcal{G}^{\text{PI}}$ denote the set of all perfect-information games.

\subsection{Behavior strategies and strategies}

Given a mechanism $G\in \mathcal{G}$, we define behavior strategies and
strategies in this subsection. A behavior strategy of agent $i$ is a $\zeta
_{i}$-measurable function $B_{i}:\phi ^{-1}\left( i\right) \longrightarrow 
\mathcal{A}$, i.e.,%
\begin{equation*}
\left[ \phi \left( h\right) =\phi \left( h^{\prime }\right) =i\text{ and }h%
\overset{G}{\sim }h^{\prime }\right] \Longrightarrow B_{i}\left( h\right)
=B_{i}\left( h^{\prime }\right) \text{, }\forall \left( h,h^{\prime }\right)
\in \left[ \mathcal{H}\diagdown \mathcal{T}\right] \times \left[ \mathcal{H}%
\diagdown \mathcal{T}\right] \text{.}
\end{equation*}%
Let $\mathcal{B}_{i}$ denote the set of all such behavior strategies of
agent $i$, and $\mathcal{B\equiv }\times _{i\in \mathcal{I}}\mathcal{B}_{i}$.

Each $B=\left( B_{i}\right) _{i\in \mathcal{I}}\in \mathcal{B}$ determines a
unique terminal history, which is described by the function $T^{G}:\mathcal{B%
}\longrightarrow \mathcal{T}$, i.e., $\left( \mathcal{B}_{\phi \left( \sigma
\right) }\left( \sigma \right) \right) \in $Path$\left[ T^{G}\left( B\right) %
\right] $, and inductively,%
\begin{equation*}
h\in \text{Path}\left[ T^{G}\left( B\right) \right] \Longrightarrow \left[ h,%
\text{ }\mathcal{B}_{\phi \left( h\right) }\left( h\right) \right] \in \text{%
Path}\left[ T^{G}\left( B\right) \right] \text{, }\forall h\in \mathcal{H}%
\diagdown \left[ \mathcal{T}\cup \left\{ \sigma \right\} \right] \text{.}
\end{equation*}%
A strategy of player $i$ in game $G$ is a function, $S_{i}:\Theta
_{i}\longrightarrow \mathcal{B}_{i}$. Let $\mathcal{S}_{i}^{G}$ denote the
set of all such strategies of agent $i$ in $G$, and $\mathcal{S}^{G}\mathcal{%
\equiv }\times _{i\in \mathcal{I}}\mathcal{S}_{i}^{G}$. We say a strategy
profile $\left( S_{i}\right) _{i\in \mathcal{I}}\in \mathcal{S}^{G}$ in $G$
implements an SCF $f:\Theta \longrightarrow \mathcal{X}$ if and only if%
\begin{equation*}
g\left[ T^{G}\left( \left[ S_{i}\left( \theta _{i}\right) \right] _{i\in 
\mathcal{I}}\right) \right] =f\left[ \left( \theta _{i}\right) _{i\in 
\mathcal{I}}\right] \text{, }\forall \left[ \left( \theta _{i}\right) _{i\in 
\mathcal{I}}\right] \in \Theta \text{.}
\end{equation*}

\subsection{Solution concepts}

Throughout this subsection, we fix any $G\in \mathcal{G}$, and define six
solution concepts.

\subsubsection{Obvious dominance}

\label{sec:def:dominance}

For any agent $i\in \mathcal{I}$, $S_{i}\in \mathcal{S}_{i}^{G}$ is
obviously dominant if and only if for any $\left( \theta ,h,h^{\prime
},B,B^{\prime }\right) \in \Theta \times \left[ \mathcal{H}\diagdown 
\mathcal{T}\right] \times \left[ \mathcal{H}\diagdown \mathcal{T}\right]
\times \mathcal{B}\times \mathcal{B}$, we have%
\begin{equation}
\left( 
\begin{array}{c}
\phi \left( h\right) =\phi \left( h^{\prime }\right) =i\text{ and }h\overset{%
G}{\sim }h^{\prime }\text{,} \\ 
h\in \text{Path}\left( T^{G}\left[ S_{i}\left( \theta _{i}\right) \text{, }%
B_{-i}\right] \right) \text{, } \\ 
h^{\prime }\in \text{Path}\left[ T^{G}\left( B^{\prime }\right) \right] 
\text{ and }S_{i}\left( \theta _{i}\right) \left[ h\right] \neq
B_{i}^{\prime }\left( h^{\prime }\right)%
\end{array}%
\right) \Longrightarrow u_{i}^{\theta _{i}}\left[ g\left( T^{G}\left[
S_{i}\left( \theta _{i}\right) \text{, }B_{-i}\right] \right) \right] \geq
u_{i}^{\theta _{i}}\left[ g\left( T^{G}\left[ B^{\prime }\right] \right) %
\right] \text{.}  \label{kkr1}
\end{equation}%
Suppose that type $\theta _{i}$ reaches the information set containing both $%
h$ and $h^{\prime }$ (i.e., $h\overset{G}{\sim }h^{\prime }$). Agent $i$
faces two options: (i) sticking to $S_{i}\left( \theta _{i}\right) $ and
(ii) deviating to $B_{i}^{\prime }$ with $S_{i}\left( \theta _{i}\right) %
\left[ h\right] \neq B_{i}^{\prime }\left( h^{\prime }\right) $. Roughly,
condition (\ref{kkr1}) means%
\begin{equation}
\min_{B_{-i}}u_{i}^{\theta _{i}}\left[ g\left( T^{G}\left[ S_{i}\left(
\theta _{i}\right) \text{, }B_{-i}\right] \right) \right] \geq
\max_{B_{-i}^{\prime }}u_{i}^{\theta _{i}}\left[ g\left( T^{G}\left[
B_{i}^{\prime }\text{, }B_{-i}^{\prime }\right] \right) \right] \text{,}
\label{kkr2}
\end{equation}%
where $B_{-i}$ and $B^{\prime }$ in (\ref{kkr2}) are restricted to those
leading agent $i$ to reach the information set containing both $h$ and $%
h^{\prime }$ (i.e., $h\in $Path$\left( T^{G}\left[ S_{i}\left( \theta
_{i}\right) \text{, }B_{-i}\right] \right) $, $h^{\prime }\in $Path$\left(
T^{G}\left[ B^{\prime }\right] \right) $).

A strategy profile $S=\left( S_{i}\right) _{i\in \mathcal{I}}\in \mathcal{S}%
^{G}$ is obviously dominant in $G$ if and only if $S_{i}$ is obviously
dominant for every $i\in \mathcal{I}$. Let $\mathcal{S}^{G\text{-OD}}$
denote the set of all such strategy profiles.

\subsubsection{Strong-obvious dominance}

For any agent $i\in \mathcal{I}$, $S_{i}\in \mathcal{S}_{i}^{G}$ is
strong-obviously dominant if and only if for any $\left( \theta ,h,h^{\prime
},B,B^{\prime }\right) \in \Theta \times \left[ \mathcal{H}\diagdown 
\mathcal{T}\right] \times \left[ \mathcal{H}\diagdown \mathcal{T}\right]
\times \mathcal{B}\times \mathcal{B}$, we have%
\begin{equation}
\left( 
\begin{array}{c}
\phi \left( h\right) =\phi \left( h^{\prime }\right) =i\text{ and }h\overset{%
G}{\sim }h^{\prime }\text{,} \\ 
h\in \text{Path}\left( T^{G}\left[ B\right] \right) \cup \text{Path}\left(
T^{G}\left[ S_{i}\left( \theta _{i}\right) \text{, }B_{-i}\right] \right) 
\text{, } \\ 
h^{\prime }\in \text{Path}\left( T^{G}\left[ B^{\prime }\right] \right) 
\text{ and }S_{i}\left( \theta _{i}\right) \left[ h\right] =B_{i}\left(
h\right) \neq B_{i}^{\prime }\left( h^{\prime }\right)%
\end{array}%
\right) \Longrightarrow u_{i}^{\theta _{i}}\left[ g\left( T^{G}\left[ B%
\right] \right) \right] \geq u_{i}^{\theta _{i}}\left[ g\left( T^{G}\left[
B^{\prime }\right] \right) \right] \text{.}  \label{kkr3}
\end{equation}%
Suppose that type $\theta _{i}$ reaches the information set containing both $%
h$ and $h^{\prime }$, and $S_{i}\left( \theta _{i}\right) \left[ h\right]
\neq B_{i}^{\prime }\left( h^{\prime }\right) $. Condition (\ref{kkr3}) means%
\begin{equation}
\min_{B}u_{i}^{\theta _{i}}\left[ g\left( T^{G}\left[ B\right] \right) %
\right] \geq \max_{B^{\prime }}u_{i}^{\theta _{i}}\left[ g\left( T^{G}\left[
B^{\prime }\right] \right) \right] \text{,}  \label{kkr4}
\end{equation}%
and we consider all of those $B$ and $B^{\prime }$ which lead agent $i$ to
reach the information set containing $h$ and $h^{\prime }$, and $S_{i}\left(
\theta _{i}\right) \left[ h\right] =B_{i}\left( h\right) \neq B_{i}^{\prime
}\left( h^{\prime }\right) $, i.e., $S_{i}\left( \theta _{i}\right) $ and $%
B_{i}$ choose the same action at the information set, though $B_{i}$ may
still deviate from $S_{i}\left( \theta _{i}\right) $ afterwards. Let $%
\mathcal{S}^{G\text{-SOD}}$ denote the set of strong-obviously dominant
strategy profiles in $G$.

\subsubsection{Weak dominance}

\label{sec:definition:dominance-weak}

For any agent $i\in \mathcal{I}$, $S_{i}\in \mathcal{S}_{i}^{G}$ is weakly
dominant if and only if for any $\left( \theta ,h,h^{\prime },B\right) \in
\Theta \times \left[ \mathcal{H}\diagdown \mathcal{T}\right] \times \left[ 
\mathcal{H}\diagdown \mathcal{T}\right] \times \mathcal{B}$, we have

\begin{equation}
\left( 
\begin{array}{c}
\phi \left( h\right) =\phi \left( h^{\prime }\right) =i\text{, }h\overset{G}{%
\sim }h^{\prime }\text{,} \\ 
h\in \text{Path}\left( T^{G}\left[ S_{i}\left( \theta _{i}\right) \text{, }%
B_{-i}\right] \right) \text{,} \\ 
h^{\prime }\in \text{Path}\left[ T^{G}\left( B\right) \right]%
\end{array}%
\right) \Longrightarrow u_{i}^{\theta _{i}}\left[ g\left( T^{G}\left[
S_{i}\left( \theta _{i}\right) \text{, }B_{-i}\right] \right) \right] \geq
u_{i}^{\theta _{i}}\left[ g\left( T^{G}\left[ B\right] \right) \right] \text{%
.}  \label{kkr1b}
\end{equation}%
Condition (\ref{kkr1b}) means%
\begin{equation}
u_{i}^{\theta _{i}}\left[ g\left( T^{G}\left[ S_{i}\left( \theta _{i}\right) 
\text{, }B_{-i}\right] \right) \right] \geq u_{i}^{\theta _{i}}\left[
g\left( T^{G}\left[ B_{i}\text{, }B_{-i}\right] \right) \right] \text{,}
\label{kkr1ba}
\end{equation}%
where $B_{i}$ and $B_{-i}$ in (\ref{kkr1ba}) are restricted to those leading
agent $i$ to reach the information set containing $h$ and $h^{\prime }$
(i.e., $h\in $Path$\left( T^{G}\left[ S_{i}\left( \theta _{i}\right) \text{, 
}B_{-i}\right] \right) $ and $h^{\prime }\in $Path$\left[ T^{G}\left(
B\right) \right] $). Let $\mathcal{S}^{G\text{-WD}}$ denote the set of
weakly dominant strategy profiles in $G$.

\subsubsection{Perfect Bayesian equilibrium}

Throughout the paper, we fix any $\mu \in \triangle \left( \Theta \right) $,
and we will use it only if we consider Perfect Bayesian equilibrium
(hereafter, PBE). For each $\left( i,\theta \right) \in \mathcal{I}\times
\Theta $, let $\mu ^{\theta _{i}}\in \triangle \left( \Theta _{-i}\right) $
denote the conditional distribution induced by $\mu $ on $\Theta _{-i}$
given $\theta _{i}$. Define%
\begin{equation*}
\Theta _{-i}^{\left( S,h\right) }\equiv \left\{ \theta _{-i}\in \Theta _{-i}:%
\begin{array}{c}
\exists \left( \theta ,h^{\prime }\right) \in \Theta \times \left[ \mathcal{H%
}\diagdown \mathcal{T}\right] \text{, }h\overset{G}{\sim }h^{\prime }\text{,}
\\ 
h^{\prime }\in \text{Path}\left( T^{G}\left[ S\left( \theta \right) \right]
\right)%
\end{array}%
\right\} \text{, }\forall \left( i,S,h\right) \in \mathcal{I}\times \mathcal{%
S}^{G}\times \left[ \mathcal{H}\diagdown \mathcal{T}\right] \text{.}
\end{equation*}%
Suppose that all of the players follow $S$. The set $\Theta _{-i}^{\left(
S,h\right) }$ contains all of $\theta _{-i}\in \Theta _{-i}$ that could
possibly reach the information set containing $h$. Thus, upon reaching this
information set, player $i$ believes her opponents' types come from $\Theta
_{-i}^{\left( S,h\right) }$.

$S=\left( S_{i}\right) _{i\in \mathcal{I}}\in \mathcal{S}^{G}$ is a $\mu $%
-PBE if and only if for any $\left( \theta ,i,h,B\right) \in \Theta \times 
\mathcal{I}\times \left[ \mathcal{H}\diagdown \mathcal{T}\right] \times 
\mathcal{B}$,%
\begin{equation*}
\left( 
\begin{array}{c}
\phi \left( h\right) =i\text{,} \\ 
S_{i}\left( \theta _{i}\right) \left[ h\right] \neq B_{i}\left( h\right)%
\end{array}%
\right) \Longrightarrow \int\limits_{\Theta _{-i}^{\left( S,h\right)
}}\left( u_{i}^{\theta _{i}}\left[ g\left( T^{G}\left[ S\left( \theta
\right) \right] \right) \right] -u_{i}^{\theta _{i}}\left[ g\left( T^{G}%
\left[ B_{i}\text{, }S_{-i}\left( \theta _{-i}\right) \right] \right) \right]
\right) \mu ^{\theta _{i}}\left[ d\theta _{-i}\right] \geq 0\text{.}
\end{equation*}%
Let $\mathcal{S}^{G\text{-}\mu \text{-PBE}}$ denote the set of $\mu $-PBEs
in $G$.

\subsubsection{Weak dominance*}

\label{sec:definition:dominance-weak*}

Following the style of PBE, we define a solution concept weaker than weak
dominance. For any agent $i\in \mathcal{I}$, $S_{i}\in \mathcal{S}_{i}^{G}$
is weakly-dominant* if and only if for any $\left( \theta ,\theta ^{\prime
},h,h^{\prime }\right) \in \Theta \times ~\Theta \times \left[ \mathcal{H}%
\diagdown \mathcal{T}\right] \times \left[ \mathcal{H}\diagdown \mathcal{T}%
\right] $, we have

\begin{equation*}
\left( 
\begin{array}{c}
\phi \left( h\right) =\phi \left( h^{\prime }\right) =i\text{, }h\overset{G}{%
\sim }h^{\prime }\text{,} \\ 
h\in \text{Path}\left( T^{G}\left[ S\left( \theta \right) \right] \right) 
\text{ } \\ 
h^{\prime }\in \text{Path}\left( T^{G}\left[ S_{i}\left( \theta _{i}^{\prime
}\right) \text{, }S_{-i}\left( \theta _{-i}\right) \right] \right)%
\end{array}%
\right) \Longrightarrow u_{i}^{\theta _{i}}\left[ g\left( T^{G}\left[
S\left( \theta \right) \right] \right) \right] \geq u_{i}^{\theta _{i}}\left[
g\left( T^{G}\left[ S_{i}\left( \theta _{i}^{\prime }\right) \text{, }%
S_{-i}\left( \theta _{-i}\right) \right] \right) \right] \text{.}
\end{equation*}%
The difference between weak dominance and weak-dominance* is illustrated as
follows.%
\begin{eqnarray*}
\text{weak dominance}\text{: } &&u_{i}^{\theta _{i}}\left[ g\left( T^{G}%
\left[ S_{i}\left( \theta _{i}\right) \text{, }B_{-i}\right] \right) \right]
\geq u_{i}^{\theta _{i}}\left[ g\left( T^{G}\left[ B_{i}\text{, }B_{-i}%
\right] \right) \right] \text{,} \\
\text{weak-dominance*}\text{: } &&u_{i}^{\theta _{i}}\left[ g\left( T^{G}%
\left[ S\left( \theta \right) \right] \right) \right] \geq u_{i}^{\theta
_{i}}\left[ g\left( T^{G}\left[ S_{i}\left( \theta _{i}^{\prime }\right) 
\text{, }S_{-i}\left( \theta _{-i}\right) \right] \right) \right] \text{.}
\end{eqnarray*}%
There are two major differences. First, upon reaching the information set
containing both $h$ and $h^{\prime }$, weak dominance requires $S_{i}\left(
\theta _{i}\right) $ be weakly better than any $B_{i}$, while
weak-dominance* requires $S_{i}\left( \theta _{i}\right) $ be weakly better
than any $S_{i}\left( \theta _{i}^{\prime }\right) $ (i.e., a special subset
of $B_{i}$). Second, weak dominance considers all possible $B_{-i}$ such
that $h\in $Path$\left( T^{G}\left[ S_{i}\left( \theta _{i}\right) \text{, }%
B_{-i}\right] \right) $ and $h^{\prime }\in $Path$\left[ T^{G}\left(
B\right) \right] $, while weak-dominance* considers all possible $%
S_{-i}\left( \theta _{-i}\right) $ such that $h\in $Path$\left( T^{G}\left[
S\left( \theta \right) \right] \right) $ and $h^{\prime }\in $Path$\left(
T^{G}\left[ S_{i}\left( \theta _{i}^{\prime }\right) \text{, }S_{-i}\left(
\theta _{-i}\right) \right] \right) $ (i.e., a special subset of $B_{-i}$).
Both differences imply that weak dominance implies weak-dominance*. Let $%
\mathcal{S}^{G\text{-WD*}}$ denote the set of weakly dominant strategy
profiles in $G$. Thus, we have 
\begin{equation}
\mathcal{S}^{G\text{-WD}}\subset \mathcal{S}^{G\text{-WD*}}\text{.}
\label{ttl2}
\end{equation}%
It is difficult to directly characterize impementation in weak dominance.
Instead, we will fully characterize implementation in weak-dominance*, which
provides an indirect way to fully characterize impementation in weak
dominance.

\subsubsection{Max-min equilibrium}

A strategy profile $S=\left( S_{i}\right) _{i\in \mathcal{I}}\in \mathcal{S}%
^{G}$ is a max-min-equilibrium if and only if for any $\left( \theta ,\theta
^{\prime },h,h^{\prime }\right) \in \Theta \times ~\Theta \times \left[ 
\mathcal{H}\diagdown \mathcal{T}\right] \times \left[ \mathcal{H}\diagdown 
\mathcal{T}\right] $, we have%
\begin{equation}
\left( 
\begin{array}{c}
\phi \left( h\right) =\phi \left( h^{\prime }\right) =i\text{, }h\overset{G}{%
\sim }h^{\prime }\text{,} \\ 
h\in \text{Path}\left( T^{G}\left[ S\left( \theta \right) \right] \right) 
\text{, } \\ 
h^{\prime }\in \text{Path}\left( T^{G}\left[ S\left( \theta ^{\prime
}\right) \right] \right) \text{,} \\ 
S_{i}\left( \theta _{i}\right) \left[ h\right] \neq S_{i}\left( \theta
_{i}^{\prime }\right) \left[ h^{\prime }\right] \text{,}%
\end{array}%
\right) \Longrightarrow \left( 
\begin{array}{c}
\exists \left( \widehat{\theta },\widehat{h}\right) \in \Theta \times \left[ 
\mathcal{H}\diagdown \mathcal{T}\right] \text{, }h\overset{G}{\sim }\widehat{%
h}\text{,} \\ 
\widehat{h}\in \text{Path}\left( T^{G}\left[ S_{i}\left( \theta _{i}^{\prime
}\right) \text{, }S_{-i}\left( \widehat{\theta }_{-i}\right) \right] \right) 
\text{, } \\ 
u_{i}^{\theta _{i}}\left[ g\left( T^{G}\left[ S\left( \theta \right) \right]
\right) \right] \geq u_{i}^{\theta _{i}}\left[ g\left( T^{G}\left[
S_{i}\left( \theta _{i}^{\prime }\right) \text{, }S_{-i}\left( \widehat{%
\theta }_{-i}\right) \right] \right) \right]%
\end{array}%
\right) \text{.}  \label{kkr1bb}
\end{equation}%
Condition (\ref{kkr1bb}) requires%
\begin{equation}
\min_{\theta _{-i}}u_{i}^{\theta _{i}}\left[ g\left( T^{G}\left[ S_{i}\left(
\theta _{i}\right) \text{, }S_{-i}\left( \theta _{-i}\right) \right] \right) %
\right] \geq \min_{\widehat{\theta }_{-i}}u_{i}^{\theta _{i}}\left[ g\left(
T^{G}\left[ S_{i}\left( \theta _{i}^{\prime }\right) \text{, }S_{-i}\left( 
\widehat{\theta }_{-i}\right) \right] \right) \right] \text{,}
\label{kkr1bc}
\end{equation}%
where $\theta _{-i}$ and $\widehat{\theta }_{-i}$ in (\ref{kkr1bc}) are
restricted to those leading agent $i$ to reach the information set
containing both $h$ and $\widehat{h}$ (i.e., $h\in $Path$\left( T^{G}\left[
S_{i}\left( \theta _{i}\right) \text{, }S_{-i}\left( \theta _{-i}\right) %
\right] \right) $ and $\widehat{h}\in $Path$\left( T^{G}\left[ S_{i}\left(
\theta _{i}^{\prime }\right) \text{, }S_{-i}\left( \widehat{\theta }%
_{-i}\right) \right] \right) $). Let $\mathcal{S}^{G\text{-MM}}$ denote the
set of $\max \min $-equilibria in $G$.

\subsection{Implementation}

\begin{define}
\label{def:OSP-implementable}Let $\varrho $ denote one of the six solution
concepts defined above. An SCF $f:\Theta \longrightarrow \mathcal{X}$ is $%
\varrho $-implementable if there exist a mechanism $G\in \mathcal{G}$ and $%
S\in \mathcal{S}^{G\text{-}\varrho }$ such that $S$ implements $f$.
\end{define}

Following the tradition, we say:%
\begin{equation*}
\left( 
\begin{array}{c}
f\text{ is SP (i.e., strategyproof) if and only if }f\text{ is
"weak-dominance"-implementable.} \\ 
f\text{ is OSP (i.e., obvious SP) if and only if }f\text{ is
"obvious-dominance"-implementable. } \\ 
f\text{ is SOSP (i.e., strong OSP) if and only if }f\text{ is
"strong-obvious-dominance"-implementable.}%
\end{array}%
\right)
\end{equation*}%
For implementation in obvious dominance and strong-obvious dominance, it
suffers no loss of generality to focus on perfect-information games only.
This is described by the following lemma, which has been proved in \cite%
{iayg2018}, \cite{mppt2023}, \cite{sbyg2017}, and \cite{am2020}.

\begin{lemma}
\label{lem:osp:perfect}Consider a solution concept $\varrho \in \left\{ 
\text{obvious dominance, strong-obvious dominance}\right\} $. An SCF $%
f:\Theta \longrightarrow Z$ is $\varrho $-implementable if and only if there
exist $G\in \mathcal{G}^{\text{PI}}$ and $S\in \mathcal{S}^{G\text{-}\varrho
}$ such that $S$ implements $f$.
\end{lemma}

The following result is immediately implied by the definitions in Sections %
\ref{sec:definition:dominance-weak} and \ref{sec:definition:dominance-weak*}
(precisely, (\ref{ttl2})).

\begin{lemma}
\label{lem:weak*}An SCF $f:\Theta \longrightarrow Z$ is
weak-dominance-implemented by a mechanism $G$ only if it is
weak-dominance*-implemented by $G$.
\end{lemma}

\section{Games and strategies simplified}

\label{sec:perfect-recall}

Let $\mathcal{G}^{\text{PC}}$ denote the set of games with perfect recall,
which is defined as follows. Clearly, $\mathcal{G}^{\text{PI}}\subsetneqq 
\mathcal{G}^{\text{PC}}$.

\begin{define}[\protect\cite{rm1997}]
\label{def:perfect-recall}$G=\left[ \mathcal{I}\text{, }\mathcal{A}\text{, }%
\mathcal{H}\text{, }\mathcal{T}\text{, }K\text{, }\phi \text{, }A\text{, }%
\left( \zeta _{i}\right) _{i\in \mathcal{I}}\text{, }g\right] \in \mathcal{G}
$ is a game with perfect recall if for any $\left( h,h^{\prime },h^{\prime
\prime },a\right) \in \mathcal{H}\times \mathcal{H}\times \mathcal{H}\times 
\mathcal{A}$, we have%
\begin{equation*}
\left( 
\begin{array}{c}
\phi \left( h\right) =\phi \left( h^{\prime }\right) =\phi \left( h^{\prime
\prime }\right) \text{ and }h^{\prime }\overset{G}{\sim }h^{\prime \prime }%
\text{,} \\ 
\left[ h,a\right] \in \text{path}\left[ h^{\prime }\right]%
\end{array}%
\right) \Longrightarrow \left( 
\begin{array}{c}
\exists \widetilde{h}\in \mathcal{H}\diagdown \mathcal{T}\text{, \ }%
\widetilde{h}\overset{G}{\sim }h\text{,} \\ 
\left[ \widetilde{h},a\right] \in \text{path}\left[ h^{\prime \prime }\right]%
\end{array}%
\right) \text{.}
\end{equation*}
\end{define}

Suppose $\phi \left( h\right) =\phi \left( h^{\prime }\right) =\phi \left(
h^{\prime \prime }\right) =i$ and $h^{\prime }\overset{G}{\sim }h^{\prime
\prime }$, i.e., player $i$ cannot distinguish between $h^{\prime }$ and $%
h^{\prime \prime }$. Upon reaching $h^{\prime }$, if player $i$ knows that
she chooses $a$ at a previous node $h$ to reach $h^{\prime }$ (i.e., $\left[
h,a\right] \in $path$\left[ h^{\prime }\right] $), then $i$ must recall the
same information upon reaching $h^{\prime \prime }$: she chooses $a$ at a
previous node $\widetilde{h}$ (with $\widetilde{h}\overset{G}{\sim }h$) to
reach $h^{\prime \prime }$.

For any $\left[ G,S\right] $ and any $h\in \mathcal{H}$, define%
\begin{eqnarray*}
\mathcal{E}^{\left[ G,S\right] \text{-}h} &\equiv &\left\{ \theta \in \Theta
:h\in \text{path}\left[ T^{G}\left( S\left( \theta \right) \right) \right]
\right\} \text{,} \\
\mathcal{E}_{i}^{\left[ G,S\right] \text{-}h} &\equiv &\left\{ \theta
_{i}\in \Theta _{i}:\exists \theta _{-i}\in \Theta _{-i}\text{, }\left(
\theta _{i},\theta _{-i}\right) \in \mathcal{E}^{\left[ G,S\right] \text{-}%
h}\right\} \text{, }\forall i\in \mathcal{I}\text{.}
\end{eqnarray*}%
In the standard revelation principle for simultaneous-move games, each pair
of $\left[ G\in \mathcal{G},\text{ }S\in \mathcal{S}^{G}\right] $ induces a 
\emph{direct mechanism} which is defined on primitives. For sequential-move
games, we will translate $\left[ G,S\right] $ into a new "device" defined on
primitives. In order to achieve this, we need to focus on the following
special class of pairs of $\left[ G,S\right] $.%
\begin{equation}
\mathcal{GS}^{\text{PC}}=\left\{ \left[ G,S\right] :%
\begin{array}{c}
\text{(A) }G\in \mathcal{G}^{\text{PC}}\text{,} \\ 
\\ 
\text{(B) }S\in \mathcal{S}^{G}\text{ such that }\left\{ T^{G}\left( S\left(
\theta \right) \right) :\theta \in \Theta \right\} =\mathcal{T}\text{,} \\ 
\\ 
\text{(C) }\mathcal{E}^{\left[ G,S\right] \text{-}h}=\mathcal{E}^{\left[ G,S%
\right] \text{-}h^{\prime }}\Longrightarrow h=h^{\prime }\text{, }\forall
\left( h,h^{\prime }\right) \in \mathcal{H}\times \mathcal{H}%
\end{array}%
\right\} \text{.}  \label{tkk1}
\end{equation}

As a comparison, \cite{iayg2018}, \cite{mppt2023}, \cite{sbyg2017}, and \cite%
{am2020} prove that it suffers no loss generality to focus on $\left[ G,S%
\right] $ in the following set, when we consider OSP and SOSP.

\begin{equation*}
\mathcal{GS}^{\text{PI}}=\left\{ \left[ G,S\right] :%
\begin{array}{c}
\text{(A) }G\in \mathcal{G}^{\text{PI}}\text{,} \\ 
\\ 
\text{(B) }S\in \mathcal{S}^{G}\text{ such that }\left\{ T^{G}\left( S\left(
\theta \right) \right) :\theta \in \Theta \right\} =\mathcal{T}\text{,} \\ 
\\ 
\text{(C) }\mathcal{E}^{\left[ G,S\right] \text{-}h}=\mathcal{E}^{\left[ G,S%
\right] \text{-}h^{\prime }}\Longrightarrow h=h^{\prime }\text{, }\forall
\left( h,h^{\prime }\right) \in \mathcal{H}\times \mathcal{H}%
\end{array}%
\right\} \text{.}
\end{equation*}%
The only difference between $\mathcal{GS}^{\text{PC}}$ and $\mathcal{GS}^{%
\text{PI}}$ lies in condition (A): $G\in \mathcal{G}^{\text{PC}}$ in $%
\mathcal{GS}^{\text{PC}}$ and $G\in \mathcal{G}^{\text{PI}}$ in $\mathcal{GS}%
^{\text{PI}}$.\footnote{%
The purposes of $\mathcal{GS}^{\text{PC}}$ and $\mathcal{GS}^{\text{PI}}$
differ. Roughly, "focusing on $\mathcal{GS}^{\text{PI}}$" is the revelation
principle established in \cite{iayg2018}, \cite{mppt2023}, \cite{sbyg2017},
and \cite{am2020}, while "focusing on $\mathcal{GS}^{\text{PC}}$" is an
intermediate step for us, and we will establish a sharper revelation
principle later.}

\subsection{Conditions (A)-(C)}

The role of perfect recall is shown by Lemma \ref{lem:perfect-recall} and
the proof is relegated to Appendix \ref{sec:lem:perfect-recall}.

\begin{lemma}
\label{lem:perfect-recall}For any $G\in \mathcal{G}^{\text{PC}}$ and any $%
S\in \mathcal{S}^{G}$, we have%
\begin{equation*}
\left( 
\begin{array}{c}
h\overset{G}{\sim }h^{\prime }\text{ and }\phi \left( h\right) =\phi \left(
h^{\prime }\right) =i\text{,} \\ 
\mathcal{E}^{\left[ G,S\right] \text{-}h}\neq \varnothing \text{ and }%
\mathcal{E}^{\left[ G,S\right] \text{-}h^{\prime }}\neq \varnothing%
\end{array}%
\right) \Longrightarrow \mathcal{E}_{i}^{\left[ G,S\right] \text{-}h}=%
\mathcal{E}_{i}^{\left[ G,S\right] \text{-}h^{\prime }}\text{, }\forall
\left( i,h,h^{\prime }\right) \in \mathcal{I}\times \left[ \mathcal{H}%
\diagdown \mathcal{T}\right] \times \left[ \mathcal{H}\diagdown \mathcal{T}%
\right] \text{.}
\end{equation*}
\end{lemma}

Condition (A) in (\ref{tkk1}) says that we focus on games with perfect
recall only. By Lemma \ref{lem:osp:perfect}, this suffers no loss of
generality, if we consider OSP or SOSP. However, this is not true for weak
dominance, $\mu $-PBE and max-min equilibria, which will be reflected
accordingly in the revelation principles established later (Theorems \ref%
{lem:osp:perfect}, \ref{theorem:revelation:mm} and \ref%
{theorem:revelation:WD}).

Consider any $\left[ G\in \mathcal{G},\text{ }S\in \mathcal{S}^{G}\right] $.
Let $\mathcal{H}^{\left[ G,S\right] }\equiv \cup _{\theta \in \Theta }$path$%
\left[ T^{G}\left( S\left( \theta \right) \right) \right] $) denote the
equilibrium-path histories. Condition (B) says that it suffers no loss
generality to delete off-equilibrium histories in $G$ (i.e., histories in $%
\mathcal{H\diagdown H}^{\left[ G,S\right] }$). This may not be true if $G\in 
\mathcal{G}\diagdown \mathcal{G}^{\text{PC}}$. To see this, consider $\left(
i,\theta ,\theta ^{\prime },h,h^{\prime }\right) \in \mathcal{I}\times
\Theta \times \Theta \times \mathcal{H}\times \mathcal{H}$ in $G$ such that%
\begin{equation*}
\left( 
\begin{array}{c}
h\overset{G}{\sim }h^{\prime }\text{ and }\phi \left( h\right) =\phi \left(
h^{\prime }\right) =i\text{,} \\ 
\theta \in \mathcal{E}^{\left[ G,S\right] \text{-}h}\text{ and }\theta
^{\prime }\in \mathcal{E}^{\left[ G,S\right] \text{-}h^{\prime }}%
\end{array}%
\right) \text{, and }\left( 
\begin{array}{c}
\text{if }G\in \mathcal{G}^{\text{PC}}\text{, Lemma \ref{lem:perfect-recall}
implies }\theta _{i}\in \mathcal{E}_{i}^{\left[ G,S\right] \text{-}h^{\prime
}}\text{;} \\ 
\text{if }G\in \mathcal{G}\diagdown \mathcal{G}^{\text{PC}}\text{, we may
have }\theta _{i}\notin \mathcal{E}_{i}^{\left[ G,S\right] \text{-}h^{\prime
}}%
\end{array}%
\right) \text{ .}
\end{equation*}%
Suppose the true state is $\theta $. At the information set containing both $%
h$ and $h^{\prime }$, player $i$ believes that $\theta ^{\prime }$ may be
the true state. If $G\in \mathcal{G}\diagdown \mathcal{G}^{\text{PC}}$, we
may have $\theta _{i}\notin \mathcal{E}_{i}^{\left[ G,S\right] \text{-}%
h^{\prime }}$, and playing $S_{i}\left( \theta \right) \left[ h^{\prime }%
\right] $ may lead to off-equilibrium histories, which may be a crucial
reason that $S_{i}\left( \theta \right) $ is a best reply for $i$ at this
information set. Therefore, it suffers loss of generality to to delete
off-equilibrium histories in $G$. However, if $G\in \mathcal{G}^{\text{PC}}$%
, Lemma \ref{lem:perfect-recall} implies that following $S_{i}\left( \theta
\right) \left[ h^{\prime }\right] $ always leads to equilibrium-path
histories (i.e., to $T^{G}\left( S_{i}\left( \theta _{i}\right)
,S_{-i}\left( \theta _{-i}^{\prime }\right) \right) $ for some $\theta
_{-i}^{\prime }\in \mathcal{E}_{i}^{\left[ G,S\right] \text{-}h^{\prime }}$
with $h\overset{G}{\sim }h^{\prime }$), and by the usual argument of
"pruning," it suffers no loss of generality to delete off-equilibrium
histories.

Given condition (B), every history is an equilibrium history, which implies
no loss of generality to impose condition (C). To see this, consider two
distinct histories $\left( h,h^{\prime }\right) \in \mathcal{H}\times 
\mathcal{H}$ such that $\mathcal{E}^{\left[ G,S\right] \text{-}h}=\mathcal{E}%
^{\left[ G,S\right] \text{-}h^{\prime }}$, which immediately implies either $%
h\in $path$\left[ h^{\prime }\right] $ or $h^{\prime }\in $path$\left[ h%
\right] $. Without loss of generality, suppose $h\in $path$\left[ h^{\prime }%
\right] $, i.e., for any $\widetilde{h}\in $path$\left[ h^{\prime }\right] $
such that $h\in $path$\left[ \widetilde{h}\right] $, we have 
\begin{equation*}
\mathcal{E}^{\left[ G,S\right] \text{-}h}=\mathcal{E}^{\left[ G,S\right] 
\text{-}\widetilde{h}}=\mathcal{E}^{\left[ G,S\right] \text{-}h^{\prime }}%
\text{, or equivalently, }\left\vert A\left( h\right) \right\vert
=\left\vert A\left( \widetilde{h}\right) \right\vert =1\text{,}
\end{equation*}%
i.e., players take non-strategic actions at $h$ and $\widetilde{h}$.
Therefore, we can identify all of such $\widetilde{h}$ with $h^{\prime }$,
until condition (C) holds.

\subsection{Solutions simplified}

Fix any $\left[ G,S\right] \in \mathcal{GS}^{\text{PC}}$. Suppose that the
true state is $\theta $, and that players have\ reached history $h\in 
\mathcal{H}\diagdown \mathcal{T}$ with $\phi \left( h\right) =i$. Recall

\begin{equation*}
\Theta _{-i}^{\left( S,h\right) }\equiv \left\{ \theta _{-i}^{\prime }\in
\Theta _{-i}:%
\begin{array}{c}
\exists \left( \theta ^{\prime },h^{\prime }\right) \in \Theta \times \left[ 
\mathcal{H}\diagdown \mathcal{T}\right] \text{, }h\overset{G}{\sim }%
h^{\prime }\text{,} \\ 
h^{\prime }\in \text{Path}\left( T^{G}\left[ S\left( \theta ^{\prime
}\right) \right] \right)%
\end{array}%
\right\} =\cup_{h^{\prime }\in \left\{ \widetilde{h}\in \mathcal{H%
}:\text{ }\widetilde{h}\text{ }\overset{G}{\sim }\text{ }h\right\} }\mathcal{%
E}_{-i}^{\left[ G,S\right] \text{-}h^{\prime }}\text{.}
\end{equation*}

Consider any $\theta _{i}^{\prime }\in \mathcal{E}_{i}^{\left[ G,S\right] 
\text{-}\left( h\right) }$. We ignore weak dominance tentatively,\footnote{%
It is not clear how to translate weak dominance to a condition similar to (%
\ref{yyi1})-(\ref{yy5}).} and for the other five solution concepts, Lemma %
\ref{lem:perfect-recall} implies that "truthfully revealing $\theta _{i}$
being better than falsely reporting $\theta _{i}^{\prime }$" can be reduced
to the following simple conditions defined on primitives.%
\begin{equation}
\mu \text{-PBE: }\int\limits_{\Theta _{-i}^{\left( S,h\right) }}\left(
u_{i}^{\theta _{i}}\left[ f\left( \theta _{i},\theta _{-i}\right) \right]
-u_{i}^{\theta _{i}}\left[ f\left( \theta _{i}^{\prime },\theta _{-i}\right) %
\right] \right) \mu ^{\theta _{i}}\left[ d\theta _{-i}\right] \geq 0\text{, }
\label{yyi1}
\end{equation}%
\begin{equation}
\text{weak-dominance*: }u_{i}^{\theta _{i}}\left[ f\left( \theta _{i},\theta
_{-i}\right) \right] \geq u_{i}^{\theta _{i}}\left[ f\left( \theta
_{i}^{\prime },\theta _{-i}\right) \right] \text{, }\forall \theta _{-i}\in
\Theta _{-i}^{\left( S,h\right) }\text{,}  \label{yyi2}
\end{equation}%
\begin{equation}
\text{max-min equilibrium: }\min_{\theta _{-i}\in \Theta _{-i}^{\left(
S,h\right) }}u_{i}^{\theta _{i}}\left[ f\left( \theta _{i},\theta
_{-i}\right) \right] \geq \min_{\theta _{-i}\in \Theta _{-i}^{\left(
S,h\right) }}u_{i}^{\theta _{i}}\left[ f\left( \theta _{i}^{\prime },\theta
_{-i}\right) \right] \text{,}  \label{yyi3}
\end{equation}%
\begin{equation}
\text{obvious dominance: }\min_{\theta _{-i}\in \Theta _{-i}^{\left(
S,h\right) }}u_{i}^{\theta _{i}}\left[ f\left( \theta _{i},\theta
_{-i}\right) \right] \geq \max_{\theta _{-i}\in \Theta _{-i}^{\left(
S,h\right) }}u_{i}^{\theta _{i}}\left[ f\left( \theta _{i}^{\prime },\theta
_{-i}\right) \right] \text{,}  \label{yyi4}
\end{equation}%
\begin{equation}
\text{strong-obvious dominance: }\min_{\widetilde{\theta }_{i}\in \mathcal{E}%
_{i}^{\left[ G,S\right] \text{-}\left[ h,\text{ }S_{i}\left( \theta
_{i}\right) \left[ h\right] \right] }}\min_{\theta _{-i}\in \Theta
_{-i}^{\left( S,h\right) }}u_{i}^{\theta _{i}}\left[ f\left( \widetilde{%
\theta }_{i},\theta _{-i}\right) \right] \geq \max_{\theta _{-i}\in \Theta
_{-i}^{\left( S,h\right) }}u_{i}^{\theta _{i}}\left[ f\left( \theta
_{i}^{\prime },\theta _{-i}\right) \right] \text{.}  \label{yy5}
\end{equation}

\section{New Tools}

\label{sec:tool}

In this section, we introduce several new tools to establish revelation
principles.

\subsection{Semi-operators and operators}

A pair of $\left[ G,S\right] $ describes how players reveal their types
sequentially. Specifically, suppose players follow $S$ in $G$, and they
reach a history $h$ with $\phi \left( h\right) =j$. Recall $\mathcal{E}^{%
\left[ G,S\right] \text{-}h}$ denotes the set of states in which $S$ reaches 
$h$. By playing $S_{j}$ at $h$, agent $j$ follows the partition below to
reveal her types.%
\begin{equation}
\left\{ \mathcal{E}_{j}^{\left[ G,S\right] \text{-}\left[ h,\text{ }%
S_{j}\left( \theta _{j}\right) \left[ h\right] \right] }\times \mathcal{E}%
_{-j}^{\left[ G,S\right] \text{-}h}:\theta _{j}\in \mathcal{E}_{j}^{\left[
G,S\right] \text{-}h}\right\} \text{.}  \label{tth1}
\end{equation}%
Or equivalently, the evolution of history is described as follows:%
\begin{equation}
h\overset{\theta _{j}}{\longrightarrow }\left[ h,\text{ }S_{j}\left( \theta
_{j}\right) \left[ h\right] \right] \text{, }\forall \theta _{j}\in \mathcal{%
E}_{j}^{\left[ G,S\right] \text{-}h}\text{,}  \label{tth1b}
\end{equation}%
and if we use the set of states $\mathcal{E}^{\left[ G,S\right] \text{-}h}$
as a proxy for $h$, (\ref{tth1b})\ becomes%
\begin{equation*}
\mathcal{E}_{j}^{\left[ G,S\right] \text{-}h}\times \mathcal{E}_{-j}^{\left[
G,S\right] \text{-}h}\overset{\theta _{j}}{\longrightarrow }\mathcal{E}_{j}^{%
\left[ G,S\right] \text{-}\left[ h,\text{ }S_{j}\left( \theta _{j}\right) %
\left[ h\right] \right] }\times \mathcal{E}_{-j}^{\left[ G,S\right] \text{-}%
h}\text{, }\forall \theta _{j}\in \mathcal{E}_{j}^{\left[ G,S\right] \text{-}%
h}\text{,}
\end{equation*}%
i.e., (\ref{tth1}). We thus propose an abstract device (called
"semi-operator") to describe this. One innovation is that we do not record
the agent attached to each history, which substantially simplifies
exposition. It will be clear that this suffers no loss of generality.

\begin{define}
\label{def:operator:semi-operator}A semi-operator is a function $\gamma
:\left( \times _{i\in \mathcal{I}}\left[ 2^{\Theta _{i}}\diagdown \left\{
\varnothing \right\} \right] \right) \times \Theta \longrightarrow \times
_{i\in \mathcal{I}}\left[ 2^{\Theta _{i}}\diagdown \left\{ \varnothing
\right\} \right] $.
\end{define}

The notion of semi-operator is too general to describe all of the details
contained in (\ref{tth1b}). For instance, (\ref{tth1b}) describes a
partition. We thus refine the idea as follows.

\begin{define}[operator]
\label{def:operator:semi}An operator is a semi-operator $\gamma $ such that $%
\gamma \left[ E,\cdot \right] |_{\theta \in E}$ forms a partition on $E$, or
equivalently, for any $\left( \theta ,\theta ^{\prime },E\right) \equiv
\Theta \times \Theta \times \left( \times _{i\in \mathcal{I}}\left[
2^{\Theta _{i}}\diagdown \left\{ \varnothing \right\} \right] \right) $ with 
$\left\{ \theta ,\theta ^{\prime }\right\} \subset E$, we have%
\begin{equation*}
\theta \in \gamma \left[ E,\theta \right] \subset E\text{ and }\theta
^{\prime }\in \gamma \left[ E,\theta \right] \Longrightarrow \gamma \left[
E,\theta \right] =\gamma \left[ E,\theta ^{\prime }\right] \text{.}
\end{equation*}
\end{define}

Throughout the paper, we consider $\gamma \left[ E,\theta \right] $ only
when $\theta \in E$, and hence, the definition of $\gamma \left[ E,\theta %
\right] $ is irrelevant if $\theta \notin E$. Let $\Gamma ^{\text{Semi}}$
and $\Gamma $ denote the set of all semi-operators and operators,
respectively.

\subsection{Operators induced by games and strategy profiles}

\label{sec:operator:semi:game}

Fix any $\left[ G,S\right] \in \mathcal{GS}^{\text{PC}}$. We will show that $%
\left[ G,S\right] $ induces a particular operator, denoted by $\gamma ^{%
\left[ G,S\right] }$. Consider%
\begin{equation*}
\Omega ^{\left[ G,S\right] }\equiv \left\{ E\in \left( \times _{i\in 
\mathcal{I}}\left[ 2^{\Theta _{i}}\diagdown \left\{ \varnothing \right\} %
\right] \right) :\exists h\in \left[ \mathcal{H}\diagdown \mathcal{T}\right]
,\text{ }E=\mathcal{E}^{\left[ G,S\right] \text{-}h}\right\} \text{.}
\end{equation*}%
By condition (C) in (\ref{tkk1}), each $E\in \Omega ^{\left[ G,S\right] }$
corresponds to a unique $h\in \left[ \mathcal{H}\diagdown \mathcal{T}\right] 
$. We now define 
\begin{gather*}
\gamma ^{\left[ G,S\right] }:\left( \times _{i\in \mathcal{I}}\left[
2^{\Theta _{i}}\diagdown \left\{ \varnothing \right\} \right] \right) \times
\Theta \longrightarrow \left( \times _{i\in \mathcal{I}}\left[ 2^{\Theta
_{i}}\diagdown \left\{ \varnothing \right\} \right] \right) \text{,} \\
\vartheta ^{\left[ G\right] }:\left( \times _{i\in \mathcal{I}}\left[
2^{\Theta _{i}}\diagdown \left\{ \varnothing \right\} \right] \right)
\longrightarrow 2^{\Theta }\text{,}
\end{gather*}%
where $\gamma ^{\left[ G,S\right] }$ is the operator induced by $\left[ G,S%
\right] $, and $\vartheta ^{\left[ G\right] }$ describes information sets in 
$G$.

First, the definition of an operator is irrelevant when $\theta \notin E$,
and furthermore, $\left[ G,S\right] $ does not provide information when $%
E\notin \Omega ^{\left[ G,S\right] }$. In such cases, we thus define them in
a trivial way: $\vartheta ^{\left[ G\right] }\left[ E\right] =E$, if $%
E\notin \Omega ^{\left[ G,S\right] }$, and $\gamma ^{\left[ G,S\right] }%
\left[ E,\theta \right] =E$, if $\theta \notin E$ or $E\notin \Omega ^{\left[
G,S\right] }$.

Second, suppose $\theta \in E\in \Omega ^{\left[ G,S\right] }$. By condition
(C) in (\ref{tkk1}), there exists a unique $\left[ j,\text{ }h\right] \in 
\mathcal{I}\times \left[ \mathcal{H}\diagdown \mathcal{T}\right] $ such that 
$E=\mathcal{E}^{\left[ G,S\right] \text{-}h}$ and $\phi \left( h\right) =j$,
i.e., $h$ is $j$'s history. Define%
\begin{equation*}
\gamma ^{\left[ G,S\right] }\left[ E,\theta \right] =\mathcal{E}^{\left[ G,S%
\right] \text{-}\left[ h,\text{ }S_{j}\left( \theta _{j}\right) \left[ h%
\right] \right] }\text{,}
\end{equation*}%
i.e., at the information set containing $h$, agent $j$ takes the action $%
S_{j}\left( \theta _{j}\right) \left[ h\right] $, and $\gamma ^{\left[ G,S%
\right] }\left[ E,\theta \right] $ is defined as the set of states in which $%
S$ leads to the history $\left[ h,\text{ }S_{j}\left( \theta _{j}\right) %
\left[ h\right] \right] $. Furthermore, define%
\begin{equation*}
\vartheta ^{\left[ G\right] }\left[ E\right] =\cup_{h^{\prime
}\in \left\{ \widetilde{h}\in \left[ \mathcal{H}\diagdown \mathcal{T}\right]
:\text{ }\widetilde{h}\text{ }\overset{G}{\sim }\text{ }h\right\} }\mathcal{E%
}^{\left[ G,S\right] \text{-}h^{\prime }}\text{,}
\end{equation*}%
i.e., at the information set containing $h$, agent $j$ cannot distinguish $h$
from $h^{\prime }\in \left\{ \widetilde{h}\in \left[ \mathcal{H}\diagdown 
\mathcal{T}\right] :\text{ }\widetilde{h}\text{ }\overset{G}{\sim }\text{ }%
h\right\} $, and hence, $\vartheta ^{\left[ G\right] }\left[ E\right] $ is
the set of states in which agent $j$ believes that $S$ leads to the
information set containing $h$. In particular, consider the degenerate
function%
\begin{equation*}
\vartheta ^{\ast }:\left( \times _{i\in \mathcal{I}}\left[ 2^{\Theta
_{i}}\diagdown \left\{ \varnothing \right\} \right] \right) \longrightarrow
2^{\Theta }\text{ such that }\vartheta ^{\ast }\left[ E\right] =E\text{, }%
\forall E\in \left( \times _{i\in \mathcal{I}}\left[ 2^{\Theta
_{i}}\diagdown \left\{ \varnothing \right\} \right] \right) \text{.}
\end{equation*}%
Clearly, we have $\vartheta ^{\left[ G\right] }=\vartheta ^{\ast }$ for any
game $G\in \mathcal{G}^{\text{PI}}$.

\subsection{Solution notions}

\label{sec:solution-concept}

We dissect a solution concept into two parts: (i) at each history, it
dictates when type $\theta _{i}$ prefers to truthfully revealing $\theta
_{i} $ to falsely reporting $\theta _{i}^{\prime }$ (Definition \ref%
{def:solution}), and (ii) it dictates when agents' sequential revealing is
acceptable (Definition \ref{def:solution:consistent}). We focus on the
former here.

\begin{define}
\label{def:solution}A solution notion is a function 
\begin{equation*}
\rho :\mathcal{X}^{\Theta }\times \Gamma ^{\text{Semi}}\times \left( \times
_{i\in \mathcal{I}}\left[ 2^{\Theta _{i}}\diagdown \left\{ \varnothing
\right\} \right] \right) \times \Theta \times \Theta \times \mathcal{I}%
\times 2^{\Theta }\longrightarrow \left\{ 0,1\right\} \text{,}
\end{equation*}%
such that 
\begin{equation}
\left( \gamma _{i}\left( E\right) ,\theta _{i},\theta _{i}^{\prime },%
\widehat{E}_{-i}\right) =\left( \widetilde{\gamma }_{i}\left( \widetilde{E}%
\right) ,\widetilde{\theta }_{i},\widetilde{\theta }_{i}^{\prime },%
\widetilde{E}_{-i}^{\prime }\right) \Longrightarrow \rho \left[ f,\left(
\gamma ,E\right) ,\theta ,\theta ^{\prime },i,\widehat{E}\right] =\rho \left[
f,\left( \widetilde{\gamma },\widetilde{E}\right) ,\widetilde{\theta },%
\widetilde{\theta }^{\prime },i,\widetilde{E}^{\prime }\right] \text{.}
\label{hht2}
\end{equation}
\end{define}

Given $\left[ G,S\right] \in \mathcal{GS}^{\text{PC}}$ and $\left[ f,\left(
\gamma ,E\right) ,\theta ,\theta ^{\prime },i,\widehat{E}\right] $ with $%
\left\{ \theta ,\theta ^{\prime }\right\} \subset E$, suppose that the true
state is $\theta $, and that the players have reached a history $h$ in $G$
with 
\begin{equation*}
E=\mathcal{E}^{\left[ G,S\right] \text{-}h}\text{ and }\widehat{E}%
=\cup_{h^{\prime }\in \left\{ \widetilde{h}\in \left[ \mathcal{H}%
\diagdown \mathcal{T}\right] :\text{ }\widetilde{h}\text{ }\overset{G}{\sim }%
\text{ }h\right\} }\mathcal{E}^{\left[ G,S\right] \text{-}h^{\prime }}\text{.%
}
\end{equation*}%
The interpretation is: 
\begin{equation*}
\rho \left[ f,\left( \gamma ,E\right) ,\theta ,\theta ^{\prime },i,\widehat{E%
}\right] =1\Longleftrightarrow \left[ 
\begin{array}{c}
\text{at a history }h\text{ with }E=\mathcal{E}^{\left[ G,S\right] \text{-}h}%
\text{ and the belief of }\widehat{E}\text{ }\left( \text{on }\Theta \right) 
\text{, } \\ 
\text{agent }i\text{ prefers truthfully revealing }\theta \text{ to falsely
reporting }\theta ^{\prime }%
\end{array}%
\right] \text{.}
\end{equation*}%
(\ref{hht2}) requires that the value of $\rho \left[ f,\left( \gamma
,E\right) ,\theta ,\theta ^{\prime },i,\widehat{E}\right] $ does not depend
on $\left( \gamma _{-i}\left( E\right) ,\theta _{-i},\theta _{-i}^{\prime },%
\widehat{E}_{i}\right) $.

Focus on $\left[ G,S\right] \in \mathcal{GS}^{\text{PC}}$, Definition \ref%
{def:solution} and (\ref{yyi1})-(\ref{yy5}) imply that the five solution
concepts defined above can be translated to the following solution notions.%
\begin{equation}
\rho ^{\mu \text{-PBE}}\left[ f,\left( \gamma ,E\right) ,\theta ,\theta
^{\prime },i,\widehat{E}\right] =\left\{ 
\begin{tabular}{ll}
$1\text{,}$ & if $\int\limits_{\widehat{E}_{-i}}\left( u_{i}^{\theta _{i}}%
\left[ f\left( \theta _{i},\widehat{\theta }_{-i}\right) \right]
-u_{i}^{\theta _{i}}\left[ f\left( \theta _{i}^{\prime },\widehat{\theta }%
_{-i}\right) \right] \right) \mu ^{\theta _{i}}\left[ d\widehat{\theta }_{-i}%
\right] \geq 0$, \\ 
&  \\ 
$0\text{,}$ & otherwise.%
\end{tabular}%
\right.  \label{yyi6}
\end{equation}%
\begin{equation}
\rho ^{\text{WD*}}\left[ f,\left( \gamma ,E\right) ,\theta ,\theta ^{\prime
},i,\widehat{E}\right] =\left\{ 
\begin{tabular}{ll}
$1\text{,}$ & if $u_{i}^{\theta _{i}}\left[ f\left( \theta _{i},\widehat{%
\theta }_{-i}\right) \right] \geq u_{i}^{\theta _{i}}\left[ f\left( \theta
_{i}^{\prime },\widehat{\theta }_{-i}\right) \right] \text{, }\forall 
\widehat{\theta }_{-i}\in \widehat{E}_{-i}$, \\ 
&  \\ 
$0\text{,}$ & otherwise.%
\end{tabular}%
\right.  \label{yyi7}
\end{equation}%
\begin{equation}
\rho ^{\text{MM}}\left[ f,\left( \gamma ,E\right) ,\theta ,\theta ^{\prime
},i,\widehat{E}\right] =\left\{ 
\begin{tabular}{ll}
$1\text{,}$ & if $\min_{\widehat{\theta }_{-i}\in \widehat{E}%
_{-i}}u_{i}^{\theta _{i}}\left[ f\left( \theta _{i},\widehat{\theta }%
_{-i}\right) \right] \geq \min_{\widehat{\theta }_{-i}\in \widehat{E}%
_{-i}}u_{i}^{\theta _{i}}\left[ f\left( \theta _{i}^{\prime },\widehat{%
\theta }_{-i}\right) \right] $, \\ 
&  \\ 
$0\text{,}$ & otherwise.%
\end{tabular}%
\right.  \label{yyi8}
\end{equation}%
\begin{equation}
\rho ^{\text{OD}}\left[ f,\left( \gamma ,E\right) ,\theta ,\theta ^{\prime
},i,\widehat{E}\right] =\left\{ 
\begin{tabular}{ll}
$1\text{,}$ & if $\min_{\widehat{\theta }_{-i}\in \widehat{E}%
_{-i}}u_{i}^{\theta _{i}}\left[ f\left( \theta _{i},\widehat{\theta }%
_{-i}\right) \right] \geq \max_{\widehat{\theta }_{-i}\in \widehat{E}%
_{-i}}u_{i}^{\theta _{i}}\left[ f\left( \theta _{i}^{\prime },\widehat{%
\theta }_{-i}\right) \right] $, \\ 
&  \\ 
$0\text{,}$ & otherwise.%
\end{tabular}%
\right.  \label{yyi9}
\end{equation}%
\begin{equation}
\rho ^{\text{SOD}}\left[ f,\left( \gamma ,E\right) ,\theta ,\theta ^{\prime
},i,\widehat{E}\right] =\left\{ 
\begin{tabular}{ll}
$1\text{,}$ & if $\left( 
\begin{array}{c}
\min_{\widehat{\theta }_{-i}\in \widehat{E}_{-i}}u_{i}^{\theta _{i}}\left[
f\left( \widetilde{\theta }_{i},\widehat{\theta }_{-i}\right) \right] \geq
\max_{\theta _{-i}\in \widehat{E}_{-i}}u_{i}^{\theta _{i}}\left[ f\left(
\theta _{i}^{\prime },\widehat{\theta }_{-i}\right) \right] , \\ 
\forall \widetilde{\theta }_{i}\in \gamma _{i}\left[ E,\theta \right]%
\end{array}%
\right) $, \\ 
&  \\ 
$0\text{,}$ & otherwise.%
\end{tabular}%
\right.  \label{yyi10}
\end{equation}

\subsection{Implementation}

For any $\left( \theta ,\gamma ,E\right) \in \Theta \times \Gamma ^{\text{%
Semi}}\times \left( \times _{i\in \mathcal{I}}\left[ 2^{\Theta
_{i}}\diagdown \left\{ \varnothing \right\} \right] \right) $, define%
\begin{equation*}
\gamma _{\left( 0\right) }\left[ E\text{ },\theta \right] =E\text{ and
inductively, }\gamma _{\left( n\right) }\left[ E\text{ },\theta \right]
=\gamma \left[ \text{ }\gamma _{\left( n-1\right) }\left[ E\text{ },\theta %
\right] ,\text{ }\theta \text{ }\right] \text{, }\forall n\in \mathbb{N}%
\text{, }
\end{equation*}

\begin{define}[achievability]
\label{def:solution:achieve}An operator $\gamma \in \Gamma $ is achievable
if there exists $N\in \mathbb{N}$ such that%
\begin{equation*}
\gamma _{\left( N\right) }\left[ \Theta \text{ },\theta \right] =\left\{
\theta \right\} \text{, }\forall \theta \in \Theta \text{.}
\end{equation*}%
Furthermore, given an SCF $f$, an operator $\gamma \in \Gamma $ is $f$%
-achievable if there exists $N\in \mathbb{N}$ such that%
\begin{equation*}
\left\{ f\left( \widetilde{\theta }\right) \in \mathcal{X}:\widetilde{\theta 
}\in \gamma _{\left( N\right) }\left[ \Theta \text{ },\theta \right]
\right\} =\left\{ f\left( \theta \right) \right\} \text{, }\forall \theta
\in \Theta \text{.}
\end{equation*}
\end{define}

Clearly, achievability is stronger than $f$-achievability.

\begin{define}[$\left( \protect\rho ,f,\protect\vartheta \right) $-consistent%
]
\label{def:solution:consistent}Given an SCF $f$, a solution notion $\rho $
and a function%
\begin{equation*}
\vartheta :\left( \times _{i\in \mathcal{I}}\left[ 2^{\Theta _{i}}\diagdown
\left\{ \varnothing \right\} \right] \right) \longrightarrow 2^{\Theta }%
\text{,}
\end{equation*}%
an operator $\gamma \in \Gamma $ is $\left( \rho ,f,\vartheta \right) $%
-consistent if 
\begin{equation}
\left( 
\begin{array}{c}
\left\{ \theta ,\theta ^{\prime }\right\} \subset E\text{,} \\ 
\rho \left[ f,\left( \gamma ,E\right) ,\theta ,\theta ^{\prime },i,\vartheta
\left( E\right) \right] =0%
\end{array}%
\right) \Longrightarrow \theta _{i}^{\prime }\in \gamma _{i}\left[ E,\theta %
\right] \text{, }\forall \left[ E,\theta ,\theta ^{\prime },i\right] \in
\left( \times _{i\in \mathcal{I}}\left[ 2^{\Theta _{i}}\diagdown \left\{
\varnothing \right\} \right] \right) \times \Theta \times \Theta \times 
\mathcal{I}\text{.}  \label{hhg3}
\end{equation}
\end{define}

\begin{define}[$\protect\rho $-implementation]
\label{def:solution:implementation}For any solution notion $\rho $, an SCF $%
f $ is $\rho $-implementable if there exists $\left[ G=\left[ \mathcal{I}%
\text{, }\mathcal{A}\text{, }\mathcal{H}\text{, }\mathcal{T}\text{, }K\text{%
, }\phi \text{, }A\text{, }\left( \zeta _{i}\right) _{i\in \mathcal{I}}\text{%
, }g\right] ,S\right] \in \mathcal{GS}^{\text{PC}}$ such that $\gamma ^{%
\left[ G,S\right] }$ is both $f$-achievable and $\left( \rho ,f,\vartheta ^{%
\left[ G\right] }\right) $-consistent.
\end{define}

For any solution concept $\varrho $ defined in Section \ref{sec:model}, let $%
\rho $ denote the induced solution notion (see (\ref{yyi6})-(\ref{yyi10})).
It is straightforward to show that an SCF $f$ is $\varrho $-implemented by $%
\left[ G,S\right] \in \mathcal{GS}^{\text{PC}}$ if and only if $f$ is $\rho $%
-implementable. To illustrate this, we adopt the solution concept of obvious
dominance. $f$ being $\rho ^{\text{OD}}$-implementable means existence of $%
\left[ G=\left[ \mathcal{I}\text{, }\mathcal{A}\text{, }\mathcal{H}\text{, }%
\mathcal{T}\text{, }K\text{, }\phi \text{, }A\text{, }\left( \zeta
_{i}\right) _{i\in \mathcal{I}}\text{, }g\right] ,S\right] \in \mathcal{GS}^{%
\text{PC}}$ such that $\gamma ^{\left[ G,S\right] }$ is both $f$-achievable
and $\left( \rho ,f,\vartheta ^{\left[ G\right] }\right) $-consistent.
First, $\gamma ^{\left[ G,S\right] }$ being $f$-achievable, which, together
with the definition of $\gamma ^{\left[ G,S\right] }$ in Section \ref%
{sec:operator:semi:game}, is equivalent to%
\begin{equation}
g\left[ T^{G}\left( \left[ S_{i}\left( \theta _{i}\right) \right] _{i\in 
\mathcal{I}}\right) \right] =f\left[ \left( \theta _{i}\right) _{i\in 
\mathcal{I}}\right] \text{, }\forall \left[ \left( \theta _{i}\right) _{i\in 
\mathcal{I}}\right] \in \Theta \text{.}  \label{ggh1}
\end{equation}%
Second, $\gamma ^{\left[ G,S\right] }$ being $\left( \rho ,f,\vartheta ^{%
\left[ G\right] }\right) $-consistent (together with (\ref{yyi9}) and
Definition \ref{def:solution:consistent}) is equivalent to%
\begin{equation}
\left( 
\begin{array}{c}
\left\{ \theta ,\theta ^{\prime }\right\} \subset \mathcal{E}^{\left[ G,S%
\right] \text{-}h}\text{,} \\ 
\phi \left( h\right) =i\text{,} \\ 
S_{i}\left( \theta _{i}\right) \left[ h\right] \neq S_{i}\left( \theta
_{i}^{\prime }\right) \left[ h\right]%
\end{array}%
\right) \Longrightarrow \min_{\widehat{\theta }_{-i}\in \vartheta _{-i}^{%
\left[ G\right] }\left( \mathcal{E}^{\left[ G,S\right] \text{-}h}\right)
}u_{i}^{\theta _{i}}\left[ f\left( \theta _{i},\widehat{\theta }_{-i}\right) %
\right] \geq \max_{\widehat{\theta }_{-i}\in \vartheta _{-i}^{\left[ G\right]
}\left( \mathcal{E}^{\left[ G,S\right] \text{-}h}\right) }u_{i}^{\theta _{i}}%
\left[ f\left( \theta _{i}^{\prime },\widehat{\theta }_{-i}\right) \right] 
\text{.}  \label{ggh2}
\end{equation}%
(\ref{ggh1}) and (\ref{ggh2}) constitute the definition of $f$ being $%
\varrho ^{\text{OD}}$-implemented by $\left[ G,S\right] $.

Note that (\ref{ggh2}) imposes the "consistency" condition only on agent $i$
with $\phi \left( h\right) =i$, while (\ref{hhg3})\ in Definition \ref%
{def:solution:consistent} requires it on all agents. The reason is that,
with $i\neq \phi \left( h\right) $, we have $\gamma _{i}^{\left[ G,S\right] }%
\left[ \mathcal{E}^{\left[ G,S\right] \text{-}h},\theta \right] =\mathcal{E}%
_{i}^{\left[ G,S\right] \text{-}h}$, and (\ref{hhg3}) holds automatically.
This is is why we do not need to keep track of the agent who is assigned to
take an action at history $h$ in $G$.

\subsection{Properties of operators and solution notions}

We consider three properties of solution notions.

\begin{define}
\label{def:solution:regular}A solution notion $\rho $ is regular if for any 
\begin{equation*}
\left[ f,\left( \gamma ,E\right) ,\theta ,\theta ^{\prime },i,\widehat{E}%
\right] \in \mathcal{X}^{\Theta }\times \Gamma ^{\text{Semi}}\times \left(
\times _{i\in \mathcal{I}}\left[ 2^{\Theta _{i}}\diagdown \left\{
\varnothing \right\} \right] \right) \times \Theta \times \Theta \times 
\mathcal{I}\times 2^{\Theta }\text{,}
\end{equation*}%
and any $\left( \overline{\gamma },\overline{E}\right) $ $\in \Gamma ^{\text{%
Semi}}\times \left( \times _{i\in \mathcal{I}}\left[ 2^{\Theta
_{i}}\diagdown \left\{ \varnothing \right\} \right] \right) $, we have%
\begin{equation*}
\rho \left[ f,\left( \gamma ,E\right) ,\theta ,\theta ^{\prime },i,\widehat{E%
}\right] =\rho \left[ f,\left( \overline{\gamma },\overline{E}\right)
,\theta ,\theta ^{\prime },i,\widehat{E}\right] \text{.}
\end{equation*}
\end{define}

\begin{define}
\label{def:solution:monotone}A solution notion $\rho $ is dissectible if for
any%
\begin{equation*}
\left[ f,\left( \gamma ,E\right) ,\theta ,\theta ^{\prime },i,\widehat{E}%
\right] \in \mathcal{X}^{\Theta }\times \Gamma ^{\text{Semi}}\times \left(
\times _{i\in \mathcal{I}}\left[ 2^{\Theta _{i}}\diagdown \left\{
\varnothing \right\} \right] \right) \times \Theta \times \Theta \times 
\mathcal{I}\times 2^{\Theta }\text{,}
\end{equation*}%
with $\left\{ \theta ,\theta ^{\prime }\right\} \subset E$, we have%
\begin{equation*}
\rho \left[ f,\left( \gamma ,E\right) ,\theta ,\theta ^{\prime },i,\widehat{E%
}\right] =0\Longleftrightarrow \left( 
\begin{array}{c}
\exists \widetilde{\theta }\in \gamma \left[ E,\text{ }\theta \right] \text{%
, } \\ 
\exists \widetilde{\gamma }\in \Gamma ^{\text{Semi}}\text{, }\widetilde{%
\gamma }\left[ E,\text{ }\theta \right] =\left\{ \theta ,\widetilde{\theta }%
\right\} \text{,} \\ 
\rho \left[ f,\left( \widetilde{\gamma },E\right) ,\theta ,\theta ^{\prime
},i,\widehat{E}\right] =0%
\end{array}%
\right) \text{,}
\end{equation*}%
\begin{equation*}
\text{or equivalently, }\rho \left[ f,\left( \gamma ,E\right) ,\theta
,\theta ^{\prime },i,\widehat{E}\right] =1\Longleftrightarrow \left( 
\begin{array}{c}
\forall \widetilde{\theta }\in \gamma \left[ E,\text{ }\theta \right] \text{,%
} \\ 
\forall \widetilde{\gamma }\in \Gamma ^{\text{Semi}}\text{ such that }%
\widetilde{\gamma }\left[ E,\text{ }\theta \right] =\left\{ \theta ,%
\widetilde{\theta }\right\} \text{,} \\ 
\rho \left[ f,\left( \widetilde{\gamma },E\right) ,\theta ,\theta ^{\prime
},i,\widehat{E}\right] =1%
\end{array}%
\right) \text{.}
\end{equation*}
\end{define}

\begin{define}
\label{def:solution:constant}A solution notion $\rho $ is normal if for any 
\begin{equation*}
\left[ f,\left( \gamma ,E\right) ,\theta ,\theta ^{\prime },i\right] \in 
\mathcal{X}^{\Theta }\times \Gamma ^{\text{Semi}}\times \left( \times _{i\in 
\mathcal{I}}\left[ 2^{\Theta _{i}}\diagdown \left\{ \varnothing \right\} %
\right] \right) \times \Theta \times \Theta \times \mathcal{I}\text{,}
\end{equation*}%
\begin{equation*}
\text{we have }\left( 
\begin{array}{c}
\left\{ \theta ,\theta ^{\prime }\right\} \subset E\text{,} \\ 
f\left( \widetilde{\theta }\right) =f\left( \theta \right) \text{, }\forall 
\widetilde{\theta }\in E\text{,}%
\end{array}%
\right) \Longrightarrow \rho \left[ f,\left( \gamma ,E\right) ,\theta
,\theta ^{\prime },i,E\right] =1\text{.}
\end{equation*}
\end{define}

Given $\rho $ being a regular solution notion, $\rho \left[ f,\left( \gamma
,E\right) ,\theta ,\theta ^{\prime },i,\widehat{E}\right] $ does not depend
on $\left( \gamma ,E\right) $. Given $\rho $ being a dissectible solution
notion, the value of $\rho \left[ f,\left( \gamma ,E\right) ,\theta ,\theta
^{\prime },i,\widehat{E}\right] $ is fully determined by the values of $\rho %
\left[ f,\left( \widetilde{\gamma },E\right) ,\theta ,\theta ^{\prime },i,%
\widehat{E}\right] $ for all $\left( \widetilde{\gamma }\text{, }\widetilde{%
\theta }\right) \in \Gamma ^{\text{Semi}}\times \gamma \left[ E,\text{ }%
\theta \right] $ with $\widetilde{\gamma }\left[ E,\text{ }\theta \right]
=\left\{ \theta ,\widetilde{\theta }\right\} $. Trivially, a regular
solution notion is also dissectible. Finally, for a normal solution notion,
truthful reporting is always a best reply on $E$, if $f$ is constant on $E$.
The following table describes the properties of the five solution notions
above (see (\ref{yyi6})-(\ref{yyi10})).\footnote{%
For strong-obvious dominance, $\rho ^{\text{SOD}}\left[ f,\left( \gamma
,E\right) ,\theta ,\theta ^{\prime },i,\widehat{E}\right] $ depends on $%
\gamma $.} 
\begin{equation*}
\begin{tabular}{|c|c|c|c|c|c|}
\hline
& $\mu \text{-PBE}$ & MM & WD* & OD & SOD \\ \hline
regular & yes & yes & yes & yes & no \\ \hline
dissectible & yes & yes & yes & yes & yes \\ \hline
normal & yes & yes & yes & yes & yes \\ \hline
\end{tabular}%
\text{.}
\end{equation*}

We consider one property of operators.

\begin{define}[increasing operators]
An operator $\gamma \in \Gamma $ is increasing if 
\begin{equation*}
\theta \in E\subset E^{\prime }\Longrightarrow \gamma \left[ E,\theta \right]
\subset \gamma \left[ E^{\prime },\theta \right] \text{, }\forall \left(
\theta ,E,E^{\prime }\right) \equiv \Theta \times \left( \times _{i\in 
\mathcal{I}}\left[ 2^{\Theta _{i}}\diagdown \left\{ \varnothing \right\} %
\right] \right) \times \left( \times _{i\in \mathcal{I}}\left[ 2^{\Theta
_{i}}\diagdown \left\{ \varnothing \right\} \right] \right) \text{.}
\end{equation*}
\end{define}

\section{Implementation by simultaneous-move games}

\label{sec:revelation-principle}

In this section, we study when implementation by a sequential-move game is
equivalent to implementation by a simultaneous-move game. Before players
take an action in a simultaneous-move game, no player has disclosed her
type, and hence, each player $i$ believes her opponents come from $\Theta
_{-i}$, and the standard revelation principle applies. For instance, $f$ can
be $\mu $-PBE-implemented by a simultaneous-move game if and only if%
\begin{equation*}
\int\limits_{\Theta _{-i}}\left( u_{i}^{\theta _{i}}\left[ f\left( \theta
_{i},\widetilde{\theta }_{-i}\right) \right] -u_{i}^{\theta _{i}}\left[
f\left( \theta _{i}^{\prime },\widetilde{\theta }_{-i}\right) \right]
\right) \mu ^{\theta _{i}}\left[ d\widetilde{\theta }_{-i}\right] \geq 0%
\text{, }\forall \left( \theta ,\theta ^{\prime },i\right) \in \Theta \times
\Theta \times \mathcal{I}\text{,}
\end{equation*}%
i.e., at every state, it is a best reply for every player to truthfully
report her type. Similarly, for any solution concept $\varrho $, we can
translate it to a solution notion $\rho $, and the standard revelation
principle defined on $\varrho $ is translated to the following condition on $%
\rho $.\footnote{%
In (\ref{yyi11}), we will consider regular solution notions only, which does
not depends on $\gamma $.}%
\begin{equation}
\rho \left[ f,\left( \gamma ,\Theta \right) ,\theta ,\theta ^{\prime
},i,\Theta \right] =1,\forall \left[ \gamma ,\theta ,\theta ^{\prime },i%
\right] \in \Gamma ^{\text{Semi}}\times \Theta \times \Theta \times \mathcal{%
I}\text{.}  \label{yyi11}
\end{equation}%
In particular, we will consider additive solution notions, which is defined
as follows.

\begin{define}[additive solution notion]
\label{def:solution:dissect}A solution notion $\rho $ is additive if for any 
\begin{equation*}
\left[ f,\gamma ,\theta ,\theta ^{\prime },i\right] \in \mathcal{X}^{\Theta
}\times \Gamma ^{\text{Semi}}\times \Theta \times \Theta \times \mathcal{I}%
\text{,}
\end{equation*}%
and any partition $P:\Theta \longrightarrow 2^{\Theta }$, (i.e., $\theta \in
P\left( \theta \right) $ and $\theta ^{\prime }\in P\left( \theta \right)
\Longrightarrow P\left( \theta \right) =P\left( \theta ^{\prime }\right) $),
we have%
\begin{equation*}
\left( 
\begin{array}{c}
\forall \widetilde{\theta }\in \Theta \text{,} \\ 
\rho \left[ f,\left( \gamma ,\Theta \right) ,\theta ,\theta ^{\prime
},i,P\left( \widetilde{\theta }\right) \right] =1%
\end{array}%
\right) \Longrightarrow \rho \left[ f,\left( \gamma ,\Theta \right) ,\theta
,\theta ^{\prime },i,\Theta \text{ }\right] =1\text{.}
\end{equation*}
\end{define}

Given an additive solution notion $\rho $ and a partition $P$ on $\Theta $,
if revealing $\theta _{i}$ is better than revealing $\theta _{i}^{\prime }$
on each $P\left( \widetilde{\theta }\right) $, then revealing $\theta _{i}$
is better than revealing $\theta _{i}^{\prime }$ on $\Theta $. It is easy to
check that $\mu $-PBE, weak-dominance* and max-min equilibrium are additive
(see (\ref{yyi6})-(\ref{yyi8})). We use the following result to establish
revelation principles for such solution concepts, and the proof is relegated
to Appendix \ref{sec:prop:revelation:additive}.

\begin{prop}
\label{prop:revelation:additive}For any solution notion $\rho $ which is
regular, normal and additive, an SCF $f:\Theta \longrightarrow \mathcal{X}$
is $\rho $-implementable if and only if%
\begin{equation*}
\rho \left[ f,\left( \gamma ,\Theta \right) ,\theta ,\theta ^{\prime
},i,\Theta \right] =1\text{, }\forall \left[ \gamma ,\theta ,\theta ^{\prime
},i\right] \in \Gamma ^{\text{Semi}}\times \Theta \times \Theta \times 
\mathcal{I}\text{.}
\end{equation*}
\end{prop}

Proposition \ref{prop:revelation:additive} immediately leads to the
following revelation principles.

\begin{theo}
\label{theorem:revelation:PBE}An SCF $f$ can be $\mu $-PBE-implemented by a
game with perfect recall if and only if%
\begin{equation*}
\int\limits_{\Theta _{-i}}\left( u_{i}^{\theta _{i}}\left[ f\left( \theta
_{i},\widetilde{\theta }_{-i}\right) \right] -u_{i}^{\theta _{i}}\left[
f\left( \theta _{i}^{\prime },\widetilde{\theta }_{-i}\right) \right]
\right) \mu ^{\theta _{i}}\left[ d\widetilde{\theta }_{-i}\right] \geq 0%
\text{, }\forall \left( \theta ,\theta ^{\prime },i\right) \in \Theta \times
\Theta \times \mathcal{I}\text{,}
\end{equation*}%
or equivalently, $f$ can be $\mu $-PBE-implemented by a simultaneous-move
game.
\end{theo}

\begin{theo}
\label{theorem:revelation:mm}An SCF $f$ can be max-min-implemented by a game
with perfect recall if and only if%
\begin{equation*}
\min_{\widetilde{\theta }_{-i}\in \Theta _{-i}}u_{i}^{\theta _{i}}\left[
f\left( \theta _{i},\widetilde{\theta }_{-i}\right) \right] \geq \min_{%
\widetilde{\theta }_{-i}\in \Theta _{-i}}u_{i}^{\theta _{i}}\left[ f\left(
\theta _{i}^{\prime },\widetilde{\theta }_{-i}\right) \right] \text{, }%
\forall \left( \theta ,\theta ^{\prime },i\right) \in \Theta \times \Theta
\times \mathcal{I}\text{,}
\end{equation*}%
or equivalently, $f$ can be max-min-implemented by a simultaneous-move game.
\end{theo}

Even though we consider weak-dominance* only (but not weak dominace), Lemma %
\ref{lem:weak*} and Proposition \ref{prop:revelation:additive} could help us
fully characterize weak-dominance-implementation.

\begin{theo}
\label{theorem:revelation:WD}An SCF $f$ can be weak-dominance-implemented
(i.e., strategyproof) by a game with perfect recall if and only if%
\begin{equation}
u_{i}^{\theta _{i}}\left[ f\left( \theta _{i},\theta _{-i}\right) \right]
\geq u_{i}^{\theta _{i}}\left[ f\left( \theta _{i}^{\prime },\theta
_{-i}\right) \right] \text{, }\forall \theta _{-i}\in \Theta _{-i}\text{,}
\label{ttl1}
\end{equation}%
or equivalently, $f$ can be WD-implemented by a simultaneous-move game.
\end{theo}

\noindent \textbf{Proof of Theorem \ref{theorem:revelation:WD}:} If (\ref%
{ttl1}) holds, the traditional direct mechanism implements $f$. If $f$ can
be weak-dominance-implemented by a game with perfect recall, Lemma \ref%
{lem:weak*} implies that it is weak-dominance*-implemented by the same game.
By Proposition \ref{prop:revelation:additive}, (\ref{ttl1}) holds.$%
\blacksquare $

\section{Monotonic solution concepts: revelation principle}

\label{sec:monotonic}

In this section, we consider non-additive solution notions (e.g., obvious
dominance, strong-obvious dominance), and it suffers loss of generality to
focus on simultaneous-move games. Nevertheless, obvious dominance and
strong-obvious dominance possess the following property.

\begin{define}[monotonic solution notion]
\label{def:solution:monotonic}A solution notion $\rho $ is monotonic if for
any%
\begin{equation*}
\left[ f,\left( \gamma ,E\right) ,\theta ,\theta ^{\prime },i,\widehat{E}%
\right] \in \mathcal{X}^{\Theta }\times \Gamma ^{\text{Semi}}\times \left(
\times _{i\in \mathcal{I}}\left[ 2^{\Theta _{i}}\diagdown \left\{
\varnothing \right\} \right] \right) \times \Theta \times \Theta \times 
\mathcal{I}\times 2^{\Theta }\text{,}
\end{equation*}%
and any $\widetilde{E}\in 2^{\Theta }$, we have $\left( 
\begin{array}{c}
\widehat{E}\subset \widetilde{E}\text{,} \\ 
\rho \left[ f,\left( \gamma ,E\right) ,\theta ,\theta ^{\prime },i,\widehat{E%
}\right] =0%
\end{array}%
\right) \Longrightarrow \rho \left[ f,\left( \gamma ,E\right) ,\theta
,\theta ^{\prime },i,\widetilde{E}\right] =0$.
\end{define}

\subsection{Canonical operator and revelation principle}

Based on any $\left( \rho ,f\right) $, we define a canonical operator $%
\gamma ^{\left( \rho ,f\right) }$. In particular, a profile $\left[ \left(
\rho ,f\right) ,\left( \gamma ,E\right) \right] $ induces a binary relation
on $\Theta _{i}$ as follows, which is denoted by $>_{i}^{\left[ \left( \rho
,f\right) ,\left( \gamma ,E\right) \right] }$.

\begin{equation}
\theta >_{i}^{\left[ \left( \rho ,f\right) ,\left( \gamma ,E\right) \right]
}\theta ^{\prime }\Longleftrightarrow \left( 
\begin{array}{c}
\left\{ \theta ,\theta ^{\prime }\right\} \subset E\text{ and} \\ 
\rho \left[ f,\left( \gamma ,E\right) ,\theta ,\theta ^{\prime },i,E\right]
=0%
\end{array}%
\right) \text{, }\forall \left( i,\theta ,\theta ^{\prime }\right) \in 
\mathcal{I}\times \Theta \times \Theta \text{.}  \label{kky1}
\end{equation}%
Furthermore, define%
\begin{equation}
\theta \sim _{i}^{\left[ \left( \rho ,f\right) ,\left( \gamma ,E\right) %
\right] }\theta ^{\prime }\Longleftrightarrow \left( 
\begin{array}{c}
\text{either }\theta _{i}=\theta _{_{i}}^{\prime }\text{,} \\ 
\text{or }\theta >_{i}^{\left[ \left( \rho ,f\right) ,\left( \gamma
,E\right) \right] }\theta ^{\prime }\text{,} \\ 
\text{or }\theta ^{\prime }>_{i}^{\left[ \left( \rho ,f\right) ,\left(
\gamma ,E\right) \right] }\theta%
\end{array}%
\right) \text{, }\forall \left( i,\theta ,\theta ^{\prime }\right) \in 
\mathcal{I}\times \Theta \times \Theta \text{.}  \label{kky4c}
\end{equation}%
Clearly, $\sim _{i}^{\left[ \left( \rho ,f\right) ,\left( \gamma ,E\right) %
\right] }$ is symmetric, i.e., 
\begin{equation}
\theta \sim ^{\left[ \left( \rho ,f\right) ,\left( \gamma ,E\right) \right]
}\theta ^{\prime }\Longleftrightarrow \theta ^{\prime }\sim ^{\left[ \left(
\rho ,f\right) ,\left( \gamma ,E\right) \right] }\theta \text{, }\forall
\left( i,\theta ,\theta ^{\prime }\right) \in \mathcal{I}\times \Theta
\times \Theta \text{,}  \label{kky4b}
\end{equation}%
For any $\left[ E,\text{ }\theta \right] \in \left( \times _{i\in \mathcal{I}%
}\left[ 2^{\Theta _{i}}\diagdown \left\{ \varnothing \right\} \right]
\right) \times \Theta $, we inductively define%
\begin{equation}
\gamma ^{\left( \rho ,f\right) \text{-}\left( 1\right) }\left[ E,\text{ }%
\theta \right] =\left\{ \theta \right\} \text{,}  \label{kky7}
\end{equation}%
\begin{equation}
\text{and }\gamma ^{\left( \rho ,f\right) \text{-}\left( n+1\right) }\left[
E,\text{ }\theta \right] =\cup_{\widetilde{\theta }\in \gamma
^{\left( \rho ,f\right) \text{-}\left( n\right) }\left[ E,\text{ }\theta %
\right] }\left\{ \theta ^{\prime }\in E:\widetilde{\theta }\sim _{i}^{\left[
\left( \rho ,f\right) ,\left( \gamma ^{\left( \rho ,f\right) \text{-}\left(
n\right) },E\right) \right] }\theta ^{\prime }\text{, }\forall i\in \mathcal{%
I}\right\} \text{, }\forall n\in \mathbb{N}\text{.}  \label{kky2}
\end{equation}%
Finally, define the canonical operator $\gamma ^{\left( \rho ,f\right) }$ as
follows.%
\begin{equation}
\gamma ^{\left( \rho ,f\right) }\left[ E,\text{ }\theta \right] =\left\{ 
\begin{tabular}{ll}
$E\text{,}$ & if $\theta \notin E$, \\ 
&  \\ 
$\cup_{n=1}^{\infty }\gamma ^{\left( \rho ,f\right) \text{-}%
\left( n\right) }\left[ E,\text{ }\theta \right] \text{,}$ & if $\theta \in
E $.%
\end{tabular}%
\right. \text{, }\forall \left[ E,\text{ }\theta \right] \in \left( \times
_{i\in \mathcal{I}}\left[ 2^{\Theta _{i}}\diagdown \left\{ \varnothing
\right\} \right] \right) \times \Theta \text{.}  \label{kky2b}
\end{equation}

$\gamma ^{\left( \rho ,f\right) }\left[ E,\text{ }\theta \right] |_{\theta
\in E}$ is aimed to be a partition on $E$, which must satisfy two necessary
conditions: (i) reflexivity: $\theta \in \gamma ^{\left( \rho ,f\right) }%
\left[ E,\text{ }\theta \right] $ (formalized in (\ref{kky7})), and (ii)
transitivity: $\theta ^{\prime \prime }\in \gamma ^{\left( \rho ,f\right) }%
\left[ E,\text{ }\theta ^{\prime }\right] $ and $\theta ^{\prime }\in \gamma
^{\left( \rho ,f\right) }\left[ E,\text{ }\theta \right] $ imply $\theta
^{\prime \prime }\in \gamma ^{\left( \rho ,f\right) }\left[ E,\text{ }\theta %
\right] $ (formalized in (\ref{kky2})). The following result establishes
revelation principles for monotonic solution notions.

\begin{prop}
\label{prop:monotone}Consider any SCF $f$ and any solution notion $\rho $
which is dissectible, monotonic and normal. Then, $f$ is $\rho $%
-implementable if and only if $\gamma ^{\left( \rho ,f\right) }$ is
achievable.
\end{prop}

\subsection{Proof of Proposition \protect\ref{prop:monotone}}

The following five lemmas describe properties of the canonical
(semi-)operator $\gamma ^{\left( \rho ,f\right) }$, which play critical
roles in our proof of Proposition \ref{prop:monotone}. Their proofs are
relegated to Appendix \ref{sec:lem:operator:canonical:operator}-\ref%
{sec:lem:operator:normal}.

\begin{lemma}
\label{lem:operator:canonical:operator}For any SCF $f$ and any dissectible
solution notion $\rho $, $\gamma ^{\left( \rho ,f\right) }$ is an operator
which is $\left( \rho ,f,\vartheta ^{\ast }\right) $-consistent.
\end{lemma}

\begin{lemma}
\label{lem:operator:canonical:increasing}For any SCF $f$ and any dissectible
and monotonic solution notion $\rho $, $\gamma ^{\left( \rho ,f\right) }$ is
an increasing operator.
\end{lemma}

\begin{lemma}
\label{lem:operator:canonical:lower}Consider any SCF $f$ and any solution
notion $\rho $ which is dissectible and monotonic. For any $\left[ G,S\right]
\in \mathcal{GS}^{\text{PC}}$ such that $\gamma ^{\left[ G,S\right] }$ is $%
\left( \rho ,f,\vartheta ^{\left[ G\right] }\right) $-consistent, we have%
\begin{equation*}
\gamma ^{\left( \rho ,f\right) }\left[ E,\text{ }\theta \right] \subset
\gamma ^{\left[ G,S\right] }\left[ E,\text{ }\theta \right] \text{, }\forall %
\left[ E,\text{ }\theta \right] \in \left( \times _{i\in \mathcal{I}}\left[
2^{\Theta _{i}}\diagdown \left\{ \varnothing \right\} \right] \right) \times
\Theta \text{.}
\end{equation*}
\end{lemma}

\begin{lemma}
\label{lem:operator:game-operator}Consider any SCF $f$ and any dissectible
solution notion $\rho $. If $\gamma ^{\left( \rho ,f\right) }$ is
achievable, there exists $\left[ G,S\right] \in \mathcal{GS}^{\text{PC}}$
such that $\gamma ^{\left[ G,S\right] }$ is both achievable and $\left( \rho
,f,\vartheta ^{\left[ G\right] }\right) $-consistent.
\end{lemma}

\begin{lemma}
\label{lem:operator:normal}For any SCF $f$ and any normal solution notion $%
\rho $, the semi-operator $\gamma ^{\left( \rho ,f\right) }$ is achievable
if and only if $\gamma ^{\left( \rho ,f\right) }$ is $f$-achievable.
\end{lemma}

\noindent \textbf{Proof of Proposition \ref{prop:monotone}: } Consider any
SCF $f$ and any solution notion $\rho $ which is dissectible, monotonic and
normal. First, suppose that $f$ is $\rho $-implementable. By Definition \ref%
{def:solution:implementation}, there exists $\left[ G=\left[ \mathcal{I}%
\text{, }\mathcal{A}\text{, }\mathcal{H}\text{, }\mathcal{T}\text{, }K\text{%
, }\phi \text{, }A\text{, }\left( \zeta _{i}\right) _{i\in \mathcal{I}}\text{%
, }g\right] \text{, }S\right] \in \mathcal{GS}^{\text{PC}}$ such that $%
\gamma ^{\left[ G,S\right] }$ is both $f$-achievable and $\left( \rho
,f,\vartheta ^{\left[ G\right] }\right) $-consistent.

$\gamma ^{\left[ G,S\right] }$ being $f$-achievable implies%
\begin{equation}
f\left( \widetilde{\theta }\right) =f\left( \theta \right) \text{, }\forall
\theta \in \Theta \text{, }\forall \widetilde{\theta }\in \gamma _{\left(
K\right) }^{\left[ G,S\right] }\left[ \Theta \text{ },\theta \right] \text{.}
\label{kkt1}
\end{equation}%
We now prove%
\begin{equation}
\gamma _{\left( k\right) }^{\left( \rho ,f\right) }\left[ \Theta ,\text{ }%
\theta \right] \subset \gamma _{\left( k\right) }^{\left[ G,S\right] }\left[
\Theta ,\text{ }\theta \right] \text{, }\forall k\in \left\{
0,1,...,K\right\} \text{, }\forall \theta \in \Theta \text{.}  \label{kkt2}
\end{equation}%
With $k=0$, we have $\gamma _{\left( 0\right) }^{\left( \rho ,f\right) }%
\left[ \Theta ,\text{ }\theta \right] =\Theta =\gamma _{\left( 0\right) }^{%
\left[ G,S\right] }\left[ \Theta ,\text{ }\theta \right] $, i.e., (\ref{kkt2}%
) holds. Suppose (\ref{kkt2}) holds for $k=n<K$. Consider any $\theta \in
\Theta $ and $k=n+1$, and we have%
\begin{equation*}
\gamma _{\left( n+1\right) }^{\left( \rho ,f\right) }\left[ \Theta ,\text{ }%
\theta \right] =\gamma ^{\left( \rho ,f\right) }\left( \gamma _{\left(
n\right) }^{\left( \rho ,f\right) }\left[ \Theta ,\text{ }\theta \right]
\right) \subset \gamma ^{\left( \rho ,f\right) }\left( \gamma _{\left(
n\right) }^{\left[ G,S\right] }\left[ \Theta ,\text{ }\theta \right] \right)
\subset \gamma ^{\left[ G,S\right] }\left( \gamma _{\left( n\right) }^{\left[
G,S\right] }\left[ \Theta ,\text{ }\theta \right] \right) =\gamma _{\left(
n+1\right) }^{\left[ G,S\right] }\left[ \Theta ,\text{ }\theta \right] \text{%
,}
\end{equation*}%
where the first $\subset $ follows from the induction hypothesis and Lemma %
\ref{lem:operator:canonical:increasing}, and the second $\subset $ follows
from $\gamma ^{\left[ G,S\right] }$ being $\left( \rho ,f,\vartheta ^{\left[
G\right] }\right) $-consistent and Lemma \ref{lem:operator:canonical:lower}.
Therefore, (\ref{kkt2}) holds.

(\ref{kkt1}) and (\ref{kkt2}) imply that $\gamma ^{\left( \rho ,f\right) }$
is $f$-achievable, which, together with Lemma \ref{lem:operator:normal},
implies that $\gamma ^{\left( \rho ,f\right) }$ is achievable. This proves
the "only if" part.

Second, suppose that $\gamma ^{\left( \rho ,f\right) }$ is achievable. By \
Lemma \ref{lem:operator:game-operator}, there exists $\left[ G,S\right] \in 
\mathcal{GS}^{\text{PC}}$ such that $\gamma ^{\left[ G,S\right] }$ is both
achievable and $\left( \rho ,f,\vartheta ^{\left[ G\right] }\right) $%
-consistent. Therefore, $f$ is $\rho $-implementable. This proves the "if"
part.$\blacksquare $

\subsection{Obvious dominance}

We apply Proposition \ref{prop:monotone} to obvious dominance which is
dissectible, monotonic and normal. Define%
\begin{equation*}
\theta ^{\prime }>_{i}^{\text{OD-}f\text{-}E}\theta \Longleftrightarrow
\left( 
\begin{array}{c}
\left\{ \theta ,\theta ^{\prime }\right\} \subset E\text{ and } \\ 
\min_{\theta _{-i}\in E_{-i}}u_{i}^{\theta _{i}}\left[ f\left( \theta
_{i},\theta _{-i}\right) \right] <\max_{\theta _{-i}\in E_{-i}}u_{i}^{\theta
_{i}}\left[ f\left( \theta _{i}^{\prime },\theta _{-i}\right) \right]%
\end{array}%
\right) \text{, }\forall \left( i,\theta ,\theta ^{\prime }\right) \in 
\mathcal{I}\times \Theta \times \Theta \text{.}
\end{equation*}%
For any $\left[ E,\text{ }\theta \right] \in \left( \times _{i\in \mathcal{I}%
}\left[ 2^{\Theta _{i}}\diagdown \left\{ \varnothing \right\} \right]
\right) \times \Theta $, define $\gamma ^{\text{OD-}f\text{-}\left( 1\right)
}\left[ E,\text{ }\theta \right] =\left\{ \theta \right\} $ and%
\begin{equation*}
\gamma ^{\text{OD-}f\text{-}\left( n+1\right) }\left[ E,\text{ }\theta %
\right] =\cup_{\widetilde{\theta }\in \gamma ^{\text{OD-}f\text{-}%
\left( n\right) }\left[ E,\text{ }\theta \right] }\left\{ \theta ^{\prime
}\in E:\forall i\in \mathcal{I}\text{, }\left( 
\begin{tabular}{l}
either $\theta _{i}^{\prime }=\widetilde{\theta }_{i}$, \\ 
or $\theta ^{\prime }>_{i}^{\text{OD-}f\text{-}E}\widetilde{\theta }$ \\ 
or $\widetilde{\theta }>_{i}^{\text{OD-}f\text{-}E}\theta ^{\prime }$%
\end{tabular}%
\right) \right\} \text{, }\forall n\in \mathbb{N}\text{.}
\end{equation*}%
Finally,%
\begin{equation*}
\gamma ^{\text{OD-}f}\left[ E,\text{ }\theta \right] \equiv \left\{ 
\begin{tabular}{ll}
$E\text{,}$ & if $\theta \notin E$, \\ 
&  \\ 
$\cup_{n=1}^{\infty }\gamma ^{\text{OD-}f\text{-}\left( n\right) }%
\left[ E,\text{ }\theta \right] \text{,}$ & if $\theta \in E$.%
\end{tabular}%
\right. \text{.}
\end{equation*}

\begin{theo}
\label{theorem:revelation:OD}An SCF $f:\Theta \longrightarrow \mathcal{X}$
is OSP if and only if $\gamma ^{\text{OD-}f}$ is achievable.
\end{theo}

\subsection{Strong-obvious dominance}

We apply Proposition \ref{prop:monotone} to strong-obvious dominance which
is dissectible, monotonic and normal. Given $\left[ f,\left( \gamma
,E\right) \right] $ and $\left( i,\theta ,\theta ^{\prime }\right) \in 
\mathcal{I}\times \Theta \times \Theta $, define

\begin{equation*}
\theta ^{\prime }>_{i}^{\text{SOD-}f\text{-}\left( \gamma ,E\right) }\theta
\Longleftrightarrow \left( 
\begin{array}{c}
\left\{ \theta ,\theta ^{\prime }\right\} \subset E\text{ and} \\ 
\min_{\widetilde{\theta }\in \gamma \left[ E,\theta \right] }\min_{\theta
_{-i}\in E_{-i}}u_{i}^{\theta _{i}}\left[ f\left( \widetilde{\theta }%
_{i},\theta _{-i}\right) \right] <\max_{\theta _{-i}\in E_{-i}}u_{i}^{\theta
_{i}}\left[ f\left( \theta _{i}^{\prime },\theta _{-i}\right) \right]%
\end{array}%
\right) \text{.}
\end{equation*}%
For any $\left[ E,\text{ }\theta \right] \in \left( \times _{i\in \mathcal{I}%
}\left[ 2^{\Theta _{i}}\diagdown \left\{ \varnothing \right\} \right]
\right) \times \Theta $, define $\gamma ^{\text{SOD-}f\text{-}\left(
1\right) }\left[ E,\text{ }\theta \right] =\left\{ \theta \right\} $ and 
\begin{equation*}
\gamma ^{\text{SOD-}f\text{-}\left( n+1\right) }\left[ E,\text{ }\theta %
\right] =\cup_{\widetilde{\theta }\in \gamma ^{\text{SOD-}f\text{-%
}\left( n\right) }\left[ E,\text{ }\theta \right] }\left\{ \theta ^{\prime
}\in E:\forall i\in \mathcal{I}\text{, }\left( 
\begin{tabular}{l}
either $\theta _{i}^{\prime }=\widetilde{\theta }_{i}$, \\ 
or $\theta ^{\prime }>_{i}^{\text{SOD-}f\text{-}\left( \gamma ^{\text{SOD-}f%
\text{-}\left( n\right) },E\right) }\widetilde{\theta }$ \\ 
or $\widetilde{\theta }>_{i}^{\text{SOD-}f\text{-}\left( \gamma ^{\text{SOD-}%
f\text{-}\left( n\right) },E\right) }\theta ^{\prime }$%
\end{tabular}%
\right) \right\} \text{, }\forall n\in \mathbb{N}\text{.}
\end{equation*}%
Finally,%
\begin{equation*}
\gamma ^{\text{SOD-}f}\left[ E,\text{ }\theta \right] \equiv \left\{ 
\begin{tabular}{ll}
$E\text{,}$ & if $\theta \notin E$, \\ 
&  \\ 
$\cup_{n=1}^{\infty }\gamma ^{\text{SOD-}f\text{-}\left( n\right)
}\left[ E,\text{ }\theta \right] \text{,}$ & if $\theta \in E$.%
\end{tabular}%
\right. \text{.}
\end{equation*}

\begin{theo}
\label{theorem:revelation:SOD}An SCF $f:\Theta \longrightarrow \mathcal{X}$
is SOSP if and only if $\gamma ^{\text{SOD-}f}$ is achievable.
\end{theo}

\section{Conclusion}

\label{sec:conclude}

Revelation principle is a pillar in traditional mechanism design which
focuses on simultaneous-move games. In this paper, we propose a general
approach to establish the analogous revelation principle for mechanism
design with sequential-move games.

\appendix

\section{Proofs}

\subsection{Proof of Lemma \protect\ref{lem:perfect-recall}}

\label{sec:lem:perfect-recall}

Consider any $G\in \mathcal{G}^{\text{PC}}$, any $S\in \mathcal{S}^{G}$, and
any $\left( i,h,h^{\prime }\right) \in \mathcal{I}\times \left[ \mathcal{H}%
\diagdown \mathcal{T}\right] \times \left[ \mathcal{H}\diagdown \mathcal{T}%
\right] $ such that 
\begin{equation*}
h\overset{G}{\sim }h^{\prime }\text{, }\phi \left( h\right) =\phi \left(
h^{\prime }\right) =i\text{, }\mathcal{E}^{\left[ G,S\right] \text{-}h}\neq
\varnothing \text{, and }\mathcal{E}^{\left[ G,S\right] \text{-}h^{\prime
}}\neq \varnothing \text{.}
\end{equation*}%
We prove $\mathcal{E}_{i}^{\left[ G,S\right] \text{-}h}=\mathcal{E}_{i}^{%
\left[ G,S\right] \text{-}h^{\prime }}$ by contradiction. Suppose $\mathcal{E%
}_{i}^{\left[ G,S\right] \text{-}h}\neq \mathcal{E}_{i}^{\left[ G,S\right] 
\text{-}h^{\prime }}$. Without loss of generality, suppose there exists $%
\theta _{i}\in \mathcal{E}_{i}^{\left[ G,S\right] \text{-}h}\diagdown 
\mathcal{E}_{i}^{\left[ G,S\right] \text{-}h^{\prime }}$. Pick any $\left[
\theta _{-i},\left( \theta _{i}^{\prime },\theta _{-i}^{\prime }\right) %
\right] \in \Theta _{-i}\times \Theta $ such that%
\begin{equation*}
\left( \theta _{i},\theta _{-i}\right) \in \mathcal{E}^{\left[ G,S\right] 
\text{-}h}\text{, and }\left( \theta _{i}^{\prime },\theta _{-i}^{\prime
}\right) \in \mathcal{E}^{\left[ G,S\right] \text{-}h^{\prime }}\text{.}
\end{equation*}%
In particular, $\theta _{i}\notin \mathcal{E}_{i}^{\left[ G,S\right] \text{-}%
h^{\prime }}$ implies that, upon reaching $h^{\prime }$, player $i$ has
already revealed that she is not of type $\theta _{i}$. Thus, there exists $%
\widehat{h}^{\prime }\in $path$\left[ h^{\prime }\right] $ such that $\phi
\left( \widehat{h}^{\prime }\right) =i$, and%
\begin{equation}
S_{i}\left( \theta _{i}\right) \left[ \widehat{h}^{\prime }\right] \neq
S_{i}\left( \theta _{i}^{\prime }\right) \left[ \widehat{h}^{\prime }\right] 
\text{.}  \label{tti1}
\end{equation}%
Since $G\in \mathcal{G}^{\text{PC}}$, Definition \ref{lem:perfect-recall}
implies existence of $\widehat{h}\in $path$\left[ h\right] $ such that $%
\widehat{h}\overset{G}{\sim }\widehat{h}^{\prime }$ and%
\begin{equation*}
\left[ \widehat{h},\text{ }S_{i}\left( \theta _{i}^{\prime }\right) \left[ 
\widehat{h}^{\prime }\right] \right] \in \text{path}\left[ h\right] \text{,}
\end{equation*}%
which, together with $\left( \theta _{i},\theta _{-i}\right) \in \mathcal{E}%
^{\left[ G,S\right] \text{-}h}$, implies%
\begin{equation}
S_{i}\left( \theta _{i}\right) \left[ \widehat{h}\right] =S_{i}\left( \theta
_{i}^{\prime }\right) \left[ \widehat{h}^{\prime }\right] \text{.}
\label{tti2}
\end{equation}%
Furthermore, $\widehat{h}\overset{G}{\sim }\widehat{h}^{\prime }$ implies%
\begin{equation}
S_{i}\left( \theta _{i}\right) \left[ \widehat{h}\right] =S_{i}\left( \theta
_{i}\right) \left[ \widehat{h}^{\prime }\right] \text{.}  \label{tti3}
\end{equation}%
Thus, (\ref{tti2}) and (\ref{tti3}) imply $S_{i}\left( \theta _{i}^{\prime
}\right) \left[ \widehat{h}^{\prime }\right] =S_{i}\left( \theta _{i}\right) %
\left[ \widehat{h}^{\prime }\right] $, contradicting (\ref{tti1}).$%
\blacksquare $

\subsection{Proof of Proposition \protect\ref{prop:revelation:additive}}

\label{sec:prop:revelation:additive}

Consider any SCF $f:\Theta \longrightarrow \mathcal{X}$ and any solution
notion $\rho $ which is regular, normal and additive. First, suppose 
\begin{equation}
\rho \left[ f,\left( \gamma ,\Theta \right) ,\theta ,\theta ^{\prime
},i,\Theta \right] =1\text{, }\forall \left[ \gamma ,\theta ,\theta ^{\prime
},i\right] \in \Gamma ^{\text{Semi}}\times \Theta \times \Theta \times 
\mathcal{I}\text{.}  \label{iit1}
\end{equation}%
Clearly, $f$ can be $\rho $-implemented by the simultaneous-move direct
mechanism $f$, and (\ref{iit1})\ implies that it is always a best reply for
every player to truthfully reveal her type. Therefore, the "if" part of
Proposition \ref{prop:revelation:additive} holds.

Second, suppose $f$ is $\rho $-implementable, i.e., there exists $\left[ G,S%
\right] \in \mathcal{GS}^{\text{PC}}$ such that $\gamma ^{\left[ G,S\right]
} $ is both $f$-achievable and $\left( \rho ,f,\vartheta ^{\left[ G\right]
}\right) $-consistent. Fix any $\left[ \gamma ,\theta ,\theta ^{\prime },i%
\right] \in \Gamma ^{\text{Semi}}\times \Theta \times \Theta \times \mathcal{%
I}$, we aim to prove%
\begin{equation}
\rho \left[ f,\left( \gamma ,\Theta \right) ,\theta ,\theta ^{\prime
},i,\Theta \right] =1\text{,}  \label{iit2}
\end{equation}%
i.e., the "only if" part of Proposition \ref{prop:revelation:additive} holds.

For any $\widetilde{\theta }_{-i}\in \Theta _{-i}$, consider%
\begin{equation*}
\mathcal{H}^{\left( \theta _{i},\theta _{i}^{\prime },\widetilde{\theta }%
_{-i}\right) }\equiv \left\{ h\in \mathcal{H}:h\in \text{Path}\left( T^{G}%
\left[ S_{i}\left( \theta _{i}\right) \text{, }S_{-i}\left( \widetilde{%
\theta }_{-i}\right) \right] \right) \cap \text{Path}\left( T^{G}\left[
S_{i}\left( \theta _{i}^{\prime }\right) \text{, }S_{-i}\left( \widetilde{%
\theta }_{-i}\right) \right] \right) \right\} \text{.}
\end{equation*}%
Clearly, $\mathcal{H}^{\left( \theta _{i},\theta _{i}^{\prime },\widetilde{%
\theta }_{-i}\right) }\neq \varnothing $, because $\sigma \in \mathcal{H}%
^{\left( \theta _{i},\theta _{i}^{\prime },\widetilde{\theta }_{-i}\right) }$%
. Since $\left[ G,S\right] \in \mathcal{GS}^{\text{PC}}$, condition (C) in (%
\ref{tkk1}) implies existence a unique maximal element in $\mathcal{H}%
^{\left( \theta _{i},\theta _{i}^{\prime },\widetilde{\theta }_{-i}\right) }$%
, i.e., 
\begin{equation*}
\exists !h\in \mathcal{H}^{\left( \theta _{i},\theta _{i}^{\prime },%
\widetilde{\theta }_{-i}\right) }\text{, }h^{\prime }\in \mathcal{H}^{\left(
\theta _{i},\theta _{i}^{\prime },\widetilde{\theta }_{-i}\right)
}\Longrightarrow h^{\prime }\in \text{Path}\left( h\right) \text{.}
\end{equation*}%
We denote this unique maximal element by $h^{\left( \theta _{i},\theta
_{i}^{\prime },\widetilde{\theta }_{-i}\right) }$. This immediately implies%
\begin{equation}
h^{\left( \theta _{i},\theta _{i}^{\prime },\widetilde{\theta }_{-i}\right)
}\in \mathcal{H}^{\left( \theta _{i},\theta _{i}^{\prime },\widetilde{\theta 
}_{-i}\right) }\diagdown \mathcal{T}\Longrightarrow \left( 
\begin{array}{c}
\phi \left( h^{\left( \theta _{i},\theta _{i}^{\prime },\widetilde{\theta }%
_{-i}\right) }\right) =i\text{,} \\ 
\text{and }S_{i}\left( \theta _{i}\right) \left[ h^{\left( \theta
_{i},\theta _{i}^{\prime },\widetilde{\theta }_{-i}\right) }\right] \neq
S_{i}\left( \theta _{i}^{\prime }\right) \left[ h^{\left( \theta _{i},\theta
_{i}^{\prime },\widetilde{\theta }_{-i}\right) }\right]%
\end{array}%
\right) \text{.}  \label{iit8}
\end{equation}%
Define $P_{-i}^{\left( \theta ,\theta ^{\prime }\right) }:\Theta
_{-i}\longrightarrow 2^{\Theta _{-i}}$ as follows.%
\begin{equation}
P_{-i}^{\left( \theta ,\theta ^{\prime }\right) }\left( \widetilde{\theta }%
_{-i}\right) =\left\{ 
\begin{array}{rl}
\mathcal{E}_{-i}^{\left[ G,S\right] \text{-}h^{\left( \theta _{i},\theta
_{i}^{\prime },\widetilde{\theta }_{-i}\right) }}\text{,} & \text{if }%
h^{\left( \theta _{i},\theta _{i}^{\prime },\widetilde{\theta }_{-i}\right)
}\in \mathcal{T}\text{,} \\ 
&  \\ 
\vartheta _{-i}^{\left[ G\right] }\left[ \mathcal{E}^{\left[ G,S\right] 
\text{-}h^{\left( \theta _{i},\theta _{i}^{\prime },\widetilde{\theta }%
_{-i}\right) }}\right] \text{,} & \text{if }h^{\left( \theta _{i},\theta
_{i}^{\prime },\widetilde{\theta }_{-i}\right) }\notin \mathcal{T}\text{,}%
\end{array}%
\right.  \label{iit7}
\end{equation}%
where $\vartheta ^{\left[ G\right] }$ is defined in Section \ref%
{sec:operator:semi:game}, and%
\begin{eqnarray*}
\mathcal{E}_{-i}^{\left[ G,S\right] \text{-}h^{\left( \theta _{i},\theta
_{i}^{\prime },\widetilde{\theta }_{-i}\right) }} &\equiv &\left\{ \widehat{%
\theta }_{-i}\in \Theta _{-i}:\exists \overline{\theta }_{i}\in \Theta _{i}%
\text{, }\left( \overline{\theta }_{i},\widehat{\theta }_{-i}\right) \in 
\mathcal{E}^{\left[ G,S\right] \text{-}h^{\left( \theta _{i},\theta
_{i}^{\prime },\widetilde{\theta }_{-i}\right) }}\right\} \text{,} \\
\vartheta _{-i}^{\left[ G\right] }\left[ \mathcal{E}^{\left[ G,S\right] 
\text{-}h^{\left( \theta _{i},\theta _{i}^{\prime },\widetilde{\theta }%
_{-i}\right) }}\right] &\equiv &\left\{ \widehat{\theta }_{-i}\in \Theta
_{-i}:\exists \overline{\theta }_{i}\in \Theta _{i}\text{, }\left( \overline{%
\theta }_{i},\widehat{\theta }_{-i}\right) \in \vartheta ^{\left[ G\right]
}\left( \mathcal{E}^{\left[ G,S\right] \text{-}h^{\left( \theta _{i},\theta
_{i}^{\prime },\widetilde{\theta }_{-i}\right) }}\right) \right\} \text{.}
\end{eqnarray*}%
That is, if $h^{\left( \theta _{i},\theta _{i}^{\prime },\widetilde{\theta }%
_{-i}\right) }$ is terminal, $P_{-i}^{\left( \theta ,\theta ^{\prime
}\right) }\left( \widetilde{\theta }_{-i}\right) $ is the set of $\widehat{%
\theta }_{-i}$ such that $S\left( \theta _{i},\widehat{\theta }_{-i}\right) $
leads to $h^{\left( \theta _{i},\theta _{i}^{\prime },\widetilde{\theta }%
_{-i}\right) }$, and if $h^{\left( \theta _{i},\theta _{i}^{\prime },%
\widetilde{\theta }_{-i}\right) }$ is non-terminal, $P_{-i}^{\left( \theta
,\theta ^{\prime }\right) }\left( \widetilde{\theta }_{-i}\right) $ is the
set of $\widehat{\theta }_{-i}$ such that $S\left( \theta _{i},\widehat{%
\theta }_{-i}\right) $ leads to the information set containing $h^{\left(
\theta _{i},\theta _{i}^{\prime },\widetilde{\theta }_{-i}\right) }$.

Since $\gamma ^{\left[ G,S\right] }$ is $\left( \rho ,f,\vartheta ^{\left[ G%
\right] }\right) $-consistent, (\ref{iit8}) implies%
\begin{equation*}
h^{\left( \theta _{i},\theta _{i}^{\prime },\widetilde{\theta }_{-i}\right)
}\notin \mathcal{T}\Longrightarrow \rho \left[ f,\left( \gamma ^{\left[ G,S%
\right] },\mathcal{E}^{\left[ G,S\right] \text{-}h^{\left( \theta
_{i},\theta _{i}^{\prime },\widetilde{\theta }_{-i}\right) }}\right) ,\theta
,\theta ^{\prime },i,\text{ }\vartheta ^{\left[ G\right] }\left( \mathcal{E}%
^{\left[ G,S\right] \text{-}h^{\left( \theta _{i},\theta _{i}^{\prime },%
\widetilde{\theta }_{-i}\right) }}\right) \right] =1\text{, }\forall 
\widetilde{\theta }\in \Theta \text{,}
\end{equation*}%
which, together with $\rho $ being regular and (\ref{hht2}) in Definition %
\ref{def:solution}, implies%
\begin{equation}
h^{\left( \theta _{i},\theta _{i}^{\prime },\widetilde{\theta }_{-i}\right)
}\notin \mathcal{T}\Longrightarrow \rho \left[ f,\left( \gamma ,\Theta
\right) ,\theta ,\theta ^{\prime },i,\text{ }\Theta _{i}\times \vartheta
_{-i}^{\left[ G\right] }\left( \mathcal{E}^{\left[ G,S\right] \text{-}%
h^{\left( \theta _{i},\theta _{i}^{\prime },\widetilde{\theta }_{-i}\right)
}}\right) \right] =1\text{, }\forall \left( \gamma ,\widetilde{\theta }%
\right) \in \Gamma ^{\text{Semi}}\times \Theta \text{.}  \label{iit3}
\end{equation}%
Since $\gamma ^{\left[ G,S\right] }$ is $f$-achievable, we have%
\begin{equation*}
h^{\left( \theta _{i},\theta _{i}^{\prime },\widetilde{\theta }_{-i}\right)
}\in \mathcal{T}\Longrightarrow \left( 
\begin{array}{c}
f\left( \widehat{\theta }\right) =f\left( \theta _{i},\widetilde{\theta }%
_{-i}\right) \text{,} \\ 
\forall \widehat{\theta }\in \mathcal{E}^{\left[ G,S\right] \text{-}%
h^{\left( \theta _{i},\theta _{i}^{\prime },\widetilde{\theta }_{-i}\right)
}}%
\end{array}%
\right) \text{, }\forall \widetilde{\theta }\in \Theta \text{,}
\end{equation*}%
which, together with $\rho $ being normal and regular, implies%
\begin{equation}
h^{\left( \theta _{i},\theta _{i}^{\prime },\widetilde{\theta }_{-i}\right)
}\in \mathcal{T}\Longrightarrow \rho \left[ f,\left( \gamma ,\Theta \right)
,\theta ,\theta ^{\prime },i,\text{ }\Theta _{i}\times \mathcal{E}_{-i}^{%
\left[ G,S\right] \text{-}h^{\left( \theta _{i},\theta _{i}^{\prime },%
\widetilde{\theta }_{-i}\right) }}\right] =1\text{, }\forall \left( \gamma ,%
\widetilde{\theta }\right) \in \Gamma ^{\text{Semi}}\times \Theta \text{.}
\label{iit4}
\end{equation}%
Define $P^{\left( \theta ,\theta ^{\prime }\right) }:\Theta \longrightarrow
2^{\Theta }$ as follows.%
\begin{equation}
P^{\left( \theta ,\theta ^{\prime }\right) }\left( \widetilde{\theta }%
\right) =\Theta _{i}\times P_{-i}^{\left( \theta ,\theta ^{\prime }\right)
}\left( \widetilde{\theta }_{-i}\right) \text{, }\forall \widetilde{\theta }%
\in \Theta \text{.}  \label{iit5}
\end{equation}%
(\ref{iit3}), (\ref{iit4}) and (\ref{iit5})\ imply%
\begin{equation}
\rho \left[ f,\left( \gamma ,\Theta \right) ,\theta ,\theta ^{\prime },i,%
\text{ }P^{\left( \theta ,\theta ^{\prime }\right) }\left( \widetilde{\theta 
}\right) \right] =1\text{, }\forall \left( \gamma ,\widetilde{\theta }%
\right) \in \Gamma \times \Theta \text{.}  \label{iit6}
\end{equation}%
We will prove that $P^{\left( \theta ,\theta ^{\prime }\right) }$ is a
partition on $\Theta $, which, together with $\rho $ being additive, implies
(\ref{iit2}), i.e., our goal.

Finally, we prove that $P^{\left( \theta ,\theta ^{\prime }\right) }$ is a
partition on $\Theta $. Clearly, the definition of $P^{\left( \theta ,\theta
^{\prime }\right) }$ implies $\widetilde{\theta }\in P^{\left( \theta
,\theta ^{\prime }\right) }\left( \widetilde{\theta }\right) $ for any $%
\widetilde{\theta }\in \Theta $. Consider any $\left( \widetilde{\theta },%
\widetilde{\theta }^{\prime }\right) \in \Theta \times \Theta $ such that $%
\widetilde{\theta }^{\prime }\in P^{\left( \theta ,\theta ^{\prime }\right)
}\left( \widetilde{\theta }\right) $, and we aim to show $P^{\left( \theta
,\theta ^{\prime }\right) }\left( \widetilde{\theta }^{\prime }\right)
=P^{\left( \theta ,\theta ^{\prime }\right) }\left( \widetilde{\theta }%
\right) $. We consider two cases. First, suppose $h^{\left( \theta
_{i},\theta _{i}^{\prime },\widetilde{\theta }_{-i}\right) }\in \mathcal{T}$%
. Then, $\widetilde{\theta }^{\prime }\in P^{\left( \theta ,\theta ^{\prime
}\right) }\left( \widetilde{\theta }\right) =\Theta _{i}\times
P_{-i}^{\left( \theta ,\theta ^{\prime }\right) }\left( \widetilde{\theta }%
_{-i}\right) $ and (\ref{iit7}) implies%
\begin{equation*}
h^{\left( \theta _{i},\theta _{i}^{\prime },\widetilde{\theta }_{-i}^{\prime
}\right) }=h^{\left( \theta _{i},\theta _{i}^{\prime },\widetilde{\theta }%
_{-i}\right) }\in \mathcal{T}\text{,}
\end{equation*}%
which further implies%
\begin{equation*}
P^{\left( \theta ,\theta ^{\prime }\right) }\left( \widetilde{\theta }%
^{\prime }\right) =\Theta _{i}\times \mathcal{E}^{\left[ G,S\right] \text{-}%
h^{\left( \theta _{i},\theta _{i}^{\prime },\widetilde{\theta }_{-i}^{\prime
}\right) }}=\Theta _{i}\times \mathcal{E}^{\left[ G,S\right] \text{-}%
h^{\left( \theta _{i},\theta _{i}^{\prime },\widetilde{\theta }_{-i}\right)
}}=P^{\left( \theta ,\theta ^{\prime }\right) }\left( \widetilde{\theta }%
\right) \text{.}
\end{equation*}%
Second, suppose $h^{\left( \theta _{i},\theta _{i}^{\prime },\widetilde{%
\theta }_{-i}\right) }\notin \mathcal{T}$. Then, $\widetilde{\theta }%
^{\prime }\in P^{\left( \theta ,\theta ^{\prime }\right) }\left( \widetilde{%
\theta }\right) =\Theta _{i}\times P_{-i}^{\left( \theta ,\theta ^{\prime
}\right) }\left( \widetilde{\theta }_{-i}\right) $ and (\ref{iit7}) implies
existence of $\widehat{h}^{\prime }\in \mathcal{H\diagdown T}$ such that%
\begin{gather}
\widehat{h}^{\prime }\sim ^{G}h^{\left( \theta _{i},\theta _{i}^{\prime },%
\widetilde{\theta }_{-i}\right) }\text{,}  \label{iit9b} \\
\text{and }\widetilde{\theta }_{-i}^{\prime }\in \mathcal{E}_{-i}^{\left[ G,S%
\right] \text{-}\widehat{h}^{\prime }}\text{,}  \label{iit9c}
\end{gather}%
which, together with $h^{\left( \theta _{i},\theta _{i}^{\prime },\widetilde{%
\theta }_{-i}\right) }\in \mathcal{H}^{\left( \theta _{i},\theta
_{i}^{\prime },\widetilde{\theta }_{-i}\right) }\mathcal{\diagdown T}$ and (%
\ref{iit8}), implies%
\begin{gather}
\left\{ \theta _{i},\theta _{i}^{\prime }\right\} \subset \mathcal{E}_{i}^{%
\left[ G,S\right] \text{-}h^{\left( \theta _{i},\theta _{i}^{\prime },%
\widetilde{\theta }_{-i}\right) }}\text{, }  \label{iit9} \\
S_{i}\left( \theta _{i}\right) \left[ \widehat{h}^{\prime }\right] \neq
S_{i}\left( \theta _{i}^{\prime }\right) \left[ \widehat{h}^{\prime }\right] 
\text{.}  \label{iit10}
\end{gather}%
By Lemma \ref{lem:perfect-recall}, (\ref{iit9b}) and (\ref{iit9}) imply%
\begin{equation*}
\left\{ \theta _{i},\theta _{i}^{\prime }\right\} \subset \mathcal{E}_{i}^{%
\left[ G,S\right] \text{-}\widehat{h}^{\prime }}\text{,}
\end{equation*}%
which, together with (\ref{iit9c}), imply%
\begin{equation}
\left\{ \left( \theta _{i},\widetilde{\theta }_{-i}^{\prime }\right) ,\left(
\theta _{i}^{\prime },\widetilde{\theta }_{-i}^{\prime }\right) \right\}
\subset \mathcal{E}^{\left[ G,S\right] \text{-}\widehat{h}^{\prime }}\text{.}
\label{iit11}
\end{equation}%
(\ref{iit9b}), (\ref{iit10}) and (\ref{iit11}) imply $h^{\left( \theta
_{i},\theta _{i}^{\prime },\widetilde{\theta }_{-i}^{\prime }\right) }=%
\widehat{h}^{\prime }$. We thus have%
\begin{equation*}
h^{\left( \theta _{i},\theta _{i}^{\prime },\widetilde{\theta }_{-i}^{\prime
}\right) }=\widehat{h}^{\prime }\sim ^{G}h^{\left( \theta _{i},\theta
_{i}^{\prime },\widetilde{\theta }_{-i}\right) }\text{,}
\end{equation*}%
and hence, $\vartheta ^{\left[ G\right] }\left[ \mathcal{E}^{\left[ G,S%
\right] \text{-}h^{\left( \theta _{i},\theta _{i}^{\prime },\widetilde{%
\theta }_{-i}^{\prime }\right) }}\right] =\vartheta ^{\left[ G\right] }\left[
\mathcal{E}^{\left[ G,S\right] \text{-}h^{\left( \theta _{i},\theta
_{i}^{\prime },\widetilde{\theta }_{-i}\right) }}\right] $, which further
implies%
\begin{equation*}
P^{\left( \theta ,\theta ^{\prime }\right) }\left( \widetilde{\theta }%
^{\prime }\right) =\Theta _{i}\times \vartheta _{-i}^{\left[ G\right] }\left[
\mathcal{E}^{\left[ G,S\right] \text{-}h^{\left( \theta _{i},\theta
_{i}^{\prime },\widetilde{\theta }_{-i}^{\prime }\right) }}\right] =\Theta
_{i}\times \vartheta _{-i}^{\left[ G\right] }\left[ \mathcal{E}^{\left[ G,S%
\right] \text{-}h^{\left( \theta _{i},\theta _{i}^{\prime },\widetilde{%
\theta }_{-i}\right) }}\right] =P^{\left( \theta ,\theta ^{\prime }\right)
}\left( \widetilde{\theta }\right) \text{.}
\end{equation*}%
That is, we have proved $P^{\left( \theta ,\theta ^{\prime }\right) }\left( 
\widetilde{\theta }^{\prime }\right) =P^{\left( \theta ,\theta ^{\prime
}\right) }\left( \widetilde{\theta }\right) $ for both cases.$\blacksquare $

\subsection{Intermediate results}

We need the following intermediate results in our proof. \label%
{sec:intermediate-results}

\begin{lemma}
\label{lem:operator:canonic:alternative:increasing}For any SCF $f$ and any
dissectible solution notion $\rho $, any%
\begin{equation*}
\left[ f,\left( \gamma ,E\right) ,\theta ,\theta ^{\prime },i,\widehat{E}%
\right] \in \mathcal{X}^{\Theta }\times \Gamma ^{\text{Semi}}\times \left(
\times _{i\in \mathcal{I}}\left[ 2^{\Theta _{i}}\diagdown \left\{
\varnothing \right\} \right] \right) \times \Theta \times \Theta \times 
\mathcal{I}\times 2^{\Theta }\text{,}
\end{equation*}%
and any $\left( \overline{\gamma },\overline{E}\right) \in \Gamma ^{\text{%
Semi}}\times \left( \times _{i\in \mathcal{I}}\left[ 2^{\Theta
_{i}}\diagdown \left\{ \varnothing \right\} \right] \right) $, we have 
\begin{equation}
\left( 
\begin{array}{c}
\gamma \left[ E,\text{ }\theta \right] \subset \overline{\gamma }\left[ 
\text{ }\overline{E},\text{ }\theta \right] \text{,} \\ 
\rho \left[ f,\left( \gamma ,E\right) ,\theta ,\theta ^{\prime },i,\widehat{E%
}\right] =0%
\end{array}%
\right) \Longrightarrow \rho \left[ f,\left( \overline{\gamma },\overline{E}%
\right) ,\theta ,\theta ^{\prime },i,\widehat{E}\right] =0\text{.}
\label{hhg1}
\end{equation}
\end{lemma}

\begin{lemma}
\label{lem:operator:canonic:n}For any SCF $f$ and any dissectible solution
notion $\rho $, we have%
\begin{equation}
\theta \sim _{i}^{\left[ \left( \rho ,f\right) ,\left( \gamma ^{\left( \rho
,f\right) \text{-}\left( n\right) },E\right) \right] }\theta ^{\prime
}\Longrightarrow \theta \sim _{i}^{\left[ \left( \rho ,f\right) ,\left(
\gamma ^{\left( \rho ,f\right) \text{-}\left( n+1\right) },E\right) \right]
}\theta ^{\prime }\text{, }\forall \left( i,\theta ,\theta ^{\prime
},n\right) \in \mathcal{I}\times \Theta \times \Theta \times \mathbb{N}\text{%
.}  \label{kky4}
\end{equation}
\end{lemma}

\begin{lemma}
\label{lem:operator:canonic:alternative}For any SCF $f$ and any dissectible
solution notion $\rho $, we have 
\begin{gather}
\left( 
\begin{array}{c}
~\theta \in E\text{,} \\ 
\theta ^{\prime }\in \gamma ^{\left( \rho ,f\right) \text{-}\left(
n+1\right) }\left[ E,\text{ }\theta \right]%
\end{array}%
\right) \Longleftrightarrow \left( 
\begin{array}{c}
~\theta \in E\text{,} \\ 
\forall i\in \mathcal{I}\text{, }\exists \left\{ \theta ^{1},...,\theta
^{n+1}\right\} \subset E\text{,} \\ 
\theta ^{1}=\theta \text{ and }\theta ^{n+1}=\theta ^{\prime }\text{,} \\ 
\theta ^{1}\sim _{i}^{\left[ \left( \rho ,f\right) ,\left( \gamma ^{\left(
\rho ,f\right) \text{-}\left( n\right) },E\right) \right] }\theta
^{2}...\sim _{i}^{\left[ \left( \rho ,f\right) ,\left( \gamma ^{\left( \rho
,f\right) \text{-}\left( n\right) },E\right) \right] }\theta ^{n+1}%
\end{array}%
\right) \text{,}  \label{kky5} \\
\forall \left( \theta ,\theta ^{\prime },E,n\right) \in \Theta \times \Theta
\times \left( \times _{i\in \mathcal{I}}\left[ 2^{\Theta _{i}}\diagdown
\left\{ \varnothing \right\} \right] \right) \times \mathbb{N}\text{.} 
\notag
\end{gather}
\end{lemma}

\noindent \textbf{Proof of Lemma \ref%
{lem:operator:canonic:alternative:increasing}:} Since $\rho $ is
dissectible, Definition \ref{def:solution:monotone} implies%
\begin{equation}
\rho \left[ f,\left( \gamma ,E\right) ,\theta ,\theta ^{\prime },i,\widehat{E%
}\right] =0\Longrightarrow \left( 
\begin{array}{c}
\exists \widetilde{\theta }\in \gamma \left[ E,\text{ }\theta \right] \text{%
, } \\ 
\exists \widetilde{\gamma }\in \Gamma ^{\text{Semi}}\text{, }\widetilde{%
\gamma }\left[ E,\text{ }\theta \right] =\left\{ \theta ,\widetilde{\theta }%
\right\} \text{,} \\ 
\rho \left[ f,\left( \widetilde{\gamma },E\right) ,\theta ,\theta ^{\prime
},i,\widehat{E}\right] =0%
\end{array}%
\right) \text{.}  \label{hhg2a}
\end{equation}%
Consider any $\widehat{\gamma }\in \Gamma ^{\text{Semi}}$ such that $%
\widehat{\gamma }\left[ \overline{E},\text{ }\theta \right] =\widetilde{%
\gamma }\left[ E,\text{ }\theta \right] =\left\{ \theta ,\widetilde{\theta }%
\right\} $, and in particular, $\widehat{\gamma }_{i}\left[ \overline{E},%
\text{ }\theta \right] =\widetilde{\gamma }_{i}\left[ E,\text{ }\theta %
\right] $. Thus, (\ref{hht2}) in Definition \ref{def:solution} implies%
\begin{equation}
\widehat{\gamma }\left[ \text{ }\overline{E},\text{ }\theta \right] =%
\widetilde{\gamma }\left[ E,\text{ }\theta \right] \Longrightarrow \rho %
\left[ f,\left( \widehat{\gamma },\overline{E}\right) ,\theta ,\theta
^{\prime },i,\widehat{E}\right] =\rho \left[ f,\left( \widetilde{\gamma }%
,E\right) ,\theta ,\theta ^{\prime },i,\widehat{E}\right] =0\text{.}
\label{hhg2b}
\end{equation}%
(\ref{hhg2a}) and (\ref{hhg2b}) imply%
\begin{equation}
\left( 
\begin{array}{c}
\gamma \left[ E,\text{ }\theta \right] \subset \overline{\gamma }\left[ 
\text{ }\overline{E},\text{ }\theta \right] \text{,} \\ 
\rho \left[ f,\left( \gamma ,E\right) ,\theta ,\theta ^{\prime },i,\widehat{E%
}\right] =0%
\end{array}%
\right) \Longrightarrow \left( 
\begin{array}{c}
\exists \widetilde{\theta }\in \gamma \left[ E,\text{ }\theta \right]
\subset \overline{\gamma }\left[ \text{ }\overline{E},\text{ }\theta \right] 
\text{, } \\ 
\exists \widehat{\gamma }\in \Gamma ^{\text{Semi}}\text{, }\widehat{\gamma }%
\left[ \text{ }\overline{E},\text{ }\theta \right] =\widetilde{\gamma }\left[
E,\text{ }\theta \right] =\left\{ \theta ,\widetilde{\theta }\right\} \text{,%
} \\ 
\rho \left[ f,\left( \widehat{\gamma },\overline{E}\right) ,\theta ,\theta
^{\prime },i,\widehat{E}\right] =0%
\end{array}%
\right) \text{.}  \label{hhg2}
\end{equation}%
Definition \ref{def:solution:monotone} implies%
\begin{equation}
\left( 
\begin{array}{c}
\exists \widetilde{\theta }\in \overline{\gamma }\left[ \text{ }\overline{E},%
\text{ }\theta \right] \text{, } \\ 
\exists \widehat{\gamma }\in \Gamma ^{\text{Semi}}\text{, }\widehat{\gamma }%
\left[ \text{ }\overline{E},\text{ }\theta \right] =\left\{ \theta ,%
\widetilde{\theta }\right\} \text{,} \\ 
\rho \left[ f,\left( \widehat{\gamma },\overline{E}\right) ,\theta ,\theta
^{\prime },i,\widehat{E}\right] =0%
\end{array}%
\right) \Longrightarrow \rho \left[ f,\left( \overline{\gamma },\overline{E}%
\right) ,\theta ,\theta ^{\prime },i,\widehat{E}\right] =0\text{.}
\label{hgg3}
\end{equation}%
Thus, (\ref{hhg2}) and (\ref{hgg3}) imply (\ref{hhg1}).$\blacksquare $

\noindent \textbf{Proof of Lemma \ref{lem:operator:canonic:n}:} (\ref{kky2}%
)\ imply that for any $\left[ E,\text{ }\theta \right] \in \left( \times
_{i\in \mathcal{I}}\left[ 2^{\Theta _{i}}\diagdown \left\{ \varnothing
\right\} \right] \right) \times \Theta $,%
\begin{equation}
\theta \in E\Longrightarrow \left\{ \theta \right\} =\gamma ^{\left( \rho
,f\right) \text{-}\left( 1\right) }\left[ E,\text{ }\theta \right] \subset
\gamma ^{\left( \rho ,f\right) \text{-}\left( n\right) }\left[ E,\text{ }%
\theta \right] \subset \gamma ^{\left( \rho ,f\right) \text{-}\left(
n+1\right) }\left[ E,\text{ }\theta \right] \subset E\text{, }\forall n\in 
\mathbb{N}\text{.}  \label{kky3a}
\end{equation}%
Since $\rho $ is dissectible, Lemma \ref%
{lem:operator:canonic:alternative:increasing} implies%
\begin{equation}
\theta >_{i}^{\left[ \left( \rho ,f\right) ,\left( \gamma ^{\left( \rho
,f\right) \text{-}\left( n\right) },E\right) \right] }\theta ^{\prime
}\Longrightarrow \theta >_{i}^{\left[ \left( \rho ,f\right) ,\left( \gamma
^{\left( \rho ,f\right) \text{-}\left( n+1\right) },E\right) \right] }\theta
^{\prime },\forall \left( i,\theta ,\theta ^{\prime },n\right) \in \mathcal{I%
}\times \Theta \times \Theta \times \mathbb{N}\text{,}  \label{kky3}
\end{equation}%
which, together with (\ref{kky4c}), implies (\ref{kky4}).$\blacksquare $

\noindent \textbf{Proof of Lemma \ref{lem:operator:canonic:alternative}:}
First, with $n=1$, (\ref{kky7}) and (\ref{kky2}) imply 
\begin{equation*}
\left( 
\begin{array}{c}
~\theta \in E\text{,} \\ 
\theta ^{\prime }\in \gamma ^{\left( \rho ,f\right) \text{-}\left( 2\right) }%
\left[ E,\text{ }\theta \right]%
\end{array}%
\right) \Longleftrightarrow \left( 
\begin{array}{c}
~\theta \in E\text{,} \\ 
\forall i\in \mathcal{I}\text{, }\theta =\theta ^{1}\sim _{i}^{\left[ \left(
\rho ,f\right) ,\left( \gamma ^{\left( \rho ,f\right) \text{-}\left(
1\right) },E\right) \right] }\theta ^{2}=\theta ^{\prime }%
\end{array}%
\right) \text{,}
\end{equation*}%
i.e., (\ref{kky5})\ holds for $n=1$. Suppose (\ref{kky5}) holds for $n=k\in 
\mathbb{N}$. Consider $n=\left( k+1\right) \in \mathbb{N}$. (\ref{kky2}) and
implies%
\begin{equation}
\left( 
\begin{array}{c}
~\theta \in E\text{,} \\ 
\theta ^{\prime }\in \gamma ^{\left( \rho ,f\right) \text{-}\left(
k+2\right) }\left[ E,\text{ }\theta \right]%
\end{array}%
\right) \Longleftrightarrow \left( 
\begin{array}{c}
~\theta \in E\text{,} \\ 
\exists \widetilde{\theta }\in \gamma ^{\left( \rho ,f\right) \text{-}\left(
k+1\right) }\left[ E,\text{ }\theta \right] \text{,} \\ 
\forall i\in \mathcal{I}\text{, }\widetilde{\theta }\sim _{i}^{\left[ \left(
\rho ,f\right) ,\left( \gamma ^{\left( \rho ,f\right) \text{-}\left(
k+1\right) },E\right) \right] }\theta ^{\prime }%
\end{array}%
\right) \text{.}  \label{kky9}
\end{equation}%
We thus have%
\begin{eqnarray}
&&\left( 
\begin{array}{c}
~\theta \in E\text{,} \\ 
\exists \widetilde{\theta }\in \gamma ^{\left( \rho ,f\right) \text{-}\left(
k+1\right) }\left[ E,\text{ }\theta \right] \text{,} \\ 
\forall i\in \mathcal{I}\text{, }\widetilde{\theta }\sim _{i}^{\left[ \left(
\rho ,f\right) ,\left( \gamma ^{\left( \rho ,f\right) \text{-}\left(
k+1\right) },E\right) \right] }\theta ^{\prime }%
\end{array}%
\right)  \label{kky8} \\
&\Longleftrightarrow &\left( 
\begin{array}{c}
~\theta \in E\text{, }\exists \widetilde{\theta }\in \gamma ^{\left( \rho
,f\right) \text{-}\left( k+1\right) }\left[ E,\text{ }\theta \right] \text{,}
\\ 
\forall i\in \mathcal{I}\text{, }\exists \left\{ \theta ^{1},...,\theta
^{k+2}\right\} \subset E\text{,} \\ 
\theta ^{1}=\theta \text{, }\theta ^{k+1}=\widetilde{\theta }\text{, and }%
\theta ^{k+2}=\theta ^{\prime }\text{,} \\ 
\theta ^{1}\sim _{i}^{\left[ \left( \rho ,f\right) ,\left( \gamma ^{\left(
\rho ,f\right) \text{-}\left( k\right) },E\right) \right] }...\sim _{i}^{%
\left[ \left( \rho ,f\right) ,\left( \gamma ^{\left( \rho ,f\right) \text{-}%
\left( k\right) },E\right) \right] }\theta ^{k+1}\sim _{i}^{\left[ \left(
\rho ,f\right) ,\left( \gamma ^{\left( \rho ,f\right) \text{-}\left(
k+1\right) },E\right) \right] }\theta ^{k+2}%
\end{array}%
\right)  \notag \\
&\Longleftrightarrow &\left( 
\begin{array}{c}
~\theta \in E\text{, }\exists \widetilde{\theta }\in \gamma ^{\left( \rho
,f\right) \text{-}\left( k+1\right) }\left[ E,\text{ }\theta \right] \text{,}
\\ 
\forall i\in \mathcal{I}\text{, }\exists \left\{ \theta ^{1},...,\theta
^{k+2}\right\} \subset E\text{,} \\ 
\theta ^{1}=\theta \text{, }\theta ^{k+1}=\widetilde{\theta }\text{, and }%
\theta ^{k+2}=\theta ^{\prime }\text{,} \\ 
\theta ^{1}\sim ^{\left[ \left( \rho ,f\right) ,\left( \gamma ^{\left( \rho
,f\right) \text{-}\left( k+1\right) },E\right) \right] }...\sim ^{\left[
\left( \rho ,f\right) ,\left( \gamma ^{\left( \rho ,f\right) \text{-}\left(
k+1\right) },E\right) \right] }\theta ^{k+1}\sim ^{\left[ \left( \rho
,f\right) ,\left( \gamma ^{\left( \rho ,f\right) \text{-}\left( k+1\right)
},E\right) \right] }\theta ^{k+2}%
\end{array}%
\right) \text{,}  \notag
\end{eqnarray}%
where the first $\Longleftrightarrow $ follows from the induction
hypothesis, and the second $\Longleftrightarrow $ follows from Lemma \ref%
{lem:operator:canonic:n}. Therefore, (\ref{kky9}) and (\ref{kky8}) imply
that (\ref{kky5})\ holds for $n=k+1$.$\blacksquare $

\subsection{Proof of Lemma \protect\ref{lem:operator:canonical:operator}}

\label{sec:lem:operator:canonical:operator}

Fix any SCF $f$ and any dissectible solution notion $\rho $.

\noindent \textbf{Proof of }$\gamma ^{\left( \rho ,f\right) }$\textbf{\
being }$\left( \rho ,f,\vartheta ^{\ast }\right) $\textbf{-consistent: }
Consider any $\left[ i,E,\theta ,\theta ^{\prime }\right] \in \mathcal{I}%
\times \left( \times _{i\in \mathcal{I}}\left[ 2^{\Theta _{i}}\diagdown
\left\{ \varnothing \right\} \right] \right) \times \Theta \times \Theta $
such that%
\begin{equation}
\left\{ \theta ,\theta ^{\prime }\right\} \subset E\text{ and }\rho \left[
f,\left( \gamma ^{\left( \rho ,f\right) },E\right) ,\theta ,\theta ^{\prime
},i,E\right] =0\text{,}  \label{tii1}
\end{equation}%
and we aim to prove $\theta _{i}^{\prime }\in \gamma _{i}^{\left( \rho
,f\right) }\left[ E,\theta \right] $.

Since $\rho $ is dissectible, (\ref{tii1}) implies%
\begin{equation}
\left( 
\begin{array}{c}
\exists \widetilde{\theta }\in \gamma ^{\left( \rho ,f\right) }\left[ E,%
\text{ }\theta \right] \text{, } \\ 
\exists \widetilde{\gamma }\in \Gamma \text{, }\widetilde{\gamma }\left[ E,%
\text{ }\theta \right] =\left\{ \theta ,\widetilde{\theta }\right\} \text{,}
\\ 
\rho \left[ f,\left( \widetilde{\gamma },E\right) ,\theta ,\theta ^{\prime
},i,E\right] =0%
\end{array}%
\right) \text{.}  \label{tii2}
\end{equation}%
Furthermore, $\widetilde{\theta }\in \gamma ^{\left( \rho ,f\right) }\left[
E,\text{ }\theta \right] $ and (\ref{kky2b}) implies%
\begin{equation}
\widetilde{\theta }\in \gamma ^{\left( \rho ,f\right) \text{-}\left(
n\right) }\left[ E,\text{ }\theta \right] \text{ for some }n\in \mathbb{N}%
\text{.}  \label{tii3}
\end{equation}%
Since $\rho $ is dissectible, (\ref{tii2}) and (\ref{tii3})\ imply%
\begin{equation*}
\rho \left[ f,\left( \gamma ^{\left( \rho ,f\right) \text{-}\left( n\right)
},E\right) ,\theta ,\theta ^{\prime },i,E\right] =0\text{,}
\end{equation*}%
and hence, 
\begin{equation*}
\left( \theta _{i}^{\prime },\theta _{-i}\right) \in \gamma ^{\left( \rho
,f\right) \text{-}\left( n+1\right) }\left[ E,\text{ }\theta \right] \text{,}
\end{equation*}%
which, together with (\ref{kky2b}), implies $\theta _{i}^{\prime }\in \gamma
_{i}^{\left( \rho ,f\right) }\left[ E,\text{ }\theta \right] $.$\blacksquare 
$

\noindent \textbf{Proof of }$\gamma ^{\left( \rho ,f\right) }$\textbf{\
being an operator: } First, (\ref{kky7}), (\ref{kky2}) and (\ref{kky2b})
imply

\begin{equation}
\theta \in E\Longrightarrow \theta \in \gamma ^{\left( \rho ,f\right) }\left[
E,\text{ }\theta \right] \subset E\text{, }\forall \left[ \theta ,E\right]
\in \Theta \times \left( \times _{i\in \mathcal{I}}\left[ 2^{\Theta
_{i}}\diagdown \left\{ \varnothing \right\} \right] \right) \text{.}
\label{itt1}
\end{equation}%
Second, for any $\left( \theta ,\theta ^{\prime },E\right) \in \Theta \times
\Theta \times \left( \times _{i\in \mathcal{I}}\left[ 2^{\Theta
_{i}}\diagdown \left\{ \varnothing \right\} \right] \right) $, we prove%
\begin{equation}
\left[ \theta \in E\text{ and }\theta ^{\prime }\in \gamma ^{\left( \rho
,f\right) }\left[ E,\theta \right] \right] \Longrightarrow \left[ \theta
^{\prime }\in E\text{ and }\theta \in \gamma ^{\left( \rho ,f\right) }\left[
E,\theta ^{\prime }\right] \right] \text{.}  \label{itt2}
\end{equation}%
Suppose $\theta \in E$ and $\theta ^{\prime }\in \gamma ^{\left( \rho
,f\right) }\left[ E,\theta \right] $. If $\theta ^{\prime }=\theta $, (\ref%
{itt2}) holds. Suppose $\theta ^{\prime }\neq \theta $. By (\ref{itt1}), we
have $\theta ^{\prime }\in \gamma ^{\left( \rho ,f\right) }\left[ E,\theta %
\right] \subset E$. Furthermore, $\theta ^{\prime }\in \gamma ^{\left( \rho
,f\right) }\left[ E,\theta \right] $ and (\ref{kky2b}) imply\ $\theta
^{\prime }\in \gamma ^{\left( \rho ,f\right) \text{-}\left( n+1\right) }%
\left[ E,\text{ }\theta \right] $ for some $n\in \mathbb{N}$. Thus, Lemma %
\ref{lem:operator:canonic:alternative} implies%
\begin{equation*}
\left( 
\begin{array}{c}
\forall i\in \mathcal{I}\text{, }\exists \left\{ \theta ^{1},...,\theta
^{n+1}\right\} \subset E\text{,} \\ 
\theta ^{1}=\theta \text{ and }\theta ^{n+1}=\theta ^{\prime }\text{,} \\ 
\theta ^{1}\sim _{i}^{\left[ \left( \rho ,f\right) ,\left( \gamma ^{\left(
\rho ,f\right) \text{-}\left( n\right) },E\right) \right] }\theta ^{2}\sim
_{i}^{\left[ \left( \rho ,f\right) ,\left( \gamma ^{\left( \rho ,f\right) 
\text{-}\left( n\right) },E\right) \right] }...\sim _{i}^{\left[ \left( \rho
,f\right) ,\left( \gamma ^{\left( \rho ,f\right) \text{-}\left( n\right)
},E\right) \right] }\theta ^{n+1}%
\end{array}%
\right) \text{,}
\end{equation*}%
which, together with symmetry of $\sim _{i}^{\left[ \left( \rho ,f\right)
,\left( \gamma ^{\left( \rho ,f\right) \text{-}\left( n\right) },E\right) %
\right] }$ (i.e., (\ref{kky4b})), implies%
\begin{equation*}
\left( 
\begin{array}{c}
\forall i\in \mathcal{I}\text{, }\exists \left\{ \theta ^{1},...,\theta
^{n+1}\right\} \subset E\text{,} \\ 
\theta ^{1}=\theta \text{ and }\theta ^{n+1}=\theta ^{\prime }\text{,} \\ 
\theta ^{n+1}\sim _{i}^{\left[ \left( \rho ,f\right) ,\left( \gamma ^{\left(
\rho ,f\right) \text{-}\left( n\right) },E\right) \right] }...\sim _{i}^{%
\left[ \left( \rho ,f\right) ,\left( \gamma ^{\left( \rho ,f\right) \text{-}%
\left( n\right) },E\right) \right] }\theta ^{2}\sim _{i}^{\left[ \left( \rho
,f\right) ,\left( \gamma ^{\left( \rho ,f\right) \text{-}\left( n\right)
},E\right) \right] }\theta ^{1}%
\end{array}%
\right) \text{,}
\end{equation*}%
i.e., $\theta \in \gamma ^{\left( \rho ,f\right) \text{-}\left( n+1\right) }%
\left[ E,\text{ }\theta ^{\prime }\right] \subset \gamma ^{\left( \rho
,f\right) }\left[ E,\text{ }\theta ^{\prime }\right] $. Therefore, (\ref%
{itt2}) holds.

Third, for any $\left( \theta ,\theta ^{\prime },E\right) \in \Theta \times
\Theta \times \left( \times _{i\in \mathcal{I}}\left[ 2^{\Theta
_{i}}\diagdown \left\{ \varnothing \right\} \right] \right) $, we prove%
\begin{equation}
\left[ \theta \in E\text{ and }\theta ^{\prime }\in \gamma ^{\left( \rho
,f\right) }\left[ E,\theta \right] \right] \Longrightarrow \gamma ^{\left(
\rho ,f\right) }\left[ E,\theta ^{\prime }\right] \subset \gamma ^{\left(
\rho ,f\right) }\left[ E,\theta \right] \text{.}  \label{itt4}
\end{equation}%
Suppose $\theta \in E$ and $\theta ^{\prime }\in \gamma ^{\left( \rho
,f\right) }\left[ E,\theta \right] $. Pick any $\theta ^{\prime \prime }\in
\gamma ^{\left( \rho ,f\right) }\left[ E,\theta ^{\prime }\right] $, and we
aim to prove $\theta ^{\prime \prime }\in \gamma ^{\left( \rho ,f\right) }%
\left[ E,\theta \right] $, i.e., (\ref{itt4}) holds.

$\theta ^{\prime }\in \gamma ^{\left( \rho ,f\right) }\left[ E,\theta \right]
$ and $\theta ^{\prime \prime }\in \gamma ^{\left( \rho ,f\right) }\left[
E,\theta ^{\prime }\right] $ imply $\theta ^{\prime }\in \gamma ^{\left(
\rho ,f\right) \text{-}\left( n+1\right) }\left[ E,\theta \right] $ and $%
\theta ^{\prime \prime }\in \gamma ^{\left( \rho ,f\right) \text{-}\left(
n^{\prime }+1\right) }\left[ E,\theta ^{\prime }\right] $ for some $\left(
n,n^{\prime }\right) \in \mathbb{N\times N}$. Thus, Lemmas \ref%
{lem:operator:canonic:n} and \ref{lem:operator:canonic:alternative} imply%
\begin{equation*}
\left( 
\begin{array}{c}
\forall i\in \mathcal{I}\text{, }\exists \left\{ \theta ^{1},...,\theta
^{n+1}\right\} \subset E\text{,} \\ 
\theta ^{1}=\theta \text{ and }\theta ^{n+1}=\theta ^{\prime }\text{,} \\ 
\theta ^{1}\sim _{i}^{\left[ \left( \rho ,f\right) ,\left( \gamma ^{\left(
\rho ,f\right) \text{-}\left( n+n^{\prime }\right) },E\right) \right]
}\theta ^{2}\sim _{i}^{\left[ \left( \rho ,f\right) ,\left( \gamma ^{\left(
\rho ,f\right) \text{-}\left( n+n^{\prime }\right) },E\right) \right]
}...\sim _{i}^{\left[ \left( \rho ,f\right) ,\left( \gamma ^{\left( \rho
,f\right) \text{-}\left( n+n^{\prime }\right) },E\right) \right] }\theta
^{n+1}%
\end{array}%
\right) \text{,}
\end{equation*}%
\begin{equation*}
\text{and }\left( 
\begin{array}{c}
\forall i\in \mathcal{I}\text{, }\exists \left\{ \theta ^{1},...,\theta
^{n+1}\right\} \subset E\text{,} \\ 
\theta ^{n+1}=\theta ^{\prime }\text{ and }\theta ^{n+n^{\prime }+1}=\theta
^{\prime \prime }\text{,} \\ 
\theta ^{n+1}\sim _{i}^{\left[ \left( \rho ,f\right) ,\left( \gamma ^{\left(
\rho ,f\right) \text{-}\left( n+n^{\prime }\right) },E\right) \right]
}\theta ^{n+2}\sim _{i}^{\left[ \left( \rho ,f\right) ,\left( \gamma
^{\left( \rho ,f\right) \text{-}\left( n+n^{\prime }\right) },E\right) %
\right] }...\sim _{i}^{\left[ \left( \rho ,f\right) ,\left( \gamma ^{\left(
\rho ,f\right) \text{-}\left( n+n^{\prime }\right) },E\right) \right]
}\theta ^{n+n^{\prime }+1}%
\end{array}%
\right) \text{,}
\end{equation*}
which, together with Lemma \ref{lem:operator:canonic:alternative}, imply $%
\theta ^{\prime \prime }\in \gamma ^{\left( \rho ,f\right) \text{-}\left(
n+n^{\prime }+1\right) }\left[ E,\theta \right] \subset \gamma ^{\left( \rho
,f\right) }\left[ E,\theta \right] $. Therefore, (\ref{itt4}) holds.

Fourth, (\ref{itt2}) and (\ref{itt4}) imply%
\begin{equation*}
\left[ \theta \in E\text{ and }\theta ^{\prime }\in \gamma ^{\left( \rho
,f\right) }\left[ E,\theta \right] \right] \Longrightarrow \left[ \theta
^{\prime }\in E\text{ and }\theta \in \gamma ^{\left( \rho ,f\right) }\left[
E,\theta ^{\prime }\right] \right] \Longrightarrow \gamma ^{\left( \rho
,f\right) }\left[ E,\theta \right] \subset \gamma ^{\left( \rho ,f\right) }%
\left[ E,\theta ^{\prime }\right] \text{,}
\end{equation*}%
which, together with (\ref{itt4}), implies%
\begin{equation}
\left[ \theta \in E\text{ and }\theta ^{\prime }\in \gamma ^{\left( \rho
,f\right) }\left[ E,\theta \right] \right] \Longrightarrow \gamma ^{\left(
\rho ,f\right) }\left[ E,\theta \right] =\gamma ^{\left( \rho ,f\right) }%
\left[ E,\theta ^{\prime }\right] \text{.}  \label{itt4c}
\end{equation}%
Finally, (\ref{itt1}) and (\ref{itt4c}) imply that $\gamma ^{\left( \rho
,f\right) }$ is an operator.$\blacksquare $

\subsection{Proof of Lemma \protect\ref{lem:operator:canonical:increasing}}

Fix any SCF $f$ and any solution notion $\rho $ which is both dissectible
and monotonic. Fix any $\left( \theta ,E,E^{\prime }\right) \in \Theta
\times \left( \times _{i\in \mathcal{I}}\left[ 2^{\Theta _{i}}\diagdown
\left\{ \varnothing \right\} \right] \right) \times \left( \times _{i\in 
\mathcal{I}}\left[ 2^{\Theta _{i}}\diagdown \left\{ \varnothing \right\} %
\right] \right) $ such that $\theta \in E\subset E^{\prime }$. Inductively,
we prove%
\begin{equation}
\left( 
\begin{array}{c}
\gamma ^{\left( \rho ,f\right) \text{-}\left( n\right) }\left[ E,\theta %
\right] \subset \gamma ^{\left( \rho ,f\right) \text{-}\left( n\right) }%
\left[ E^{\prime },\theta \right] \text{,} \\ 
\text{and }\forall i\in \mathcal{I}\text{,} \\ 
\theta ^{\prime }>_{i}^{\left[ \left( \rho ,f\right) ,\left( \gamma ^{\left(
\rho ,f\right) \text{-}\left( n\right) },E\right) \right] }\theta ^{\prime
\prime }\Longrightarrow \theta ^{\prime }>_{i}^{\left[ \left( \rho ,f\right)
,\left( \gamma ^{\left( \rho ,f\right) \text{-}\left( n\right) },E^{\prime
}\right) \right] }\theta ^{\prime \prime }%
\end{array}%
\right) \text{, }\forall n\in \mathbb{N}\text{, }\forall \left( \theta
,\theta ^{\prime },\theta ^{\prime \prime }\right) \equiv \Theta \times
\Theta \times \Theta \text{.}  \label{itt8}
\end{equation}%
With $n=1$, (\ref{kky7}) implies%
\begin{equation*}
\gamma ^{\left( \rho ,f\right) \text{-}\left( 1\right) }\left[ E,\theta %
\right] =\left\{ \theta \right\} \subset \left\{ \theta \right\} =\gamma
^{\left( \rho ,f\right) \text{-}\left( 1\right) }\left[ E^{\prime },\theta %
\right] \text{.}
\end{equation*}%
Furthermore, for any $i\in \mathcal{I}$, we have%
\begin{gather*}
\theta ^{\prime }>_{i}^{\left[ \left( \rho ,f\right) ,\left( \gamma ^{\left(
\rho ,f\right) \text{-}\left( 1\right) },E\right) \right] }\theta ^{\prime
\prime }\Longrightarrow \left( 
\begin{array}{c}
\left\{ \theta ^{\prime },\theta ^{\prime \prime }\right\} \subset E\text{
and} \\ 
\rho \left[ f,\left( \gamma ^{\left( \rho ,f\right) \text{-}\left( 1\right)
},E\right) ,\theta ,\theta ^{\prime },i,E\right] =0%
\end{array}%
\right) \\
\Longrightarrow \left( 
\begin{array}{c}
\left\{ \theta ^{\prime },\theta ^{\prime \prime }\right\} \subset E^{\prime
}\text{ and} \\ 
\rho \left[ f,\left( \gamma ^{\left( \rho ,f\right) \text{-}\left( 1\right)
},E^{\prime }\right) ,\theta ,\theta ^{\prime },i,E^{\prime }\right] =0%
\end{array}%
\right) \\
\Longrightarrow \theta ^{\prime }>^{\left[ \left( \rho ,f\right) ,\left(
\gamma ^{\left( \rho ,f\right) \text{-}\left( 1\right) },E^{\prime }\right) %
\right] }\theta ^{\prime \prime }\text{,}
\end{gather*}%
where the first and third "$\Longrightarrow $" follow from (\ref{kky1}), and
the second "$\Longrightarrow $" follow from $E\subset E^{\prime }$, $\gamma
^{\left( \rho ,f\right) \text{-}\left( 1\right) }\left[ E,\theta \right]
\subset \gamma ^{\left( \rho ,f\right) \text{-}\left( 1\right) }\left[
E^{\prime },\theta \right] $, Lemma \ref%
{lem:operator:canonic:alternative:increasing} and $\rho $ being monotonic.
Therefore (\ref{itt8}) holds for $n=1$.

Suppose (\ref{itt8}) holds for $n=k\in \mathbb{N}$. Consider $n=\left(
k+1\right) \in \mathbb{N}$. We have%
\begin{eqnarray*}
\gamma ^{\left( \rho ,f\right) \text{-}\left( k+1\right) }\left[ E,\theta %
\right] &=&\cup_{\widetilde{\theta }\in \gamma ^{\left( \rho
,f\right) \text{-}\left( k\right) }\left[ E,\text{ }\theta \right] }\left\{
\theta ^{\prime }\in E:\forall i\in \mathcal{I}\text{, }\left( 
\begin{tabular}{l}
either $\theta _{i}^{\prime }=\widetilde{\theta }_{i}$, \\ 
or $\theta ^{\prime }>_{i}^{\left[ \left( \rho ,f\right) ,\left( \gamma
^{\left( \rho ,f\right) \text{-}\left( k\right) },E\right) \right] }%
\widetilde{\theta }$ \\ 
or $\widetilde{\theta }>_{i}^{\left[ \left( \rho ,f\right) ,\left( \gamma
^{\left( \rho ,f\right) \text{-}\left( k\right) },E\right) \right] }\theta
^{\prime }$%
\end{tabular}%
\right) \right\} \\
&\subset &\cup_{\widetilde{\theta }\in \gamma ^{\left( \rho
,f\right) \text{-}\left( k\right) }\left[ E^{\prime },\text{ }\theta \right]
}\left\{ \theta ^{\prime }\in E:\forall i\in \mathcal{I}\text{, }\left( 
\begin{tabular}{l}
either $\theta _{i}^{\prime }=\widetilde{\theta }_{i}$, \\ 
or $\theta ^{\prime }>_{i}^{\left[ \left( \rho ,f\right) ,\left( \gamma
^{\left( \rho ,f\right) \text{-}\left( k\right) },E^{\prime }\right) \right]
}\widetilde{\theta }$ \\ 
or $\widetilde{\theta }>_{i}^{\left[ \left( \rho ,f\right) ,\left( \gamma
^{\left( \rho ,f\right) \text{-}\left( k\right) },E^{\prime }\right) \right]
}\theta ^{\prime }$%
\end{tabular}%
\right) \right\} \\
&=&\gamma ^{\left( \rho ,f\right) \text{-}\left( k+1\right) }\left[
E^{\prime },\theta \right]
\end{eqnarray*}%
where the two equalities follow from (\ref{kky2}), and "$\subset $" follows
from the induction hypothesis.

Furthermore, for any $i\in \mathcal{I}$, we have

\begin{gather*}
\theta ^{\prime }>_{i}^{\left[ \left( \rho ,f\right) ,\left( \gamma ^{\left(
\rho ,f\right) \text{-}\left( k+1\right) },E\right) \right] }\theta ^{\prime
\prime }\Longrightarrow \left( 
\begin{array}{c}
\left\{ \theta ^{\prime },\theta ^{\prime \prime }\right\} \subset E\text{
and} \\ 
\rho \left[ f,\left( \gamma ^{\left( \rho ,f\right) \text{-}\left(
k+1\right) },E\right) ,\theta ,\theta ^{\prime },i,E\right] =0%
\end{array}%
\right) \\
\Longrightarrow \left( 
\begin{array}{c}
\left\{ \theta ^{\prime },\theta ^{\prime \prime }\right\} \subset E^{\prime
}\text{ and} \\ 
\rho \left[ f,\left( \gamma ^{\left( \rho ,f\right) \text{-}\left(
k+1\right) },E^{\prime }\right) ,\theta ,\theta ^{\prime },i,E^{\prime }%
\right] =0%
\end{array}%
\right) \\
\Longrightarrow \theta ^{\prime }>_{i}^{\left[ \left( \rho ,f\right) ,\left(
\gamma ^{\left( \rho ,f\right) \text{-}\left( k+1\right) },E^{\prime
}\right) \right] }\theta ^{\prime \prime }\text{,}
\end{gather*}%
where the first and third "$\Longrightarrow $" follow from (\ref{kky1}), and
the second "$\Longrightarrow $" follow from $E\subset E^{\prime }$, $\gamma
^{\left( \rho ,f\right) \text{-}\left( k+1\right) }\left[ E,\theta \right]
\subset \gamma ^{\left( \rho ,f\right) \text{-}\left( k+1\right) }\left[
E^{\prime },\theta \right] $, Lemma \ref%
{lem:operator:canonic:alternative:increasing} and $\rho $ being monotonic.
Therefore (\ref{itt8}) holds for $n=k+1$.

Finally, (\ref{itt8}) implies that for any $\left( \theta ,E,E^{\prime
}\right) \equiv \Theta \times \left( \times _{i\in \mathcal{I}}\left[
2^{\Theta _{i}}\diagdown \left\{ \varnothing \right\} \right] \right) \times
\left( \times _{i\in \mathcal{I}}\left[ 2^{\Theta _{i}}\diagdown \left\{
\varnothing \right\} \right] \right) $,%
\begin{equation*}
\theta \in E\subset E^{\prime }\Longrightarrow \gamma ^{\left( \rho
,f\right) }\left[ E,\text{ }\theta \right] =\cup _{n=1}^{\infty }\gamma
^{\left( \rho ,f\right) \text{-}\left( n\right) }\left[ E,\text{ }\theta %
\right] \subset \cup _{n=1}^{\infty }\gamma ^{\left( \rho ,f\right) \text{-}%
\left( n\right) }\left[ E^{\prime },\text{ }\theta \right] =\gamma ^{\left(
\rho ,f\right) }\left[ E^{\prime },\text{ }\theta \right] \text{.}
\end{equation*}%
Therefore, $\gamma ^{\left( \rho ,f\right) }$ is an increasing operator.$%
\blacksquare $

\subsection{Proof of Lemma \protect\ref{lem:operator:canonical:lower}}

\noindent Fix any SCF $f$ and any solution notion $\rho $ which is
dissectible and monotonic. Fix any $\left[ G,S\right] \in \mathcal{GS}^{%
\text{PC}}$ such that $\gamma ^{\left[ G,S\right] }$ is $\left( \rho
,f,\vartheta ^{G}\right) $-consistent, and we aim to show%
\begin{equation*}
\gamma ^{\left( \rho ,f\right) }\left[ E,\text{ }\theta \right] \subset
\gamma ^{\left[ G,S\right] }\left[ E,\text{ }\theta \right] \text{, }\forall %
\left[ E,\text{ }\theta \right] \in \left( \times _{i\in \mathcal{I}}\left[
2^{\Theta _{i}}\diagdown \left\{ \varnothing \right\} \right] \right) \times
\Theta \text{.}
\end{equation*}%
Recall%
\begin{equation*}
\Omega ^{\left[ G,S\right] }\equiv \left\{ E\in \left( \times _{i\in 
\mathcal{I}}\left[ 2^{\Theta _{i}}\diagdown \left\{ \varnothing \right\} %
\right] \right) :\exists h\in \left[ \mathcal{H}\diagdown \mathcal{T}\right]
,\text{ }E=\mathcal{E}^{\left[ G,S\right] \text{-}h}\right\} \text{.}
\end{equation*}%
First, if $\theta \notin E$ or $E\notin \Omega ^{\left[ G,S\right] }$, we
have%
\begin{equation*}
\gamma ^{\left( \rho ,f\right) }\left[ E,\text{ }\theta \right] \subset
E=\gamma ^{\left[ G,S\right] }\left[ E,\text{ }\theta \right] \text{.}
\end{equation*}%
From now, we assume $\theta \in E\in \Omega ^{\left[ G,S\right] }$. Consider
any $h\in \left[ \mathcal{H}\diagdown \mathcal{T}\right] $, and we aim to
show%
\begin{equation}
\gamma ^{\left( \rho ,f\right) }\left[ \mathcal{E}^{\left[ G,S\right] \text{-%
}h},\text{ }\theta \right] \subset \gamma ^{\left[ G,S\right] }\left[ 
\mathcal{E}^{\left[ G,S\right] \text{-}h},\text{ }\theta \right] \text{.}
\label{rry1}
\end{equation}%
Inductively, we prove%
\begin{equation}
\gamma ^{\left( \rho ,f\right) \text{-}\left( k\right) }\left[ \mathcal{E}^{%
\left[ G,S\right] \text{-}h},\text{ }\theta \right] \subset \gamma ^{\left[
G,S\right] }\left[ \mathcal{E}^{\left[ G,S\right] \text{-}h},\text{ }\theta %
\right] \text{, }\forall k\in \mathbb{N}\text{,}  \label{rry2}
\end{equation}%
which implies (\ref{rry1}) because of the definition of $\gamma ^{\left(
\rho ,f\right) }$ (see (\ref{kky2b})). With $k=1$, we have%
\begin{equation*}
\gamma ^{\left( \rho ,f\right) \text{-}\left( 1\right) }\left[ \mathcal{E}^{%
\left[ G,S\right] \text{-}h},\text{ }\theta \right] =\left\{ \theta \right\}
\subset \gamma ^{\left[ G,S\right] }\left[ \mathcal{E}^{\left[ G,S\right] 
\text{-}h},\text{ }\theta \right] \text{,}
\end{equation*}%
i.e., (\ref{rry2}) holds. Suppose (\ref{rry2}) holds for any $k=n\in \mathbb{%
N}$. We show (\ref{rry2}) for $k=n+1$. Consider any $\theta ^{\prime }\in
\gamma ^{\left( \rho ,f\right) \text{-}\left( n+1\right) }\left[ \mathcal{E}%
^{\left[ G,S\right] \text{-}h},\text{ }\theta \right] $, and we aim to prove 
$\theta ^{\prime }\in \gamma ^{\left[ G,S\right] }\left[ \mathcal{E}^{\left[
G,S\right] \text{-}h},\text{ }\theta \right] $. By (\ref{kky1}), (\ref{kky4c}%
) and (\ref{kky2}), $\theta ^{\prime }\in \gamma ^{\left( \rho ,f\right) 
\text{-}\left( n+1\right) }\left[ \mathcal{E}^{\left[ G,S\right] \text{-}h},%
\text{ }\theta \right] $ implies existence of $\widetilde{\theta }\in \gamma
^{\left( \rho ,f\right) \text{-}\left( n\right) }\left[ \mathcal{E}^{\left[
G,S\right] \text{-}h},\text{ }\theta \right] $ such that for any $i\in 
\mathcal{I}$, one of the following three conditions holds:%
\begin{eqnarray}
\theta _{i}^{\prime } &=&\widetilde{\theta }_{i}\text{,}  \label{rry3} \\
\rho \left[ f,\left( \gamma ^{\left( \rho ,f\right) \text{-}\left( n\right)
},\mathcal{E}^{\left[ G,S\right] \text{-}h}\right) ,\widetilde{\theta }%
,\theta ^{\prime },i,\mathcal{E}^{\left[ G,S\right] \text{-}h}\right] &=&0%
\text{,}  \label{rry4} \\
\rho \left[ f,\left( \gamma ^{\left( \rho ,f\right) \text{-}\left( n\right)
},\mathcal{E}^{\left[ G,S\right] \text{-}h}\right) ,\theta ^{\prime },%
\widetilde{\theta },i,\mathcal{E}^{\left[ G,S\right] \text{-}h}\right] &=&0%
\text{.}  \label{rry5}
\end{eqnarray}%
Fix any $i\in \mathcal{I}$, we now prove%
\begin{equation}
\theta _{i}^{\prime }\in \gamma _{i}^{\left[ G,S\right] }\left[ \mathcal{E}^{%
\left[ G,S\right] \text{-}h},\text{ }\theta \right] \text{,}  \label{rry66}
\end{equation}%
and as a result, we have $\theta ^{\prime }\in \gamma ^{\left[ G,S\right] }%
\left[ \mathcal{E}^{\left[ G,S\right] \text{-}h},\text{ }\theta \right] $.

By the induction hypothesis, $\widetilde{\theta }\in \gamma ^{\left( \rho
,f\right) \text{-}\left( n\right) }\left[ \mathcal{E}^{\left[ G,S\right] 
\text{-}h},\text{ }\theta \right] $ implies%
\begin{equation}
\widetilde{\theta }\in \gamma ^{\left[ G,S\right] }\left[ \mathcal{E}^{\left[
G,S\right] \text{-}h},\text{ }\theta \right] \text{,}  \label{rry6}
\end{equation}%
which, together with $\gamma ^{\left[ G,S\right] }\left[ \mathcal{E}^{\left[
G,S\right] \text{-}h},\text{ }\cdot \right] $ being a partition on $\mathcal{%
E}^{\left[ G,S\right] \text{-}h}$ (due to $\gamma ^{\left[ G,S\right] }$
being an operator), implies%
\begin{equation}
\gamma ^{\left[ G,S\right] }\left[ \mathcal{E}^{\left[ G,S\right] \text{-}h},%
\text{ }\widetilde{\theta }\right] =\gamma ^{\left[ G,S\right] }\left[ 
\mathcal{E}^{\left[ G,S\right] \text{-}h},\text{ }\theta \right] \text{.}
\label{rry6a}
\end{equation}

First, if (\ref{rry3}) holds, (\ref{rry6}) implies (\ref{rry66}). Second, if
(\ref{rry4}) holds, Lemma \ref{lem:operator:canonic:alternative:increasing}
and the induction hypothesis imply%
\begin{equation*}
\rho \left[ f,\left( \gamma ^{\left[ G,S\right] },\mathcal{E}^{\left[ G,S%
\right] \text{-}h}\right) ,\widetilde{\theta },\theta ^{\prime },i,\mathcal{E%
}^{\left[ G,S\right] \text{-}h}\right] =0\text{,}
\end{equation*}%
which, together with $\rho $ being monotonic, implies%
\begin{equation}
\rho \left[ f,\left( \gamma ^{\left[ G,S\right] },\mathcal{E}^{\left[ G,S%
\right] \text{-}h}\right) ,\widetilde{\theta },\theta ^{\prime },i,\vartheta
^{\left[ G\right] }\left( \mathcal{E}^{\left[ G,S\right] \text{-}h}\right) %
\right] =0\text{.}  \label{rry7}
\end{equation}%
Since $\gamma ^{\left[ G,S\right] }$ is $\left( \rho ,f,\vartheta
^{G}\right) $-consistent, (\ref{rry7}) implies $\theta _{i}^{\prime }\in
\gamma _{i}^{\left[ G,S\right] }\left[ \mathcal{E}^{\left[ G,S\right] \text{-%
}h},\text{ }\widetilde{\theta }\right] $, which, together with (\ref{rry6a}%
), implies (\ref{rry66}). Third, if (\ref{rry5}) holds, a similar argument
as above shows%
\begin{equation*}
\rho \left[ f,\left( \gamma ^{\left[ G,S\right] },\mathcal{E}^{\left[ G,S%
\right] \text{-}h}\right) ,\theta ^{\prime },\widetilde{\theta },i,\vartheta
^{\left[ G\right] }\left( \mathcal{E}^{\left[ G,S\right] \text{-}h}\right) %
\right] =0\text{,}
\end{equation*}%
which, together with (\ref{hht2}) in Definition \ref{def:solution}, implies%
\begin{equation}
\rho \left[ f,\left( \gamma ^{\left[ G,S\right] },\mathcal{E}^{\left[ G,S%
\right] \text{-}h}\right) ,\left( \theta _{i}^{\prime },\widetilde{\theta }%
_{-i}\right) ,\widetilde{\theta },i,\vartheta ^{\left[ G\right] }\left( 
\mathcal{E}^{\left[ G,S\right] \text{-}h}\right) \right] =0\text{.}
\label{rry7a}
\end{equation}%
Since $\gamma ^{\left[ G,S\right] }$ is $\left( \rho ,f,\vartheta
^{G}\right) $-consistent, (\ref{rry7a}) implies 
\begin{equation}
\widetilde{\theta }_{i}\in \gamma _{i}^{\left[ G,S\right] }\left[ \mathcal{E}%
^{\left[ G,S\right] \text{-}h},\text{ }\left( \theta _{i}^{\prime },%
\widetilde{\theta }_{-i}\right) \right] \text{.}  \label{rry7ab}
\end{equation}%
Since $\left( \theta _{i}^{\prime },\widetilde{\theta }_{-i}\right) \in
\gamma ^{\left[ G,S\right] }\left[ \mathcal{E}^{\left[ G,S\right] \text{-}h},%
\text{ }\left( \theta _{i}^{\prime },\widetilde{\theta }_{-i}\right) \right] 
$, we have%
\begin{equation}
\widetilde{\theta }_{j}\in \gamma _{j}^{\left[ G,S\right] }\left[ \mathcal{E}%
^{\left[ G,S\right] \text{-}h},\text{ }\left( \theta _{i}^{\prime },%
\widetilde{\theta }_{-i}\right) \right] \text{, }\forall j\in \mathcal{I}%
\diagdown \left\{ i\right\} \text{.}  \label{rry7ac}
\end{equation}%
(\ref{rry7ab}) and (\ref{rry7ac}) imply 
\begin{equation}
\widetilde{\theta }\in \gamma ^{\left[ G,S\right] }\left[ \mathcal{E}^{\left[
G,S\right] \text{-}h},\text{ }\left( \theta _{i}^{\prime },\widetilde{\theta 
}_{-i}\right) \right] \text{.}  \label{rry7ad}
\end{equation}%
We thus have 
\begin{equation}
\left( \theta _{i}^{\prime },\widetilde{\theta }_{-i}\right) \in \gamma ^{%
\left[ G,S\right] }\left[ \mathcal{E}^{\left[ G,S\right] \text{-}h},\text{ }%
\left( \theta _{i}^{\prime },\widetilde{\theta }_{-i}\right) \right] =\gamma
^{\left[ G,S\right] }\left[ \mathcal{E}^{\left[ G,S\right] \text{-}h},\text{ 
}\widetilde{\theta }\right] =\gamma ^{\left[ G,S\right] }\left[ \mathcal{E}^{%
\left[ G,S\right] \text{-}h},\text{ }\theta \right] \text{,}  \label{rry7ae}
\end{equation}%
where $\in $ follows from $\gamma ^{\left[ G,S\right] }\left[ \mathcal{E}^{%
\left[ G,S\right] \text{-}h},\text{ }\cdot \right] $ being a partition on $%
\mathcal{E}^{\left[ G,S\right] \text{-}h}$, and the first equality follows
from (\ref{rry7ad}) and $\gamma ^{\left[ G,S\right] }\left[ \mathcal{E}^{%
\left[ G,S\right] \text{-}h},\text{ }\cdot \right] $ being a partition on $%
\mathcal{E}^{\left[ G,S\right] \text{-}h}$, and the second equality follows
from (\ref{rry6a}). Finally, (\ref{rry7ae}) implies (\ref{rry66}).$%
\blacksquare $

\subsection{Proofs of Lemma \protect\ref{lem:operator:game-operator}}

\label{sec:lem:operator:game-operator}

Consider any SCF $f$ and any solution notion $\rho $ which is dissectible.
Suppose $\gamma ^{\left( \rho ,f\right) }$ is achievable, i.e., there exists 
$N\in \mathbb{N}$ such that%
\begin{equation}
\gamma _{\left( N\right) }^{\left( \rho ,f\right) }\left[ \Theta \text{ }%
,\theta \right] =\left\{ \theta \right\} \text{, }\forall \theta \in \Theta 
\text{. }  \label{kkt4}
\end{equation}%
For any $\theta \in \Theta $ and any $n\in \mathbb{N}$, recall%
\begin{eqnarray*}
\gamma _{\left( 0\right) }^{\left( \rho ,f\right) }\left[ \Theta \text{ }%
,\theta \right]  &=&\Theta \text{,} \\
\gamma _{\left( n\right) }^{\left( \rho ,f\right) }\left[ \Theta \text{ }%
,\theta \right]  &=&\gamma ^{\left( \rho ,f\right) }\left[ \gamma _{\left(
n-1\right) }^{\left( \rho ,f\right) }\left[ \Theta \text{ },\theta \right]
,\theta \right] \text{.}
\end{eqnarray*}%
For any $\theta \in \Theta $ and any $n\in \mathbb{N}$, define%
\begin{equation*}
\gamma _{\left( n\right) \text{-}i}^{\left( \rho ,f\right) }\left[ \Theta 
\text{ },\theta \right] =\gamma _{i}^{\left( \rho ,f\right) }\left[ \gamma
_{\left( n-1\right) }^{\left( \rho ,f\right) }\left[ \Theta \text{ },\theta %
\right] ,\theta \right] \text{, }\forall \theta \in \Theta \text{, }\forall
n\in \mathbb{N}\text{.}
\end{equation*}%
and%
\begin{eqnarray}
\mathcal{I}_{\left( \theta ,n\right) } &=&\left\{ i\in \mathcal{I}:%
\begin{tabular}{l}
$\exists \theta ^{\prime }\in \gamma _{\left( n-1\right) }^{\left( \rho
,f\right) }\left[ \Theta \text{ },\theta \right] $, \\ 
$\gamma _{\left( n\right) \text{-}i}^{\left( \rho ,f\right) }\left[ \Theta 
\text{ },\theta ^{\prime }\right] \neq \gamma _{\left( n\right) \text{-}%
i}^{\left( \rho ,f\right) }\left[ \Theta \text{ },\theta \right] $%
\end{tabular}%
\right\} \text{,}  \notag \\
\mathcal{I}\diagdown \mathcal{I}_{\left( \theta ,n\right) } &=&\left\{ i\in 
\mathcal{I}:%
\begin{tabular}{l}
$\forall \theta ^{\prime }\in \gamma _{\left( n-1\right) }^{\left( \rho
,f\right) }\left[ \Theta \text{ },\theta \right] $, \\ 
$\gamma _{\left( n\right) \text{-}i}^{\left( \rho ,f\right) }\left[ \Theta 
\text{ },\theta ^{\prime }\right] =\gamma _{\left( n\right) \text{-}%
i}^{\left( \rho ,f\right) }\left[ \Theta \text{ },\theta \right] $%
\end{tabular}%
\right\} \text{,}  \notag \\
N_{\left( \theta \right) } &=&\max \left\{ n\in \mathbb{N}:\mathcal{I}%
_{\left( \theta ,n\right) }\neq \varnothing \right\} \text{.}  \label{hhi1}
\end{eqnarray}%
Since $\rho $ is dissectible, Lemma \ref{lem:operator:canonical:operator}
implies that $\gamma ^{\left( \rho ,f\right) }$ is an operator, i.e., given
any $E$, every $\gamma _{i}^{\left( \rho ,f\right) }\left[ E,\cdot \right] $
is a partition on $E_{i}$. The set $\mathcal{I}_{\left( \theta ,n\right) }$
includes all of the player $i$ such that $\gamma _{i}^{\left( \rho ,f\right)
}\left[ \gamma _{\left( n-1\right) }^{\left( \rho ,f\right) }\left[ \Theta 
\text{ },\theta \right] ,\cdot \right] $ is a non-trivial partition of $%
\gamma _{\left( n-1\right) }^{\left( \rho ,f\right) }\left[ \Theta \text{ }%
,\theta \right] $, i.e., player $i$ reveals non-trivial information at the $n
$-th round. The set $\mathcal{I}\diagdown \mathcal{I}_{\left( \theta
,n\right) }$ includes all of the players who reveal no additional
information at the $n$-th round.

We are ready to define $\left[ G,S\right] \in \mathcal{GS}^{\text{PC}}$ such
that $\gamma ^{\left[ G,S\right] }$ is both achievable and $\left( \rho
,f,\vartheta ^{\left[ G\right] }\right) $-consistent. In $G$, there are at
most $N$ rounds. At each round, we invite a group of players to partially
disclose their types simultaneously, and the disclosure is made public at
the end of each round.

At the first period, we invite players in $\mathcal{I}_{\left( \theta
,1\right) }$, and each $i\in \mathcal{I}_{\left( \theta ,1\right) }$ chooses
one element in the following partition of $\Theta _{i}$:%
\begin{equation*}
\left\{ \gamma _{i}^{\left( \rho ,f\right) }\left[ \Theta ,\theta \right]
:\theta \in \Theta \right\} \text{.}
\end{equation*}%
$S$ denotes the strategy of (partially) truth revealing, i.e., at the true
state $\theta $, player $i\in \mathcal{I}_{\left( \theta ,1\right) }$
chooses $\gamma _{i}^{\left( \rho ,f\right) }\left[ \Theta ,\theta \right] $
in the first round. By following $S$, players disclose that the true state
is in the following set at the end of the first round:%
\begin{eqnarray*}
\gamma _{\left( 1\right) }^{\left( \rho ,f\right) }\left[ \Theta \text{ }%
,\theta \right]  &=&\left( \times _{i\in \mathcal{I}_{\left( \theta
,1\right) }}\gamma _{i}^{\left( \rho ,f\right) }\left[ \Theta ,\theta \right]
\right) \times \left( \times _{i\in \mathcal{I}\diagdown \mathcal{I}_{\left(
\theta ,1\right) }}\gamma _{i}^{\left( \rho ,f\right) }\left[ \Theta ,\theta %
\right] \right)  \\
&=&\left( \times _{i\in \mathcal{I}_{\left( \theta ,1\right) }}\gamma
_{i}^{\left( \rho ,f\right) }\left[ \Theta ,\theta \right] \right) \times
\left( \times _{i\in \mathcal{I}\diagdown \mathcal{I}_{\left( \theta
,1\right) }}\Theta _{i}\right) \text{.}
\end{eqnarray*}%
Inductively, suppose that players have disclosed $\gamma _{\left( k\right)
}^{\left( \rho ,f\right) }\left[ \Theta,\theta \right] $ at the
end of $k$-th round with $k\leq N$. If either $k=N$ or $\mathcal{I}_{\left(
\theta ,k+1\right) }=\varnothing $, the game ends. If $k<N$ and $\mathcal{I}%
_{\left( \theta ,k+1\right) }\neq \varnothing $, we proceed to the $\left(
k+1\right) $-th round. We invite players in $\mathcal{I}_{\left( \theta
,k+1\right) }$ and each $i\in \mathcal{I}_{\left( \theta ,k+1\right) }$
chooses one element in the following partition of $\gamma _{\left( k\right)
}^{\left( \rho ,f\right) }\left[ \Theta \text{ },\theta \right] $:%
\begin{equation*}
\left\{ \gamma _{i}^{\left( \rho ,f\right) }\left[ \gamma _{\left( k\right)
}^{\left( \rho ,f\right) }\left[ \Theta \text{ },\theta \right] ,\text{ }%
\theta \right] :\theta \in \gamma _{\left( k\right) }^{\left( \rho ,f\right)
}\left[ \Theta \text{ },\theta \right] \right\} \text{.}
\end{equation*}%
Recall that $S$ denotes the strategy of (partially) truth revealing, i.e.,
at the true state $\theta $, player $i\in \mathcal{I}_{\left( \theta
,k+1\right) }$ chooses $\gamma _{i}^{\left( \rho ,f\right) }\left[ \gamma
_{\left( k\right) }^{\left( \rho ,f\right) }\left[ \Theta \text{ },\theta %
\right] ,\text{ }\theta \right] $ in the $\left( k+1\right) $-th round. By
following $S$, players disclose that the state is in the following set at
the end of the $\left( k+1\right) $-th round:%
\begin{eqnarray*}
\gamma _{\left( k+1\right) }^{\left( \rho ,f\right) }\left[ \Theta \text{ }%
,\theta \right]  &=&\left( \times _{i\in \mathcal{I}_{\left( \theta
,k+1\right) }}\gamma _{i}^{\left( \rho ,f\right) }\left[ \gamma _{\left(
k\right) }^{\left( \rho ,f\right) }\left[ \Theta \text{ },\theta \right] ,%
\text{ }\theta \right] \right) \times \left( \times _{i\in \mathcal{I}%
\diagdown \mathcal{I}_{\left( \theta ,k+1\right) }}\gamma _{i}^{\left( \rho
,f\right) }\left[ \gamma _{\left( k\right) }^{\left( \rho ,f\right) }\left[
\Theta \text{ },\theta \right] ,\text{ }\theta \right] \right)  \\
&=&\left( \times _{i\in \mathcal{I}_{\left( \theta ,k+1\right) }}\gamma
_{i}^{\left( \rho ,f\right) }\left[ \gamma _{\left( k\right) }^{\left( \rho
,f\right) }\left[ \Theta \text{ },\theta \right] ,\text{ }\theta \right]
\right) \times \left( \times _{i\in \mathcal{I}\diagdown \mathcal{I}_{\left(
\theta ,k+1\right) }}\gamma _{\left( k\right) \text{-}i}^{\left( \rho
,f\right) }\left[ \Theta \text{ },\theta \right] \right) \text{.}
\end{eqnarray*}

At each round, player move simultaneously, and as usual, we pick any order
of the players and translate it into a traditional extensive-form game
(i.e., players move sequentially following the order, but does not observe
previous moves in this round).

We will show that $\gamma ^{\left[ G,S\right] }$ is both achievable and $%
\left( \rho ,f,\vartheta ^{\left[ G\right] }\right) $-consistent. Consider
any true state $\theta \in \Theta $. By following $S$, we reach the
following history:%
\begin{equation*}
T^{G}\left[ S\left( \theta \right) \right] =\left( a^{1},...,a^{n^{\theta
}}\right) \text{, where }n^{\theta }=\sum\limits_{n=1}^{N_{\left( \theta
\right) }}\left\vert \mathcal{I}_{\left( \theta ,n\right) }\right\vert
<N\times \left\vert \mathcal{I}\right\vert \text{,}
\end{equation*}%
where $N_{\left( \theta \right) }$ (as defined in (\ref{hhi1})) is the
largest $k$ such that $\gamma ^{\left( \rho ,f\right) }\left[ \gamma
_{\left( k-1\right) }^{\left( \rho ,f\right) }\left[ \Theta \text{ },\theta %
\right] ,\cdot \right] $ is a non-trivial partition, which implies%
\begin{equation}
\gamma _{\left( n\right) }^{\left( \rho ,f\right) }\left[ \Theta \text{ }%
,\cdot \right] =\gamma _{\left( N_{\left( \theta \right) }\right) }^{\left(
\rho ,f\right) }\left[ \Theta \text{ },\cdot \right] \text{, }\forall n\geq
N_{\left( \theta \right) }\text{.}  \label{hhi2}
\end{equation}%
In particular, for each $k\leq N_{\left( \theta \right) }$, 
\begin{equation*}
h_{\left( k\right) }=\left( a^{1},...,a^{k^{\theta }}\right) \in \text{path}%
\left( T^{G}\left[ S\left( \theta \right) \right] \right) \text{, where }%
k^{\theta }=\sum\limits_{n=1}^{k}\left\vert \mathcal{I}_{\left( \theta
,n\right) }\right\vert \text{,}
\end{equation*}%
is the history at the end of $k$-th round, and we have 
\begin{equation*}
\mathcal{E}^{\left[ G,S\right] \text{-}h_{\left( k\right) }}=\gamma _{\left(
k\right) }^{\left( \rho ,f\right) }\left[ \Theta \text{ },\theta \right] 
\text{, }\forall k\leq N_{\left( \theta \right) }\text{,}
\end{equation*}%
By (\ref{kkt4}), we have $N_{\left( \theta \right) }\leq N$, and in
particular, we have%
\begin{equation*}
\gamma _{\left( n^{\theta }\right) }^{\left[ G,S\right] }\left[ \Theta \text{
},\theta \right] =\mathcal{E}^{\left[ G,S\right] \text{-}T^{G}\left[ S\left(
\theta \right) \right] }=\mathcal{E}^{\left[ G,S\right] \text{-}h_{\left(
N_{\left( \theta \right) }\right) }}=\gamma _{\left( N_{\left( \theta
\right) }\right) }^{\left( \rho ,f\right) }\left[ \Theta \text{ },\theta %
\right] \text{,}
\end{equation*}%
which, together with (\ref{kkt4}), (\ref{hhi2}) and $N_{\left( \theta
\right) }\leq N$, implies%
\begin{equation*}
\gamma _{\left( n^{\theta }\right) }^{\left[ G,S\right] }\left[ \Theta \text{
},\theta \right] =\gamma _{\left( N_{\left( \theta \right) }\right)
}^{\left( \rho ,f\right) }\left[ \Theta \text{ },\theta \right] =\gamma
_{\left( N\right) }^{\left( \rho ,f\right) }\left[ \Theta \text{ },\theta %
\right] =\left\{ \theta \right\} \text{.}
\end{equation*}%
Since $n^{\theta }=\sum\limits_{n=1}^{N_{\left( \theta \right) }}\left\vert 
\mathcal{I}_{\left( \theta ,n\right) }\right\vert <N\times \left\vert 
\mathcal{I}\right\vert $ for every $\theta \in \Theta $, we thus have%
\begin{equation*}
\gamma _{\left( N\times \left\vert \mathcal{I}\right\vert \right) }^{\left[
G,S\right] }\left[ \Theta \text{ },\theta \right] =\left\{ \theta \right\} 
\text{, }\forall \theta \in \Theta \text{,}
\end{equation*}%
i.e., $\gamma ^{\left[ G,S\right] }$ is achievable.

Finally, we show $\gamma ^{\left[ G,S\right] }$ is $\left( \rho ,f,\vartheta
^{\left[ G\right] }\right) $-consistent. Consider any non-terminal history $h
$. Suppose that $h$ is player $j$'s history at the $k$-th round.
Furthermore,\ suppose that players have disclosed $\gamma _{\left(
k-1\right) }^{\left( \rho ,f\right) }\left[ \Theta \text{ },\theta \right] $
at the end of $\left( k-1\right) $-th round for some $\theta \in \Theta $.
We thus have%
\begin{eqnarray}
\gamma _{j}^{\left[ G,S\right] }\left[ \mathcal{E}^{\left[ G,S\right] \text{-%
}h},\theta \right]  &=&\gamma _{j}^{\left( \rho ,f\right) }\left[ \gamma
_{\left( k-1\right) }^{\left( \rho ,f\right) }\left[ \Theta \text{ },\theta %
\right] ,\theta \right] \text{,}  \label{kkt5} \\
\vartheta ^{\left[ G\right] }\left( \mathcal{E}^{\left[ G,S\right] \text{-}%
h}\right)  &=&\gamma _{\left( k-1\right) }^{\left( \rho ,f\right) }\left[
\Theta \text{ },\theta \right] \text{.}  \label{kkt6}
\end{eqnarray}%
We now prove%
\begin{equation}
\left( 
\begin{array}{c}
\left\{ \theta ,\theta ^{\prime }\right\} \subset \mathcal{E}^{\left[ G,S%
\right] \text{-}h}\text{,} \\ 
\rho \left[ f,\left( \gamma ^{\left[ G,S\right] },\mathcal{E}^{\left[ G,S%
\right] \text{-}h}\right) ,\theta ,\theta ^{\prime },j,\vartheta ^{\left[ G%
\right] }\left( \mathcal{E}^{\left[ G,S\right] \text{-}h}\right) \right] =0%
\end{array}%
\right) \Longrightarrow \theta _{j}^{\prime }\in \gamma _{j}^{\left[ G,S%
\right] }\left[ \mathcal{E}^{\left[ G,S\right] \text{-}h},\theta \right] 
\text{,}  \label{kkt7}
\end{equation}%
i.e., $\gamma ^{\left[ G,S\right] }$ is $\left( \rho ,f,\vartheta
^{G}\right) $-consistent. Suppose $\left\{ \theta ,\theta ^{\prime }\right\}
\subset \mathcal{E}^{\left[ G,S\right] \text{-}h}$ and 
\begin{equation}
\rho \left[ f,\left( \gamma ^{\left[ G,S\right] },\mathcal{E}^{\left[ G,S%
\right] \text{-}h}\right) ,\theta ,\theta ^{\prime },j,\vartheta ^{\left[ G%
\right] }\left( \mathcal{E}^{\left[ G,S\right] \text{-}h}\right) \right] =0%
\text{.}  \label{kkt8}
\end{equation}%
We thus have%
\begin{eqnarray}
&&\rho \left[ f,\left( \gamma ^{\left( \rho ,f\right) },\gamma _{\left(
k-1\right) }^{\left( \rho ,f\right) }\left[ \Theta \text{ },\theta \right]
\right) ,\theta ,\theta ^{\prime },j,\gamma _{\left( k-1\right) }^{\left(
\rho ,f\right) }\left[ \Theta \text{ },\theta \right] \right]   \label{kkt9}
\\
&=&\rho \left[ f,\left( \gamma ^{\left( \rho ,f\right) },\gamma _{\left(
k-1\right) }^{\left( \rho ,f\right) }\left[ \Theta \text{ },\theta \right]
\right) ,\theta ,\theta ^{\prime },j,\vartheta ^{\left[ G\right] }\left( 
\mathcal{E}^{\left[ G,S\right] \text{-}h}\right) \right]   \notag \\
&=&\rho \left[ f,\left( \gamma ^{\left[ G,S\right] },\mathcal{E}^{\left[ G,S%
\right] \text{-}h}\right) ,\theta ,\theta ^{\prime },j,\vartheta ^{\left[ G%
\right] }\left( \mathcal{E}^{\left[ G,S\right] \text{-}h}\right) \right]  
\notag \\
&=&0\text{,}  \notag
\end{eqnarray}%
where the first equality follows from (\ref{kkt6}), and the second equality
follows from (\ref{kkt5}) and (\ref{hht2}) in Definition \ref{def:solution},
and the third equality follows from (\ref{kkt8}). By Lemma \ref%
{lem:operator:canonical:operator}, $\gamma ^{\left( \rho ,f\right) }$ is $%
\left( \rho ,f,\vartheta ^{\ast }\right) $-consistent, which, together with (%
\ref{kkt9}), implies%
\begin{equation}
\theta _{j}^{\prime }\in \gamma _{j}^{\left( \rho ,f\right) }\left[ \gamma
_{\left( k-1\right) }^{\left( \rho ,f\right) }\left[ \Theta \text{ },\theta %
\right] ,\theta \right] \text{.}  \label{kkt10}
\end{equation}%
(\ref{kkt5}) and (\ref{kkt10}) imply $\theta _{j}^{\prime }\in \gamma _{j}^{%
\left[ G,S\right] }\left[ \mathcal{E}^{\left[ G,S\right] \text{-}h},\theta %
\right] $, i.e., (\ref{kkt7}) holds.$\blacksquare $

\subsection{Proofs of Lemma \protect\ref{lem:operator:normal}}

\label{sec:lem:operator:normal}

Consider any SCF $f$ and any normal solution notion $\rho $. The "only if"
part of Lemma \ref{lem:operator:normal} is implied by Definition \ref%
{def:solution:achieve}. To prove the "if" part, suppose $\gamma ^{\left(
\rho ,f\right) }$ is $f$-achievable. i.e., there exists $N\in \mathbb{N}$,%
\begin{equation}
f\left( \widetilde{\theta }\right) =f\left( \theta \right) \text{, }\forall
\theta \in \Theta \text{, }\forall \widetilde{\theta }\in \gamma _{\left(
N\right) }^{\left( \rho ,f\right) }\left[ \Theta \text{ },\theta \right] 
\text{, }  \label{oot1}
\end{equation}%
and we aim to $\gamma ^{\left( \rho ,f\right) }$ is achievable. Since $\rho $
is normal, (\ref{oot1}) and Definition \ref{def:solution:constant} imply%
\begin{equation*}
\rho \left[ f,\left( \gamma ^{\left( \rho ,f\right) },\gamma _{\left(
N\right) }^{\left( \rho ,f\right) }\left[ \Theta \text{ },\theta \right]
\right) ,\theta ,\theta ^{\prime },i,\gamma _{\left( N\right) }^{\left( \rho
,f\right) }\left[ \Theta \text{ },\theta \right] \right] =1\text{, }\forall
\theta \in \Theta \text{, }\forall \theta ^{\prime }\in \gamma _{\left(
N\right) }^{\left( \rho ,f\right) }\left[ \Theta \text{ },\theta \right] 
\text{,}
\end{equation*}%
which, together with (\ref{kky7}) and (\ref{kky2}), implies%
\begin{equation*}
\gamma ^{\left( \rho ,f\right) \text{-}\left( n\right) }\left[ \gamma
_{\left( N\right) }^{\left( \rho ,f\right) }\left[ \Theta \text{ },\theta %
\right] ,\text{ }\theta \right] =\left\{ \theta \right\} \text{, }\forall
\theta \in \Theta \text{, }\forall n\in \mathbb{N}\text{, }
\end{equation*}%
and hence,%
\begin{equation*}
\gamma ^{\left( \rho ,f\right) }\left[ \gamma _{\left( N\right) }^{\left(
\rho ,f\right) }\left[ \Theta \text{ },\theta \right] ,\text{ }\theta \right]
=\cup_{n=1}^{\infty }\left( \gamma ^{\left( \rho ,f\right) \text{-%
}\left( n\right) }\left[ \gamma _{\left( N\right) }^{\left( \rho ,f\right) }%
\left[ \Theta \text{ },\theta \right] ,\text{ }\theta \right] \right)
=\left\{ \theta \right\} \text{, }\forall \theta \in \Theta \text{.}
\end{equation*}%
Therefore, 
\begin{equation*}
\gamma _{\left( N+1\right) }^{\left( \rho ,f\right) }\left[ \Theta \text{ }%
,\theta \right] =\gamma ^{\left( \rho ,f\right) }\left[ \gamma _{\left(
N\right) }^{\left( \rho ,f\right) }\left[ \Theta \text{ },\theta \right] ,%
\text{ }\theta \right] =\left\{ \theta \right\} ,\forall \theta \in \Theta 
\text{,}
\end{equation*}%
i.e., $\gamma ^{\left( \rho ,f\right) }$ is achievable.$\blacksquare $

\bibliographystyle{ier}
\bibliography{OSP}

\end{document}